\documentclass[12pt,english]{article}
\usepackage{setspace}
\usepackage[margin=1.25in]{geometry}

\usepackage{lmodern}
\usepackage[T1]{fontenc}
\usepackage[utf8]{inputenc}

\usepackage{color}
\definecolor{document_fontcolor}{rgb}{0, 0, 0}
\color{document_fontcolor}
\usepackage{babel}
\usepackage{amsmath}
\usepackage{dsfont}
\usepackage{xcolor}
\usepackage{amssymb}
\usepackage{graphicx}
\usepackage{setspace}
\usepackage[pdftitle={Identification and Estimation in Many-to-one Two-sided Matching without Transfers},
pdfauthor={YingHua He, Shruti Sinha, and Xiaoting Sun},
colorlinks=true,linkcolor=black,citecolor=black,urlcolor=black]{hyperref}
\usepackage{verbatim}  

\PassOptionsToPackage{hyphens}{url}

\usepackage{booktabs}
\newenvironment{tabnotes}[2][1]{\begin{minipage}[t]{#1\textwidth}\vspace{0.1cm}\scriptsize{\emph{Notes:} #2}}{\end{minipage}}

\usepackage{bbm}

\newcommand{\bfI}{\mathbf{I}}
\newcommand{\bfC}{\mathbf{C}}

\usepackage{booktabs} 
\usepackage{threeparttable} 
\usepackage{multirow}
\usepackage{float}
\usepackage{caption}

\usepackage{comment}

\definecolor{maroon}{RGB}{128,0,0}

\makeatletter

\usepackage{amsthm}
\usepackage[capitalise]{cleveref}

\theoremstyle{plain}
\newtheorem{thm}{Theorem}[section]
  \theoremstyle{definition}

\theoremstyle{plain}
\newtheorem{prop}[thm]{\protect\propositionname}
\theoremstyle{plain}

\theoremstyle{plain}
\newtheorem{cor}[thm]{\protect\corollaryname}
\theoremstyle{plain}
\newtheorem{assm}[thm]{\protect\assumptionname}
\theoremstyle{plain}
\newtheorem{cond}[thm]{\protect\conditionname}
\theoremstyle{plain}
\newtheorem{example}[thm]{\protect\examplename}
\theoremstyle{plain}
\newtheorem{remark}[thm]{\protect\remarkname}

\usepackage{babel}
\providecommand{\corollaryname}{Corollary}
\providecommand{\definitionname}{Definition}
\providecommand{\lemmaname}{Lemma}
\providecommand{\propositionname}{Proposition}

\providecommand{\assumptionname}{Assumption}
\providecommand{\conditionname}{Condition}
\providecommand{\examplename}{Example}
\providecommand{\remarkname}{Remark}

\ifx\proof\undefined\
\newenvironment{proof}[1][\proofname]{\par
	\normalfont\topsep6\p@\@plus6\p@\relax
	\trivlist
	\itemindent\parindent
	\item[\hskip\labelsep
	\scshape
	#1]\ignorespaces
}{%
	\endtrivlist\@endpefalse
}
\providecommand{\proofname}{Proof}
\fi

\usepackage{mathtools}

\usepackage{enumitem}

\usepackage{pgfplots}
\usepackage{tikz}
\pgfplotsset{width=10cm, ticks=none}
\usetikzlibrary{shapes.arrows}
\usetikzlibrary{arrows}
\usetikzlibrary{patterns}
\usepgflibrary{patterns}
\usetikzlibrary{datavisualization.formats.functions}

\usepgfplotslibrary{fillbetween}
\usepackage{graphicx}
\usepackage{subcaption}
\usepackage{mwe}

\usepackage{appendix}

\usepackage{float}

\usepackage{natbib}

\setcitestyle{authoryear,round} 

\usepackage{tabularx}
\everypar{\looseness=-1}

\expandafter\def\expandafter\normalsize\expandafter{%
	\normalsize
	\setlength\abovedisplayskip{2.5pt}
	\setlength\belowdisplayskip{2.5pt}
	\setlength\abovedisplayshortskip{-1pt}
	\setlength\belowdisplayshortskip{-1pt}
}

\newenvironment{mfignotesin}[4][1]{
	\begin{figure}[!ht]\begin{center}
			\begin{minipage}{#1\textwidth}\begin{center}#3\caption{#2}\end{center}\end{minipage}
			\if!#4!\empty \else \\
			\begin{footnotesize}\begin{minipage}{\textwidth}\scriptsize{\medskip\par
						\emph{Notes:} #4}\end{minipage}\end{footnotesize} \fi }
		{\end{center}\end{figure}}

\usepackage{pgfplots}
\usepackage{tikz}
\pgfplotsset{width=10cm, ticks=none}
\usetikzlibrary{shapes.arrows}
\usetikzlibrary{arrows}
\usetikzlibrary{patterns}
\usepgflibrary{patterns}
\usetikzlibrary{datavisualization.formats.functions}

\usepgfplotslibrary{fillbetween}
\usepackage{graphicx}
\usepackage{subcaption}
\usepackage{mwe}

\begin{document}

\title{{Identification and Estimation in Many-to-one Two-sided Matching without Transfers\thanks{\scriptsize This paper subsumes Sun's job market paper ``Identification and Estimation of Many-to-One Matching with an Application to the U.S. College Admissions'' (first draft: 2018) and He and Sinha's previously circulated paper ``Identification and Estimation in Many-to-One Two-Sided Matching without Transfers'' (first draft: 2020). We thank Nikhil Agarwal, Cristina Gualdani, Thierry Magnac, Rosa Matzkin, Ismael Mourifie, Shuyang Sheng, and Xun Tang, for their useful comments. We would also like to thank the seminar and conference participants at 2018 California Econometrics Conference, 2018 Midwest Econometrics Group Conference, 2019 Canadian Econometric Study Group Meetings, 2019 Network Econometrics Junior's conference at Northwestern University, 2019 Seattle-Vancouver Econometrics Conference, Auburn University,  Econometric Society World Congress 2020, EEA-ESEM 2019, ICEF Moscow, Jinan University, Johns Hopkins University, National University of Singapore, Peking University, Rice University, Simon Fraser University, Toulouse School of Economics, UCLA, UC Riverside, University of Florida, University of Toronto, UNSW Sydney, and Western University. Thanks also to Al\'ipio Ferreira Cantisani for excellent research assistance. Sinha acknowledges funding from the French National Research Agency (ANR) under the Investments for the Future (Investissements d'Avenir) program, grant ANR-17-EURE-0010.}}}
\vspace{-2mm}
\author{YingHua He\thanks{\scriptsize Email: \href{mailto:yinghua.he@rice.edu}{yinghua.he@rice.edu}, Rice University, Houston, Texas, USA.} \text{ } \text{ } Shruti Sinha\thanks{\scriptsize Email: \href{mailto:shruti.sinha@tse-fr.eu}{shruti.sinha@tse-fr.eu}, Toulouse School of Economics, University of Toulouse Capitole, Toulouse, France.} \text{ } \text{ } Xiaoting Sun\thanks{\scriptsize Email: \href{mailto:xiaoting\_sun@sfu.ca}{xiaoting\_sun@sfu.ca}, Simon Fraser University, Vancouver, Canada.}}

\setstretch{1.05}
\date{\today}

\maketitle

\begin{abstract}
In a setting of many-to-one two-sided matching with non-transferable utilities, e.g., college admissions, we study conditions under which preferences of both sides are identified with data on one single market. Regardless of whether the market is centralized or decentralized, assuming that the observed matching is stable, we show nonparametric identification of preferences of both sides under certain  exclusion restrictions. To take our results to the data, we use Monte Carlo simulations to evaluate different estimators, including the ones that are directly constructed from the identification. We find that a parametric Bayesian approach with a Gibbs sampler works well in realistically sized problems. Finally, we illustrate our methodology in decentralized admissions to public and private schools in Chile and conduct a counterfactual analysis of an affirmative action policy.
\vspace{2mm}
\end{abstract}

\setstretch{1.5}
\newpage

\setstretch{1.3}

\section{Introduction}\label{sec:intro}
    In a many-to-one two-sided matching market, agents are categorized into two sides; everyone on one side has preferences over those on the other side; an agent on only one of the two sides can have  multiple match partners from the other side.  
Many real-life markets fit this description, for example,  
the medical resident match \citep{roth1984evolution,agarwal_empirical_2015} in the US, school admissions in Chile \citep{gazmuri2017school} and Hungary \citep{aue2020happens}, college admissions in the US, and graduate program admissions in France \citep{he2020application}. Such markets often exclude personalized transfers, even though limited monetary exchanges may exist.  Hence, the literature defines it as matching without transfers or matching with non-transferable utility. 

While the literature has extensively studied this type of matching theoretically \citep[see, e.g.,][]{roth1992two,azevedo_supply_2016}, its econometrics is less explored. Our paper aims to make a contribution by answering the following questions:  Are the preferences of both sides identified from data on who matches with whom? If so, how can the preferences be estimated? 

To fix ideas, we proceed in the language of college admissions.   We derive a set of sufficient conditions under which both student and college preferences are nonparametrically identified. Our results are obtained from a single market in which there are a continuum of students and a fixed number of colleges. We use that to approximate a single large market. Further, we provide an estimation procedure that is practical even in settings with many agents, allowing for rich observed and unobserved heterogeneity.  Understanding agent preferences is often crucial for policymaking, and one may analyze a wide range of counterfactual policies with estimated preferences. Potentially, our results open a new avenue of research on such matching markets. 
 
The main challenge in identifying student preferences is that each student's {\it actual} choice set is unobservable to the researcher.  For student~$i$ to be able to enroll at college~$c$, college~$c$ needs to accept $i$. The same difficulty exists in the identification of college preferences. Moreover, each student's and each college's choice sets are endogenously determined in equilibrium without  market-clearing prices. 

In our continuum setting, we assume that an observed matching is stable. 
That is, no college prefers to reject any of its currently matched students to vacate a seat, and no student prefers to leave her current match to become unmatched or matched with a college that is willing to accept her and, if necessary, reject one of its currently matched students.  Stability is often imposed in the study of various matching markets \citep[see, for a survey,][]{chiappori_econometrics_2016} and is satisfied in equilibrium in our setting in certain game-theoretical models \citep{ACH2017,fack_beyond_2019}.   

Importantly, there is generically a unique stable matching that is characterized by the colleges' admission cutoffs  \citep{azevedo_supply_2016}.  When college preferences over individual students are represented by utility functions, a college's cutoff is the lowest utility level among its matched students.  Cutoffs further define a student's actual choice set in equilibrium, called {\it feasible set}. A college is in a student's feasible set if the college's utility of being matched with her is higher than its cutoff.  Stability implies that a student is matched with her most-preferred feasible college, similar to a discrete choice problem, except that feasible sets are unobservable and heterogeneous.   

A simple equation, called the $i$-$c$ match probability, is the key to understanding our identification result. Specifically, the conditional probability of student~$i$ being matched with college~$c$ is the sum of conditional probabilities of $i$ choosing $c$ from a given feasible set $L$ weighted by the conditional probability of facing $L$:  
\vskip-0.8cm
\begin{footnotesize}\begin{align*}
		~& \mathbb{{P}}(\text{student } i \text{ is matched with college } c \mid x_{i}) \\
	= \quad & \sum_{\text{all possible feasible sets, } L}
	\underbrace{{\mathbb{{P}}(L\,\text{{is}}\, i \text{{'s feasible}}\,\text{{set}} \mid x_{i})}}_{\equiv A\; \text{(college preferences)}}\cdot\underbrace{{\mathbb{{P}}( c\,\text{{is}}\,i \text{{'s most-preferred college}}\,\text{{in}}\,L \mid L,x_{i})}}_{\equiv B\; \text{(student preferences)}}, 
\end{align*}\end{footnotesize}%
where $x_{i}$ consists of all observed characteristics of student $i$ (e.g., pair-specific characteristics like distance to colleges). The equation provides a decomposition of the preferences of the two sides: for each given $L$, piece $A$ only depends on the preferences of {\it all} colleges given cutoffs, while $B$ only depends on $i$'s preferences over all colleges. 

We then detail a set of exclusion restrictions, among other regularity conditions, such that the excluded variables act as ``demand shifters'' and ``feasible-set shifters'' (or supply shifters). Sufficient variation in these excluded variables identifies the preferences of colleges and students using the $i$-$c$ match probability described above. 

Here are some intuitions. For a college $d$, an $(i,d)$-specific demand shifter traces out how $i$'s preferences for $d$ affects the $i$-$c$ match probability. Similarly, an $(i,d)$-specific feasible-set shifter traces out how $d$'s preference for $i$ affects the $i$-$c$ match probability. A non-excluded variable affects the $i$-$c$ match probability through preferences on both sides for all colleges. By taking derivatives of the $i$-$c$ match probability with respect to (w.r.t.) all the variables, excluded and non-excluded, we derive systems of linear equations that link the effects of variations in demand and supply. Hence, the identification problem reduces to setting up  systems of linear equations and ensuring the existence of a unique solution. 

The second objective of our paper is to provide practical methods that can be used to analyze real-life markets. We achieve this by deriving theoretical guidelines and showcasing a practical estimation method. 

When taken to the data,  the requirement of a large number of excluded variables may be difficult to meet. To address this, we theoretically characterize the tradeoff between exclusion restrictions and the degree of identifiable preference heterogeneity (\cref{deg-of-heter}). The researcher can use this result as a guideline for empirical studies when having insufficient excluded variables.  

We also need a practical estimation method to take these identification results to the data. In fact, our identification arguments are constructive, leading to nonparametric and semiparametric estimators. Monte Carlo simulations suggest that estimating the matrices of partial derivatives in the linear systems using the average derivative estimators of \cite{powell_semiparametric_1989} performs well in finite samples only when the curse of dimensionality is not severe.  In a reasonably sized problem, we resort to a parametric Bayesian approach with a Gibbs sampler \citep{rossi2012bayesian}, resembling applications such as \cite{logan_two-sided_2008} for one-to-one two-sided matching and \cite{abdulkadiroglu_welfare_2017} for a one-sided problem. We demonstrate its good performance in Monte Carlo simulations with high dimensionality. 

As an empirical application, we consider the decentralized admissions to secondary schools in Chile. To the best of our knowledge, this is {one of the first attempts to estimate the preferences of both sides in a {\it decentralized} market of many-to-one two-sided matching without transfers.} There is no clearinghouse, and students do not submit rank-order lists of schools. By allowing flexible preference heterogeneity in the Bayesian approach, we estimate the preferences of students and schools. We also consider a counterfactual policy in which students from low-income families are prioritized for admissions to all schools.  Segregation in terms of ability and income decreases, albeit slightly. We find that simply giving low-income students access to schools may not significantly change matching outcomes due to student preferences.

\paragraph{Related Literature.} This paper is related to the literature on the identification of matching models; \cref{tab:sumlit} provides an incomplete summary. 

\begin{table}[h!]
\caption{Identification Results of Matching Models\label{tab:sumlit}}
\centering\footnotesize
\resizebox{\textwidth}{!}{
\begin{tabular}{p{0.15\textwidth}p{0.47\textwidth}p{0.47\textwidth}}
\toprule 
 & Transferable Utility (TU) & Non-Transferable Utility (NTU) \tabularnewline
\midrule
One-to-one & The match surplus is identified \citep[see, e.g.,][]{choo_who_2006,fox_identification_2010,chiappori2017partner,galichon_cupids_2020}. & The match surplus is identified \citep[see, e.g.,][]{dagsvik_aggregation_2000,menzel_large_2015}. \\
[0.2cm]
Many-to-one   &  The utility function and the
distribution of the unobservables of both sides are identified in a homogeneous setting \citep[see, e.g.,][]{diamond_latent_2017}. &  The utility function and the
distribution of the unobservables of both sides are identified in a homogeneous setting \citep[see, e.g.,][]{diamond_latent_2017}. \\
 &   &  {\it Our paper}:  The utility functions of both sides
with heterogeneity and the distribution of the unobservables are identified.\\
[0.2cm]
Many-to-many &  The match surplus  and/or the distribution of the unobservables are identified \citep[see, e.g.,][]{fox_identification_2010,fox_unobserved_2018}. & The match surplus  is identified \citep[see, e.g.,][]{menzel_strategic_2017}.\\
\bottomrule 
\end{tabular}}
\end{table}

This literature is split into several strands depending on the preference structures of the agents -- transferable utility (TU) and non-transferable utilities (NTU);\footnote{See \cite{galichon2019costly} for an example of imperfectly transferable utility (ITU) models.} and the maximum number of links an agent is permitted to form across sides -- one-to-one, many-to-one, and many-to-many \citep[see][for a survey]{chiappori_econometrics_2016}. 

There is also a close relationship between the one-to-one TU matching model \citep{choo_who_2006,fox_identification_2010,fox_estimating_2018,GRAHAM2011965,sinha_identification_2015,chiappori2017partner,diamond_latent_2017,galichon_cupids_2020,gualdani_identification_2020} and the many-to-one NTU matching model considered here. Market-clearing college cutoffs in our setting play the role of market-clearing shadow prices, although the endogenous cutoffs do not determine how the surplus is split among the agents.

Most of the work on identification within the NTU framework focuses on one-to-one markets \citep{dagsvik_aggregation_2000,menzel_large_2015,uetake2020entry}. Allowing for infinitely many agents on {\it both} sides of a many-to-one matching market, {\cite{agarwal_empirical_2015}} and \cite{diamond_latent_2017} prove identification under a homogeneity restriction on the preferences,\footnote{When studying the medical resident match, \cite{agarwal_empirical_2015} discusses some intuitions of using exclusion restrictions to identify heterogeneous preferences on each side of the market.}  and \cite{ederer2022} shows identification by relaxing the homogeneity assumption but still restricting preferences. \cite{as_2020} provide a recent survey on empirical models of NTU matching. 

Many-to-one NTU matching has been empirically studied in the context of secondary school admissions in Hungary \citep{aue2020happens} and graduate program admissions in France \citep{he2020application}. Their data include information on the preferences of both sides that is reported to a centralized mechanism. Therefore, they can independently identify and estimate the preferences of each side, essentially reducing the two-sided matching to two separate one-sided problems. 

Centralized many-to-one NTU matching in the context of school choice has been studied extensively, both theoretically since \cite{abdulkadiroglu_school_2003} and empirically \citep[e.g.,][]{abdulkadiroglu_welfare_2017,agarwal_demand_2018,calsamiglia2020structural,fack_beyond_2019,he_gaming_2017,kapor2020heterogeneous}. 
In this literature, school preferences are (assumed to be) known because schools rank students according to certain pre-specified rules. The problem then reduces to identifying and estimating student preferences. See  \cite{agarwal2020revealed} for a survey. 


Feasible sets in our setting resemble endogenous consideration sets that arise in one-sided decision problems. In this sense, our paper relates to the growing strand of literature that studies the econometrics of decision problems under consideration set formation \citep[e.g.,][]{abaluck_what_2018, barseghyan_heterogeneous_2019, barseghyan2021discrete,cattaneo2020random}. Our contribution here is that we provide a structural two-sided setting where the consideration probabilities in a student's decision problem are entirely determined by the college (supply side) preferences. Along similar lines to ours, \cite{agarwal2022demand} study consumer choice models with latent choice-set constraints. Their identification conditions and ours are non-nested (see Section~\ref{id}).

The remaining paper is organized as follows: \cref{model} describes
the model and data generating process; \cref{id} discusses the identification of preferences of the agents on both sides of the market; \cref{sec:empirical} illustrates an empirical analysis of the match between students and secondary schools in Chile; and \cref{sec:conclusion} concludes. 

\section{Model}\label{model}
    For the sake of exposition, our model is set up as a college admissions problem. Consider a single market with a continuum of students and finitely many colleges.  The set of all students is  $\textbf{I}$, with a probability measure $Q$ defined over it,\footnote{Our probability space is $(\mathbf{I},\mathcal{B}(\mathbf{I}),Q)$ where $\mathcal{B}(\mathbf{I})$ is the Borel set of $\mathbf{I}$, $Q:\mathcal{B}(\mathbf{I})\rightarrow[0,1]$, $Q(\mathbf{I})=1$.} and   
the set of all colleges is $\mathbf{C}=\{1,2,\dots,C\}$. College $c \in \mathbf{C}$ has a capacity $q_{c}\in (0,1)$. 

For $i\in\mathbf{I}$, the utility of being matched with college $c$ is $u_{ic}$; for $c\in\mathbf{C}$,  the utility of being matched with $i$ is $v_{ci}$. To prepare for  our identification results in \cref{id}, we assume additive separability and excluded variables in the utility functions,\footnote{\label{fn:ns_id}See Online Appendix \ref{app:ns_id} for our identification results in a more general nonseparable utility specification.} although the rest of the current section applies to more general models. 

The utilities $u_{ic}$ and $v_{ci}$ depend on the vector $(z_{i},y_{i},w_{i})$, which is observable to the researcher, and $\epsilon_{i} =(\epsilon_{i1}\dots,\epsilon_{iC})$ and $\eta_{i} =(\eta_{1i}\dots,\eta_{Ci})$, which are unobservable to the researcher.  Specifically, $z_{i}\in\mathcal{Z}\subseteq\mathbb{R}^{d_{z}}$, $y_{i}=(y_{i1}\dots,y_{iC})\in\mathcal{{Y}}= \mathcal{Y}_1 \times \dots \times \mathcal{Y}_C \subseteq\mathbb{{R}}^{C}$, and $w_{i}=(w_{1i}\dots,w_{Ci})\in\mathcal{{W}}\subseteq\mathbb{{R}}^{C}$. Further, for any $c\in\mathbf{C}$, $\epsilon_{ic}$ and $\eta_{ci}$ are scalar random variables. $(\epsilon_{i},\eta_{i})$ are independent and identically distributed (i.i.d.) draws from a joint distribution $F$. We make no restrictions on this joint distribution, allowing for arbitrary correlations within $(\epsilon_{i},\eta_{i})$.\par

\begin{assm}\label{assm_struc}
Let $u^{c}: \mathcal{Z}\rightarrow\mathbb{R}$, $r^c : \mathcal{Y}_c\rightarrow\mathbb{R}$, and $v^{c}: \mathcal{Z}\rightarrow\mathbb{R}$ be nonparametric functions, such that
\begin{align}\label{NP_specification}
u_{ic}  =  \tau_{ic}+\epsilon_{ic} & \mbox{ and } 
v_{ci}  =  \iota_{ci}+\eta_{ci}, \forall c\in\mathbf{C}, 
\end{align}
where $\tau_{ic} = u^{c}(z_{i})+r^{c}\left(y_{ic}\right)$  and $\iota_{ci} =  v^{c}(z_{i})+w_{ci}$. 
\end{assm}%

Scalar $y_{ic}$ is a demand shifter and scalar $w_{ci}$ is a supply shifter: $y_{ic}$ enters only $u_{ic}$ and is excluded from all other utility functions, and $w_{ci}$ enters only $v_{ci}$. 

We impose scale normalization on each side. For students, there exists a known value $\overline{y}_{c}$ in the interior of its support such that $\frac{\partial r^{c}\left(\overline{y}_{c}\right)}{\partial y_{ic}}=1$.\footnote{If $y_{ic}$ is known to have a negative effect on student preferences, the partial is normalized to $-1$. The same applies to $w_i$. } This holds trivially if $r^{c}\left(y_{ic}\right)=y_{ic}$. 
{For colleges, we assume $w_{ci}$ enters $v_{ci}$ linearly with a coefficient normalized to one. As detailed below, this linearity assumption allows us to  vary $w_i$ to construct a sufficient number of equations without increasing the number of unknowns.\footnote{\label{fn:nonlinear} The linearity in $w_{ci}$ can be relaxed. We can allow $w_{ci}$ to enter $v_{ci}$ through a nonlinear function, which is known for some colleges. Namely, assume $v_{ci}  = v^{c}(z_{i})+s^c(w_{ci})+\eta_{ci}$, where $s^c$: $\mathcal{W}_{c}\to\mathbb{R}$ is a nonparametric function such that (i) for each $c\in \overline{\bfC} \subset \bfC$, $s^c$ is \textit{unknown} with $\frac{\partial s^{c}\left(\overline{w}_{c}\right)}{\partial w_{ci}}=1$ for some known value $\overline{w}_{c}$; and (ii) for each $c\in \bfC \setminus \overline{\bfC}$, $s^c$ is \textit{known}. In this specification, our identification relies on varying $w_{ci}$ for $c\in \bfC \setminus \overline{\bfC}$ instead of all $c\in \bfC$.}  The assumptions on $y_i$ and $w_i$ can be switched, that is, having nonlinearity in $w_i$ and linearity in $y_i$.}

Students can remain unmatched or, equivalently, be matched with an outside option denoted by ``$0$.'' The utility of the outside option is normalized to 0, $u_{i0}=0$ $\forall i\in\mathbf{I}$. 
We assume that colleges have responsive preferences.\footnote{For $\varepsilon>0$, let $N_\varepsilon (i)$ and $N_\varepsilon (i')$ be a neighborhood of students around $v_{ci}$ and $v_{ci'}$, respectively, such that $Q(N_\varepsilon (i)) = Q(N_\varepsilon (i'))$. Responsive preferences imply that, for any $\mathbf{I}^c \subset \mathbf{I}$ with $Q(\mathbf{I}^c) \leq q_c - Q(N_\varepsilon (i))$, $N_\varepsilon (i) \subset  \mathbf{I}\setminus \mathbf{I}^c$, and $N_\varepsilon (i') \subset  \mathbf{I}\setminus \mathbf{I}^c$, college~$c$ prefers $\mathbf{I}^c \cup N_\varepsilon (i) $ to  $\mathbf{I}^c \cup N_\varepsilon (i') $ if and only if $v_{ci} > v_{ci'}$. See \cite{roth1992two} for a definition in a case with discrete students.}  This implies that the total utility of a college from being matched with a subset of students (up to its capacity) is increasing in its utility from each student; for example, the total utility is the sum of the utility from each of its matched students. College~$c$ has an acceptability threshold, $T_c$, and finds student~$i$ unacceptable if $v_{ci}<T_c$.


With the data from one such continuum market on $\{(z_i,y_i,w_i)\}_{i}$, $\{q_c\}_{c}$, and who matches with whom, we aim to identify student and college preferences by identifying $\{u^{c}, r^c, v^{c}, T_c\}_{c}$ and $F$, although, as we shall see, $\{T_c\}_{c}$ are not always point identified. Note that $(z_i,y_i,w_i)$ does not include college-specific variables that are constant across students, as these will be absorbed by the college-specific utility functions, $(u^c,r^c,v^c)$.

We use the continuum market to approximate a data generating process in a large finite market as follows: $(z_i, y_i, w_i,\epsilon_{i},\eta_i)$ is an i.i.d.\ draw from its joint distribution, college~$c$'s capacity is a $q_c$-fraction of the total number of students, and $\{u_{ic}, v_{ci},T_c\}_{i,c}$ determines both sides' preferences. This approximation is close to the matching outcomes when we use the equilibrium concept that will be introduced in \cref{sec:stable}.\footnote{\label{fn:approx}In a typical identification argument, the number of observations is taken to infinity while  the ``game'' is kept constant. Our setting contains one single matching game that changes with the market size. Proposition~3 of \cite{azevedo_supply_2016}, Proposition~4 of \cite{fack_beyond_2019}, and Corollary 2 of \cite{ACH2023} imply that, under certain conditions, the equilibrium outcome in the continuum approximates well an equilibrium outcome in a large finite market. Such an approximation is also used in the network literature \citep[e.g.,][]{menzel2022strategic}.}

 \begin{remark}\label{rm:location}
The location normalization in the functions is worth highlighting because the joint distribution of $(\epsilon_i,\eta_i)$, $F$, is fully nonparametric. 
	In student preferences, we already impose the normalization, $u_{i0}=0$, but we need another location normalization for each $c$ on either $u^{c}(z_{i})+r^{c}\left(y_{ic}\right)$ or $\epsilon_{ic}$ to separately identify the two. 	Similarly, for each college~$c$'s preferences, we need to location-normalize two of the three model primitives  $(v^c(z_i), \eta_{ci}, T_c)$ to pin down the third. 
	
In \cref{sec:derivative}, we identify the derivatives of ($u^c, r^c, v^c$), so this additional location normalization is not needed. However, it is necessary for identifying $T_c$ and $F$  in \cref{sec:id_dist}, where we location-normalize $u^{c}(z_{i})+r^{c}\left(y_{ic}\right)$, $v^c(z_i)$, and $\eta_{ci}$.
\end{remark}

\subsection{Matching and Stable Matching}\label{sec:stable}

We define a matching function or, simply, a matching, $\mu:\mathbf{{I}}\to\mathbf{{C}}\cup\{0\}$, such that (i) $\mu(i) = c \iff i \in\mu^{-1}(c)$, and (ii) $\forall\,c\in\mathbf{{C}},\,\mu^{-1}(c)\subseteq{\bf {I}},$ where $0 \leq Q(\mu^{-1}(c))\leq q_{c}$.

The following concepts are important for our analysis: individual rationality, blocking pairs, and stability. For notational reasons, we define them in the case with discrete students, corresponding to our empirical application. The precise definitions for a model with a continuum of students can be found in \cite{azevedo_supply_2016}, with measure-zero sets of students appropriately dealt with. 

A matching $\mu$ is {\it individually rational} if $u_{i\mu(i)} \geq u_{i0}$ and $v_{\mu(i)i} \geq T_{\mu(i)}$ for all $i \in \bfI$. A student-college pair $(i,c)\in\mathbf{{I}}\times\mathbf{C}$ \textit{blocks}
a matching $\mu$ if (i) student $i$ strictly prefers college $c$ to her current match $\mu(i)$, $u_{ic}>u_{i\mu(i)}$; and (ii) either college $c$ has excess capacity, $ Q(\mu^{-1}(c))<q_{c}$, or college $c$ prefers $i$ to one of its matched students, $\exists\,i^{\prime}\in\mu^{-1}(c),\,\text{{s.t.}},\,v_{ci}>v_{ci^{\prime}}$. Finally, a matching is \textit{stable}
if it is individually rational and not blocked by any pair $(i,c)\in\mathbf{{I}}\times\mathbf{{C}}$.

We assume that the matching in the data is stable.\footnote{\label{fn:stable}Stability can be achieved in certain equilibrium if students apply to all acceptable colleges and if a stable mechanism, e.g.,  the deferred acceptance \citep{gale_college_1962}, is used to find the matching. Theoretically, provided that students know what criteria colleges use to rank them, stability can still be satisfied in equilibrium, even if students choose not to apply to all acceptable colleges due to application costs \citep{fack_beyond_2019} or if students make certain application mistakes \citep{ACH2017}.  Importantly, achieving stability does not require the market to be centralized, as shown in laboratory experiments \cite[see, e.g.,][]{Pais-decentralized}, and steps in mechanisms such as the deferred acceptance can be implemented in a decentralized fashion \citep{GHK}. }  A stable matching exists and is generically unique \citep{azevedo_supply_2016}.\footnote{We need the regularity condition that the set $\{i\in\mathbf{I}:\mu(i)\,\text{is strictly less preferred than}\,c\}$  is open $\forall\, c\in \mathbf{C}$. This condition implies that a stable matching always allows an extra measure zero set of students into a college when this can be done without compromising stability.}
Moreover, a stable matching is characterized by college cutoffs. College~$c$'s cutoff is determined by its least-preferred matched student when its capacity constraint is binding; otherwise, it coincides with the  acceptability threshold. Let $\delta_{c}$ be college $c$'s cutoff. Then, $\forall\,c\in\mathbf{{C}}$, 
\begin{equation}\label{cutoff}
 \delta_{c}=\inf_{j\in\mu^{-1}(c)}v_{cj} \text{ \ if \ } Q(\mu^{-1}(c))=q_c; \; \delta_{c}=T_c  \text{ \ if \ } Q(\mu^{-1}(c))<q_c.
\end{equation}
By definition, $\delta_{c} \geq T_c$. Under the assumption of responsive college preferences, to determine if student~$i$ can be accepted by $c$, we just need to compare $v_{ci}$ and $\delta_{c}$. With non-responsive preferences, how $c$ ranks $i$ and $j$ would depend on who else $c$ accepts, and $\delta_{c}$ alone would not be sufficient to determine if $i$ could have been accepted by $c$.

In a stable matching of the continuum market, $\{u^{c}, r^{c},v^{c},T_c\}_{c\in\mathbf{C}}$, and $F$ imply a unique vector of cutoffs, $\{\delta_c\}_c$. Therefore, $\{\delta_c\}_c$ is merely a shorthand notation for the expression in equation~\eqref{cutoff} rather than additional parameters.\footnote{\label{fn:approx_cutoff}As mentioned earlier, especially, in footnote~\ref{fn:approx}, the continuum approximates a large finite market. The literature cited therein shows that equilibrium cutoffs, hence matching outcomes, in the large market can be close to $\{\delta_c\}_c$ with agents being practically ``cutoff-takers.'' That is, each agent's realized preferences have a negligible effect on cutoffs in large markets.} 


\subsection{Two-sided Discrete Choice Problem in a Stable Matching}\label{sec:notation}
College~$c$ is said to be \textit{feasible} for student~$i$ if and only if $v_{ci} \geq \delta_c$. 
A student can ``choose'' to match with any of her feasible colleges, but not any infeasible college. We call the set of all feasible colleges of a student her {\it feasible set}. 
Let $\mathcal{{L}}$ be the collection of the $2^{C}$ possible feasible sets, $\mathcal{{L}}\equiv\{L:\,0\,\in L,\,L\setminus\{0\}\subseteq\mathbf{{C}}\}$. By construction, the outside option always belongs to every feasible set.  A matching is stable if and only if every student is matched with her most-preferred feasible college \citep{fack_beyond_2019}. A classic issue in two-sided matching is that students' feasible sets are determined endogenously, unobserved by the researcher, and heterogeneous across students. 

For any given matching $\mu$, the probability that $L\in\mathcal{{L}}$ is student $i$'s feasible set conditional on $(z_i,w_i)$ is
\begin{align}\label{lambda} 
	\mathbb{{P}}\left(\text{{feasible set is }} L | z_i, w_i\right) &=\mathbb{P}\left(v_{ci}\geq\delta_{c}\;\forall\,c\in L;\,v_{di}<\delta_{d}\;\forall\,d\notin L | z_i, w_i\right)\nonumber\\
&\equiv \lambda_{L}(\iota_{1i},\dots,\iota_{Ci}).
\end{align}
If each student's feasible set was observed,  $ \lambda_L$ could be identified from the data, and recovering student and college preferences would follow from standard arguments in the discrete choice literature. However, we do not observe the feasible sets. 

Similarly, for students, $\mathbb{{P}}\left(c=\arg\max_{d\in L}u_{id}|L, z_{i}, y_{i}\right)$ is the probability that utility-maximizing  students with observables $(z_{i}, y_{i})$ ``choose'' $c$ from feasible set $L$. Conditional on $L$, this probability only depends on student preferences. We define 
\begin{equation}\label{g}
 g_{c,L}(\tau_{ic};\tau_{id},d\ne c) \equiv \mathbb{{P}}\left(c=\arg\max_{d\in L}u_{id}|L, z_{i}, y_{i}\right) , \forall c\in\mathbf{C}, 
\end{equation}
where the first argument of $g_{c,L}$ is always $\tau_{ic}$. If $d\notin L$, $g_{c,L}$ does not vary with $\tau_{id}$.


\section{Nonparametric Identification}\label{id}
    
We now turn to nonparametrically identifying the distribution of student and college preferences given a stable matching $\mu$ and covariates $(z_{i},y_{i},w_{i})$ in {\it one} market. 

A matching is stable if and only if every student is matched with the most-preferred college in her feasible set. Thus, stability implies, for $c\in\mathbf{{C}}\cup\{0\}$,  
\begin{align}
\sigma_{c}(z_{i},y_{i},w_{i}) & \equiv  \mathbb{{P}}(\mu(i)=c|z_{i},y_{i},w_{i}) \notag  \\
&=\sum_{L\in\mathcal{{L}}}\mathbb{{P}}(\text{{feasible set is }} L |z_i, w_i) \cdot \mathbb{{P}}\left(c=\arg\max_{d\in L}u_{id}|L,z_{i}, y_{i}\right)\nonumber\\
&= \sum_{L\in\mathcal{{L}}}\lambda_{L}(\iota_{1i},\dots,\iota_{Ci})\cdot g_{c,L}(\tau_{ic};\tau_{id},d\ne c).\label{ideq2}
\end{align}
Hence, the conditional match probability, $\sigma_{c}(z_{i},y_{i},w_{i})$, which is the fraction of students with $(z_{i},y_{i},w_{i})$ matched with $c$ and known in the population data, is linked to the model through student and college preferences. 

Below, we first study the conditions under which the functions $\{u^{c},\,r^{c},\,v^{c}\}_c$ are nonparametrically identified; we then identify the joint distribution of $(\epsilon_{i},\eta_{i})$, $F$. Later in Section~\ref{sec:practical}, we present results that impose fewer requirements on the data.

\subsection{Identifying the Derivatives of the Utility Functions \label{sec:derivative}}

We nonparametrically identify the derivatives of the functions $\{u^{c},r^{c}, v^{c}\}_c$ w.r.t.\ the
observables. With these derivatives identified, the functions are identified up to a constant, provided that $z_{i}$ and $y_i$ have full support. The idea is to use the variation in the excluded variables to trace out how each argument in equation~\eqref{ideq2} affects the conditional match probabilities. Specifically, the excluded variables in student preferences $(y_i)$ only shift demand, while the excluded variables in college preferences $(w_i)$ shift supply, or feasible sets. The effect of other variables $(z_i)$ that enter both demand and supply can be written as a combination of the effects of $y_i$ and $w_i$. This leads to  a  system of linear equations in the derivatives of the conditional match probabilities, whose solution  is our parameters of interest.  

\subsubsection{\label{id_eg}A Simple Example with One College}
We describe the intuition for identification in a one-college example, $\mathbf{C}=\{1\}$. Student utility functions are $u_{i1}=u^1\left(z_{i}\right)+r^1\left(y_{i1}\right)+\epsilon_{i1}$ for college~$1$ and $u_{i0}=0$ for the outside option. Here, $z_{i}$ is a scalar and $\frac{\partial r^{1}\left(\overline{y}_{1}\right)}{\partial y_1}=1$ for a known value, $\overline{y}_{1}$. 
College~$1$'s utility function is $v_{1i}=v^1(z_{i})+w_{1i}+\eta_{1i}$, and the (unobserved) cutoff is $\delta_1$.

To identify $\frac{\partial u^1}{\partial z_i}$ and $\frac{\partial v^1}{\partial z_i}$, we fix $y_{i1}=\overline{y}_1$ and consider any value $(z,w_1)$ in the interior of its support. Figure~\ref{partition}(a) shows that the space of $(\epsilon_{i1},\eta_{1i})$
is partitioned into four parts based on the feasibility of college~1 and
student~$i$'s preferences (i.e., the acceptability of college~$1$ to $i$). Moreover, $\mu\left(i\right)=1$ if and only if $\epsilon_{i1}>-u^1\left(z\right)-r^1\left(\overline{y}_1\right)$ (college~$1$ is acceptable to $i$) and $\eta_{1i}>\delta_1-v^1\left(z\right)-w_1$ (college $1$ is feasible to $i$).

\pgfplotsset{ticks=none} 
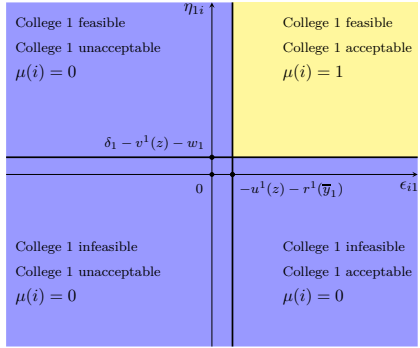
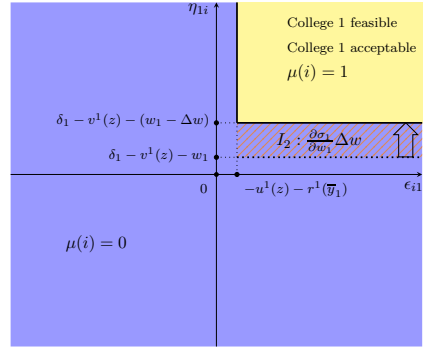
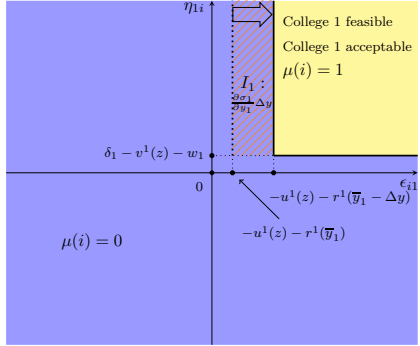
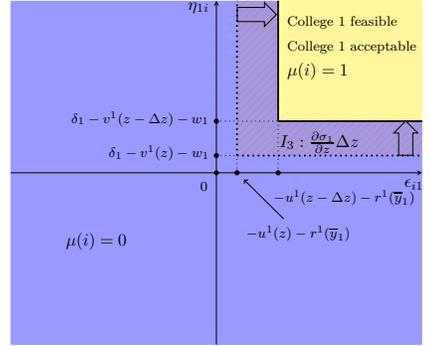
\begin{figure}[h!]
	\centering
	\begin{minipage}[t]{.45\linewidth}            
		\centering
		
		\begin{tikzpicture}[scale=0.65]  
			
			\tikzset{every pin edge/.style={draw=black,<-, very near end}}

			\begin{axis} 
				[
				ymin=-5,
				xmin=-5,
				ymax=5,
				xmax=5,
				width=10cm, 
				axis on top=true, 
				axis x line=middle, 
				axis y line=middle,
				yticklabels={,,},
				xticklabels={,,}
				] 
				\draw[fill=yellow!60,opacity=0.8, draw=none]      (axis cs:0.5,0.5) -- (axis cs:0.5,5) -- (axis cs:5,5) -- (axis cs:5,0.5) -- cycle ;
				\draw[fill=blue!50,opacity=0.8, draw=none]      (axis cs:0.5,0.5) -- (axis cs:5,0.5) -- (axis cs:5,-5) -- (axis cs:-5,-5) --  (axis cs:-5,5) -- (axis cs:0.5,5) --cycle ;

				\addplot+ [mark=none, black, line width = 1, domain=-5:5] {0.5}; 
				\addplot +[mark=none, black, line width = 1] coordinates {(0.5,-5) (0.5, 5)};

				\node [below, label={[align=left] \scriptsize College 1  feasible \\ \scriptsize College 1  acceptable \\ \footnotesize $\mu(i)=1$}] at (axis cs: 3.3,2.5){};
\node [below, label={[align=left] \scriptsize College 1  feasible \\ \scriptsize College 1 unacceptable \\ \footnotesize $\mu(i)=0$}] at (axis cs: -3,2.5){};
\node [below, label={[align=left] \scriptsize College 1  infeasible \\ \scriptsize College 1 unacceptable \\ \footnotesize $\mu(i)=0$}] at (axis cs: -3,-4){};
\node [below, label={[align=left] \scriptsize College 1  infeasible \\ \scriptsize College 1  acceptable \\ \footnotesize $\mu(i)=0$}] at (axis cs: 3.3,-4){};

				\node [below] at (axis cs: 4.8,0) {\footnotesize $\epsilon_{i1}$};
				\node [left] at (axis cs: 0,4.8) {\footnotesize $\eta_{1i}$};

            \node [below right] at (axis cs: 0.5,0) {\scriptsize $-u^1(z)-r^1(\overline{y}_1)$};
            \node [above left] at (axis cs: 0,0.5) {\scriptsize $\delta_1-v^1(z)-w_1$};
            \addplot[mark=*, mark size=1.2pt] coordinates {(0,0.5)}; 
            \addplot[mark=*, mark size=1.2pt] coordinates {(0.5,0)} ;
            \node[label={230:{\scriptsize 0}},circle,fill,inner sep=1pt] at (axis cs:0,0) {};

			\end{axis} 
			
		\end{tikzpicture}
		\subcaption[]%
		{{\footnotesize Feasible sets, student preferences, \& $\mu(i)$} }                
		\label{fig:panela}

		\begin{tikzpicture} [scale=0.65]  
			
			\tikzset{every pin edge/.style={draw=black, <-,very near end}}
			
		\begin{axis} 
			[
			ymin=-5,
			xmin=-5,
			ymax=5,
			xmax=5,
			width=10cm, 
			axis on top=true, 
			axis x line=middle, 
			axis y line=middle,
			yticklabels={,,},
			xticklabels={,,}
			] 
			\draw[fill=yellow!60,opacity=0.8, draw=none]      (axis cs:1.5,0.5) -- (axis cs:1.5,5) -- (axis cs:5,5) -- (axis cs:5,0.5) -- cycle ;
			\draw[fill=blue!50,opacity=0.8, draw=none]      (axis cs:1.5,0.5) -- (axis cs:5,0.5) -- (axis cs:5,-5) -- (axis cs:-5,-5) --  (axis cs:-5,5) -- (axis cs:1.5,5) --cycle ;
			\draw[fill=blue!50,opacity=0.8, draw=none, pattern=north east lines, pattern color=orange] (axis cs:0.5,0.5) -- (axis cs:1.5,0.5) -- (axis cs:1.5,5) -- (axis cs:0.5,5)  --cycle ;
			
			
			\addplot+ [mark=none, black, line width = 1, domain=1.5:5] {0.5}; 
			\addplot +[mark=none, black, line width = 1, dotted] coordinates {(0.5,0.5) (0.5, 5)};
			\addplot +[mark=none, black, line width = 1] coordinates {(1.5,0.5) (1.5, 5)};

			
			\addplot +[mark=none, black, line width = 0.5, dotted] coordinates {(0,0.5) (1.5, 0.5)};
			\addplot +[mark=none, black, line width = 0.5, dotted] coordinates {(1.5,0) (1.5, 0.5)};
			\addplot +[mark=none, black, line width = 0.5, dotted] coordinates {(0.5,0) (0.5, 0.5)};			
			
			\node [below, label={[align=left] \scriptsize College 1  feasible \\ \scriptsize College 1  acceptable \\ \footnotesize $\mu(i)=1$}] at (axis cs: 3.3,2.5){};			
			
			\node [left] at (axis cs: -2,-2) {\footnotesize $\mu(i)=0$};
			\node [below] at (axis cs: 1,3) {\footnotesize $I_1:$};
			\node [below] at (axis cs: 0.96,2.5) {\tiny $\frac{\partial\sigma_{1}}{\partial y_{1}}\Delta y$};
			\node [below] at (axis cs: 4.8,0) {\footnotesize $\epsilon_{i1}$};
			\node [left] at (axis cs: 0,4.8) {\footnotesize $\eta_{1i}$};

		    \node [above left] at (axis cs: 0,0.15) {\scriptsize $\delta_1-v^1(z)-w_1$};
	     	\addplot[mark=*, mark size=1.2pt] coordinates {(0,0.5)}; 
	    	\addplot[mark=*, mark size=1.2pt] coordinates {(1.5,0)};
			
			\node[label={[label distance=0.18cm]275:{\scriptsize $-u^1(z)-r^1(\overline{y}_1-\Delta y)$}},inner sep=0pt] at (axis cs:1.2,0) {};
			\addplot[mark=*, mark size=1.2pt] coordinates {(0.5,0)} node[pin={[pin distance=0.8cm]273:{\scriptsize $-u^1(z)-r^1(\overline{y}_1)$}}]{};
			\node[label={230:{\scriptsize 0}},circle,fill,inner sep=1pt] at (axis cs:0,0) {};

			
			\node[draw, single arrow,
			minimum height=8mm, minimum width=1mm,
			single arrow head extend=1mm,
			anchor=west, rotate=0] at (axis cs: 0.5,4.6) {};
		\end{axis} 
		\end{tikzpicture}
		\subcaption[]%
		{{\footnotesize Changes in $\mu(i)$  when $y_{i1} \downarrow$ by ${\Delta}y$}}                 
		\label{fig:panelb}

	\end{minipage}         
	\hspace*{\fill}
	\begin{minipage}[t]{.45\linewidth}        
		
		\centering

				\begin{tikzpicture} [scale=0.65] 
			
			\tikzset{every pin edge/.style={draw=black, <-,very near end}}
			
			\begin{axis} 
				[
				ymin=-5,
				xmin=-5,
				ymax=5,
				xmax=5,
				width=10cm, 
				axis on top=true, 
				axis x line=middle, 
				axis y line=middle,
				yticklabels={,,},
				xticklabels={,,}
				] 
				\draw[fill=yellow!60,opacity=0.8, draw=none]      (axis cs:0.5,1.5) -- (axis cs:0.5,5) -- (axis cs:5,5) -- (axis cs:5,1.5) -- cycle ;
				\draw[fill=blue!50,opacity=0.8, draw=none]      (axis cs:0.5,1.5) -- (axis cs:5,1.5) -- (axis cs:5,-5) -- (axis cs:-5,-5) --  (axis cs:-5,5) -- (axis cs:0.5,5) --cycle ;
				\draw[fill=blue!50,opacity=0.8, draw=none, pattern=north east lines, pattern color=orange] (axis cs:0.5,0.5) -- (axis cs:5,0.5) -- (axis cs:5,1.5) -- (axis cs:0.5,1.5)  -- (axis cs:0.5,5) -- (axis cs:0.5,5)--cycle ;
				
				
				\addplot+ [mark=none, black, line width = 1, domain=0.5:5, dotted] {0.5}; 
				\addplot +[mark=none, black, line width = 1] coordinates {(0.5,1.5) (0.5, 5)};
				\addplot+ [mark=none, black, line width = 1, domain=0.5:5] {1.5};

				
				\addplot +[mark=none, black, line width = 0.5, dotted] coordinates {(0,0.5) (0.5, 0.5)};
				\addplot +[mark=none, black, line width = 0.5, dotted] coordinates {(0,1.5) (0.5, 1.5)};
				\addplot +[mark=none, black, line width = 0.5, dotted] coordinates {(0.5,0) (0.5, 1.5)};
				
				\node [below, label={[align=left] \scriptsize College 1  feasible \\ \scriptsize College 1  acceptable \\ \footnotesize $\mu(i)=1$}] at (axis cs: 3.3,2.5){};
								
				\node [left] at (axis cs: -2,-2) {\footnotesize $\mu(i)=0$};
				\node [below] at (axis cs: 2.5,1.5) {\footnotesize $I_2:\frac{\partial\sigma_{1}}{\partial w_{1}}\Delta w$};
				\node [below] at (axis cs: 4.8,0) {\footnotesize $\epsilon_{i1}$};
				\node [left] at (axis cs: 0,4.8) {\footnotesize $\eta_{1i}$};

				\node [below right] at (axis cs: 0.5,0) {\scriptsize $-u^1(z)-r^1(\overline{y}_1)$};
				\node [above left] at (axis cs: 0,0.15) {\scriptsize $\delta_1-v^1(z)-w_1$};
				\node [above left] at (axis cs: 0,1.15) {\scriptsize $\delta_1-v^1(z)-(w_1-\Delta w)$};
				\addplot[mark=*, mark size=1.2pt] coordinates {(0,0.5)}; 
				\addplot[mark=*, mark size=1.2pt] coordinates {(0.5,0)};
				\addplot[mark=*, mark size=1.2pt] coordinates {(0,1.5)};

				\node[label={230:{\scriptsize 0}},circle,fill,inner sep=1pt] at (axis cs:0,0) {};


				\node[draw, single arrow,
				minimum height=7mm, minimum width=1mm,
				single arrow head extend=1mm,
				anchor=west, rotate=90] at (axis cs: 4.6,0.5) {};
			\end{axis} 
		\end{tikzpicture}
		\subcaption[]%
		{{\footnotesize Changes in $\mu(i)$ when $w_{1i} \downarrow$ by $\Delta w$}}                 
		\label{fig:panelc}

		\begin{tikzpicture} [scale=0.65] 
			
			\tikzset{every pin edge/.style={draw=black, <-,very near end}}
			
		\begin{axis} 
			[
			ymin=-5,
			xmin=-5,
			ymax=5,
			xmax=5,
			width=10cm, 
			axis on top=true, 
			axis x line=middle, 
			axis y line=middle,
			yticklabels={,,},
			xticklabels={,,}
			] 
			\draw[fill=yellow!60,opacity=0.8, draw=none]      (axis cs:1.5,1.5) -- (axis cs:1.5,5) -- (axis cs:5,5) -- (axis cs:5,1.5) -- cycle ;
			\draw[fill=blue!50,opacity=0.8, draw=none]      (axis cs:1.5,1.5) -- (axis cs:5,1.5) -- (axis cs:5,-5) -- (axis cs:-5,-5) --  (axis cs:-5,5) -- (axis cs:1.5,5) --cycle ;
			\draw[fill=blue!50,opacity=0.8, draw=none, pattern=north east lines, pattern color=orange] (axis cs:0.5,0.5) -- (axis cs:5,0.5) -- (axis cs:5,1.5) -- (axis cs:1.5,1.5)  -- (axis cs:1.5,5) -- (axis cs:0.5,5)--cycle ;
			
			\addplot+ [mark=none, black, line width = 1, domain=0.5:5, dotted] {0.5}; 
			\addplot +[mark=none, black, line width = 1, dotted] coordinates {(0.5,0.5) (0.5, 5)};
			\addplot +[mark=none, black, line width = 1] coordinates {(1.5,1.5) (1.5, 5)};
			\addplot+ [mark=none, black, line width = 1, domain=1.5:5] {1.5};

			
			\addplot +[mark=none, black, line width = 0.5, dotted] coordinates {(0,0.5) (0.5, 0.5)};
			\addplot +[mark=none, black, line width = 0.5, dotted] coordinates {(1.5,0) (1.5, 1.5)};
			\addplot +[mark=none, black, line width = 0.5, dotted] coordinates {(0.5,0) (0.5, 0.5)};
			\addplot +[mark=none, black, line width = 0.5, dotted] coordinates {(0,1.5) (1.5, 1.5)};

			\node [below, label={[align=left]	\scriptsize College 1  feasible \\ \scriptsize College 1  acceptable \\ \footnotesize $\mu(i)=1$}] at (axis cs: 3.3,2.5){};
			\node [left] at (axis cs: -2,-2) {\footnotesize $\mu(i)=0$};
			\node [below] at (axis cs: 2.5,1.4) {\footnotesize $I_3: \frac{\partial\sigma_{1}}{\partial z}\Delta z$};
			\node [below] at (axis cs: 4.8,0) {\footnotesize $\epsilon_{i1}$};
			\node [left] at (axis cs: 0,4.8) {\footnotesize $\eta_{1i}$};

			\node[label={[label distance=0.18cm]275:{\scriptsize $-u^1(z-\Delta z)-r^1(\overline{y}_1)$}},inner sep=0pt] at (axis cs:1.2,0) {};
			\node [above left] at (axis cs: 0,0.15) {\scriptsize $\delta_1-v^1(z)-w_1$};
			\node [above left] at (axis cs: 0,1.15) {\scriptsize $\delta_1-v^1(z-\Delta z)-w_1$};
			\addplot[mark=*, mark size=1.2pt] coordinates {(0.5,0)} node[pin={[pin distance=0.8cm]273:{\scriptsize $-u^1(z)-r^1(\overline{y}_1)$}}]{};
			\addplot[mark=*, mark size=1.2pt] coordinates {(0,0.5)}; 
			\addplot[mark=*, mark size=1.2pt] coordinates {(1.5,0)};
			\addplot[mark=*, mark size=1.2pt] coordinates {(0,1.5)};

			\node[label={230:{\scriptsize 0}},circle,fill,inner sep=1pt] at (axis cs:0,0) {};

			
			\node[draw, single arrow,
			minimum height=8mm, minimum width=1mm,
			single arrow head extend=1mm,
			anchor=west, rotate=0] at (axis cs: 0.5,4.6) {};
			
			\node[draw, single arrow,
			minimum height=7mm, minimum width=1mm,
			single arrow head extend=1mm,
			anchor=west, rotate=90] at (axis cs: 4.6,0.5) {};
		\end{axis} 
		\end{tikzpicture}
		\subcaption[]%
		{{\footnotesize Changes in $\mu(i)$  when $z_i \downarrow$ by $\Delta z$}}                 
		\label{fig:paneld}         
		
	\end{minipage} 
	\caption{Partitioning the Space of Unobservables in the One-College Case}\label{partition}
	\begin{flushleft}
		{\footnotesize\noindent \textit{Notes:} Panel (a) describes the partition of the $(\epsilon_{i1},\eta_{1i})$ space given $(z_{i}, y_{i1}, w_{1i}) = (z,\overline{y}_1,w_1)$ by student $i$'s feasible set and preferences. The other panels show the changes in $\mu(i)$ when $y_{i}$ decreases by  $\Delta y$ and affects only student preferences (panel~b), when $w_{1i}$ decreases by $\Delta w$ and affects only feasible set (panel~c), and when $z_{i}$ decreases by $\Delta z$ (panel~d).}						
	\end{flushleft}    
\end{figure}

Figures~\ref{partition}(b)--(d) depict how the marginal effect of $z_{i}$ on match probability is linked to the marginal effects of the excluded variables, $y_{i1}$ and $w_{1i}$. 
Panel (b) describes the marginal effect of $y_{i1}$. Specifically, decreasing $y_{i1}$ from $\overline{y}_1$ to $\overline{y}_1-\Delta y$ makes college~$1$ less attractive to student~$i$, and the region in which $\mu(i)=1$ shrinks along the horizontal $\epsilon_{i1}$-axis.  The area $I_{1}$ depicts the set of students whose match differs when $y_{i1}$ decreases. The induced change in the match probability, $\sigma_1=\mathbb{P}\left(\mu(i)=1|(z_{i},y_{i1},w_{1i}) = (z,\overline{y}_1,w_1)\right)\equiv\mathbb{P}(\mu(i)=1|z,\overline{y}_1,w_1)$, is the mass that the density of $\left(\epsilon_{i1},\eta_{1i}\right)$ puts on $I_{1}$, or, for  $(z_{i},y_{i1},w_{1i}) = (z,\overline{y}_1,w_1)$, 
\begin{equation}
	\frac{\partial\sigma_{1}(z,\overline{y}_{1},w_{1})}{\partial y_{i1}}=\frac{\partial r^1(\overline{y}_{1})}{\partial y_{i1}}\cdot\sum_{L\in\mathcal{{L}}}\lambda_{L}(\iota_{1})\cdot \frac{\partial g_{1,L}(\tau_{1})}{\partial \tau_{i1}}=\sum_{L\in\mathcal{{L}}}\lambda_{L}(\iota_{1})\cdot \frac{\partial g_{1,L}(\tau_{1})}{\partial \tau_{i1}}, 
	\label{diffy}
\end{equation}
where $\tau_1 \equiv u^1(z)+r^1(\overline{y}_{1})$ and $\iota_{1} \equiv v^1(z)+w_{1}$. By scale normalization, $\frac{\partial r^{1}\left(\overline{y}_{1}\right)}{\partial y_1}=1$. 

Panel~(c) shows a similar graph in which decreasing $w_{1i}$ from $w_1$ to $w_1 - \Delta w$ makes college~$1$ less likely to be feasible to student~$i$. Hence, the region $\mu(i) = 1$ shrinks along the vertical $\eta_{1i}$-axis. 
The change in the match probability induced by the decrease in $w_{1i}$ is the mass that
the density of $\left(\epsilon_{i1},\eta_{1i}\right)$ puts on the area $I_{2}$, or,   
\begin{equation}
\frac{\partial\sigma_1(z,\overline{y}_{1},w_{1})}{\partial w_{1i}}=\sum_{L\in\mathcal{{L}}}\frac{\partial \lambda_{L}(\iota_{1})}{\partial \iota_{1i}}\cdot g_{1,L}(\tau_{1}). \label{diffw}
\end{equation}

In panel (d), the decrease in $z_{i}$ reduces the attractiveness and feasibility of college~$1$ for student~$i$, because $z_{i}$ enters both student and college preferences. Besides, how
$z_{i}$ changes the region of $\mu(i)=1$ relies on the shape of the functions $u^1$ and $v^1$. The change in the match probability
induced by the change in $z_{i}$ corresponds to the area
$I_{3}$, or, 
\begin{align}
\frac{{\partial\sigma_1(z,\overline{y}_{1},w_{1})}}{\partial z_{i}} & =\frac{\partial u^1(z)}{\partial z_{i}}\cdot\sum_{L\in\mathcal{{L}}}\lambda_{L}(\iota_{1})\cdot \frac{\partial g_{1,L}(\tau_{1})}{\partial \tau_{i1}}+\frac{\partial v^1(z)}{\partial z_{i}}\cdot\sum_{L\in\mathcal{{L}}}\frac{\partial \lambda_{L}(\iota_{1})}{\partial \iota_{1i}}\cdot g_{1,L}(\tau_{1}). \label{diffz}
\end{align}
Our identification result relies on the changes caused by $z_{i}$, $y_{i1}$, and $w_{1i}$. Plugging equations \eqref{diffy}
and \eqref{diffw} into equation \eqref{diffz}, we have,   for  $(z_{i},y_{i1},w_{1i}) = (z,\overline{y}_1,w_1)$, 
\begin{equation}
\frac{{\partial\sigma_1(z,\overline{y}_{1},w_{1})}}{\partial z_{i}}=\frac{\partial u^1(z)}{\partial z_{i}}\cdot\frac{{\partial\sigma_1(z,\overline{y}_{1},w_{1})}}{\partial y_{i1}}+\frac{\partial v^1(z)}{\partial z_{i}}\cdot\frac{{\partial\sigma_1(z,\overline{y}_{1},w_{1})}}{\partial w_{1i}}.\label{eq:indsim}
\end{equation}
This equation reflects the chain rule: the effect of $z_{i}$ on
the  match probability, $\frac{\partial\sigma_1}{\partial z_{i}}$, is
realized through its effects on utilities $u_{i1}$ and $v_{1i}$,
captured by $\frac{\partial u^1}{\partial z_{i}}$ and
$\frac{\partial v^1}{\partial z_{i}}$, and the effects of the utilities on the  match probability, captured by $\frac{\partial {\sigma_1}}{\partial y_{i1}}$
and $\frac{\partial {\sigma_1}}{\partial w_{1i}}$.
In equation~\eqref{eq:indsim}, the derivatives of the match probability can be recovered from the population data, and the two unknowns, $\frac{\partial {u^1}}{\partial z_{i}}$ and $\frac{\partial v^1}{\partial z_{i}}$ are the parameters of interest. 

Importantly, when $w_{1i}$ varies, the conditional match probability changes, but the two unknowns remain constant, a consequence of $w_{1i}$ entering $v_{1i}$ in a known way. If, for any $z$, $(\epsilon_{i1},\eta_{1i})$ has enough
variation such that two distinct values of $w_{1i}$ produce two linearly independent equations that have a unique solution, we identify the unknowns. This requirement is formalized as \cref{cond} later, which imposes a mild restriction on the distribution of $(\epsilon_{i1},\eta_{1i})$ as detailed in Online Appendix \ref{app:rank_onecollege}. 

We then identify $\frac{\partial r^1}{\partial y_{i1}}$, which is not one when $y_{i1}\neq \overline{y}_1$.  For any value $(z,y_{1},w_{1})$, plugging equations \eqref{diffy}
	and \eqref{diffw} into equation \eqref{diffz}, we rearrange and obtain
\begin{equation}
\left(\frac{\partial\sigma_{1}(z,y_{1},w_{1})}{\partial z_{i}}-\frac{\partial v^{1}(z)}{\partial z_{i}}\cdot\frac{\partial\sigma_{1}(z,y_{1},w_{1})}{\partial w_{1i}}\right)\cdot\frac{\partial r^{1}(y_{1})}{\partial y_{i1}}=\frac{\partial u^{1}(z)}{\partial z_{i}}\cdot\frac{\partial\sigma_{1}(z,y_{1},w_{1})}{\partial y_{i1}}, \label{eq:indsim2}
\end{equation}
where all the terms except for $\frac{\partial r^1}{\partial y_{i1}}$ are either identified or known. For any $y_{1}$, if $\frac{\partial\sigma_{1}}{\partial z_{i}}-\frac{\partial v^{1}}{\partial z_{i}}\cdot\frac{\partial\sigma_{1}}{\partial w_{1i}} \neq 0$ for some value $(z,w_{1})$, 
$\frac{\partial r^1}{\partial y_{i1}}$ is identified; otherwise, equation~\eqref{eq:indsim2} implies that $\frac{\partial\sigma_{1}}{\partial y_{i1}}=0$ for all $(z,w_{1})$ and thus $\frac{\partial r^1}{\partial y_{i1}}$ is also identified and equal to zero.

Below, we extend this example to the case with multiple colleges. We derive equation~\eqref{eq:indsim} for each college, in which the marginal effect of $z_i$ on the probability of being matched with each college is the sum of its marginal effects on $(u^c,v^c)$ for all $c$. By varying $\{w_{ci}\}_c$, we form a system of equations in $\{\frac{\partial u^{c}}{\partial z_{i}}, \frac{\partial v^{c}}{\partial z_{i}}\}_c$.
 The identification of $\frac{\partial r^{c}}{\partial y_{ic}}$ is the same as above and relies on a generalized version of equation~\eqref{eq:indsim2} because we can identify $\frac{\partial r^{c}}{\partial y_{ic}}$ for each $c$ separately by holding $\frac{\partial r^{d}}{\partial y_{id}}=1$ for $d\neq c$.

\subsubsection{Formal Identification Results}\label{sec:id_results}

Our nonparametric identification of $\{\frac{\partial u^{c}}{\partial z_{i}}, \frac{\partial r^{c}}{\partial y_{ic}}, \frac{\partial v^{c}}{\partial z_{i}}\}_{c}$ extends \cite{matzkin_constructive_2019} who uses excluded variables to identify nonparametric nonseparable discrete choice models. 
\begin{assm} \label{assm_nonpara}
(i) $z_i$, $y_i$, and $w_i$ are continuously distributed;  (ii) for each $c\in\mathbf{C}$,  the functions, $(u^{c}, r^{c}, v^{c})$, are continuously differentiable; and (iii) $F$ is continuously differentiable. 
\end{assm}

Part (i) of Assumption \ref{assm_nonpara} requires that all covariates are continuous (but not necessarily have full-support), which is relaxed in \cref{sec:practical}. 

\begin{assm}\label{indep} $(\epsilon_{i},\eta_{i})$ is distributed
independently of $(z_i,y_i,w_i)$. \end{assm}

This exogeneity assumption is made for simplicity. One way to relax this assumption is to adopt a control function approach \citep{heckman1985alternative,blundell2004endogeneity, imbens2009identification}. See Online Appendix~\ref{app:cf} for a discussion.

Additionally, we need a condition on the derivatives of match probabilities w.r.t.\ the excluded variables. Let $\Pi_{y}(z_{i},y_{i},w_{i})$ be a $C\times C$ Jacobian matrix of the match probabilities w.r.t.\ the excluded variable $y_{i}$, whose $(c,d)$ element is $\frac{\partial \sigma_{c}(z_{i},y_{i},w_{i})}{\partial y_{id}}$. Similarly, let $\Pi_{w}(z_{i},y_{i},w_{i})$ be a $C\times C$ Jacobian matrix w.r.t.\ the excluded variable $w_{i}$. Fix $y_i=\overline{y},$ where $\overline{y}=(\overline{y}_1,\cdots,\overline{y}_C)$. We then consider a pair of distinct values of $w_{i}$, $\widehat{w}$ and $\widetilde{w}$, and define a $2C\times2C$ matrix evaluated at $\left(z_{i},y_{i}\right)=\left(z,\overline{y}\right)$, 
	\[
	\Pi(z,\overline{y},\widehat{w},\widetilde{w})\equiv\begin{pmatrix}\Pi_{y}(z,\overline{y},\widehat{w}) & \Pi_{w}(z,\overline{y},\widehat{w})\\
		\Pi_{y}(z,\overline{y},\widetilde{w}) & \Pi_{w}(z,\overline{y},\widetilde{w})
	\end{pmatrix}.
	\]
We impose the following testable condition on $\Pi(z,\overline{y},\widehat{w},\widetilde{w})$.\footnote{\label{fn:test}For a statistical test of \cref{cond} given a value of $z_i$ and a pair of values of $w_i$, one may use the method proposed by \cite{chen_improved_2019}. Testing $H_{0}$: rank$(\Pi(z,\overline{y},\widehat{w},\widetilde{w}))\leq2C-1$ against $H_{1}$: rank$(\Pi(z,\overline{y},\widehat{w},\widetilde{w}))>2C-1$ is a special case of setup (1) in \citet[p.1788]{chen_improved_2019}. In practice, one needs to find two values of $w_i$ satisfying \cref{cond}, which can be achieved by the following procedure: (i) choose $m$ pairs of $w_i$ values, (ii) for each pair, apply this test and calculate the $p$ values, and (iii) use the Holm–Bonferroni method to control the overall size of this multiple hypothesis testing problem and then determine which null hypothesis, if any, is rejected. The pair of values of $w_i$ associated with any rejected hypothesis satisfies \cref{cond}.}
\begin{cond}\label{cond}
For any $z$ in the interior of $\mathcal{Z} $, there exist two values of $w_i$, $\widehat{w}$ and $\widetilde{w}$, in $w_i$'s support conditional on $(z_{i},y_{i})=(z,\overline{y})$ such that $\Pi(z,\overline{y},\widehat{w},\widetilde{w})$ has rank $2C$.
\end{cond}

Note that, for any value of $z_i$, \cref{cond} only needs {\it two} values of $w_i$ at which $\Pi(z,\overline{y},\widehat{w},\widetilde{w})$ is full-rank.  
In other words, \cref{cond} can hold even when there are \textit{infinitely many} values of $w_i$ at which $\Pi(z,\overline{y},\widehat{w},\widetilde{w})$ is \textit{not} full-rank. The condition fails if, for example, for the given value $(z,\overline{y})$, the student's probability of matching with a certain college is always zero for a neighborhood around $\overline{y}$ and all values of $w_i$.\footnote{This may occur if the college is infeasible or unaccaptable to the student with probability one.}
 
 For a set of sufficient, yet weak and testable, conditions for \cref{cond}, consider $\widehat{w}$ that is large enough such that all colleges are feasible when $w_i=\widehat{w}$. In this case, the upper half of $\Pi(z,\overline{y},\widehat{w},\widetilde{w})$ corresponds to a one-sided discrete choice model, where the feasible-set shifter $w_i$ has no impact on the matching probabilities and thus $\Pi_{w}(z,\overline{y},\widehat{w})=\boldsymbol{0}_{C\times C}$. This makes $\Pi(z,\overline{y},\widehat{w},\widetilde{w})$ a lower triangular block matrix. Then, \cref{cond} holds if and only if both $\Pi_{y}(z,\overline{y},\widehat{w})$ and $\Pi_{w}(z,\overline{y},\widetilde{w})$ are invertible, which can be achieved under the connected substitutes conditions \citep[ Theorem 2]{berry2013connected}.\footnote{The invertibility of $\Pi_{y}(z,\overline{y},\widehat{w})$ holds under the following two assumptions: (i) for each $d\in\mathbf{C}\cup\{0\}$, $\frac{\partial \sigma_{d}(z_i,y_i,w_i)}{\partial y_{ic}}\leq 0$ for all $c\in\mathbf{C}\backslash\{d\}$; (ii) for any nonempty $\overline{\mathbf{C}}\subseteq\mathbf{C}$, there exists $c\in\overline{\mathbf{C}}$ and $d\notin\overline{\mathbf{C}}$ such that $\frac{\partial\sigma_{d}(z,\overline{y},\widehat{w})}{\partial y_{ic}}<0$. The invertibility of $\Pi_{w}(z,\overline{y},\widetilde{w})$ holds under similar assumptions.}\textsuperscript{, }\footnote{\label{fn:as2022}This sufficient condition for \cref{cond} provides us with a clear comparison with a related study, \cite{agarwal2022demand}. Their conditions for the identification of a similar model and our conditions are not nested. Specifically, they assume large support on $w_i$, impose a substitution condition on $\Pi_{y}(z,y,w)$ stronger than the one in \cite{berry2013connected}, and require the substitution condition to hold for \textit{all} but a finite set of $y_i$ values. In contrast, in this sufficient condition for \cref{cond}, with large support of $w_i$, we need a substitution condition on both $\Pi_{y}(z,\overline{y},w)$ and $\Pi_{w}(z,\overline{y},w)$ \`a la \cite{berry2013connected}, but only for $y_i = \overline{y}$.} This suggests that \cref{cond} holds under common logit or probit models. It is worth noting that the large support of $w_i$ above is mainly used to facilitate a clear connection with the connected substitutes conditions, but is  stronger than necessary. In Online Appendix \ref{app:rank_onecollege}, we show that in a one-college case, \cref{cond} holds for all distribution of $\eta_{1i}$ except for the exponential distribution. In Online Appendix \ref{app:rank_para}, we analyze common logit and probit models with two or more colleges. The results suggest that \cref{cond} is non-restrictive and satisfied for a wide range of values of $(\widehat{w},\widetilde{w})$; in fact, the \textit{failure} of \cref{cond} imposes strict restrictions on the supply, or the conditional probabilities of different feasible sets.

\begin{prop} \label{prop_nonpara} Under Assumptions \ref{assm_struc},  \ref{assm_nonpara}, \ref{indep}, and Condition \ref{cond}, for $k= 1,\dots,d_{z}$, 
$\{\frac{\partial u^{c}(z)}{\partial z_{i}^{k}},\frac{\partial v^{c}(z)}{\partial z_{i}^{k}},\frac{\partial r^{c}(y_c)}{\partial y_{ic}}\}_{c\in\mathbf{C}}$ are identified for all $\left(z,y\right)$ in the interior of $\mathcal{Z} \times \mathcal{Y}$. 
\end{prop}

Our identification arguments proceed in two steps. First, to identify $\frac{\partial u^{c}(z)}{\partial z_{i}^{k}}$ and  $\frac{\partial v^{c}(z)}{\partial z_{i}^{k}}$, we use the variation in the excluded variables $(y_{i},w_{i})$ to derive a system of linear equations that generalizes equation~\eqref{eq:indsim}. Fixing $y_i=\overline{y}$, for any value $(z,w)$ in the interior of $\mathcal{Z}\times\mathcal{W}$, for each college $d \in \mathbf{C}$, and for any component of $z_{i}$, $z_{i}^{k}$,
\begin{equation}
	\frac{\partial\sigma_{d}\left(z,\overline{y},w\right)}{\partial z_{i}^{k}}=\sum_{c\in\mathbf{C}}\frac{\partial\sigma_{d}\left(z,\overline{y},w\right)}{\partial y_{ic}} \cdot \frac{\partial u^{c}(z)}{\partial z_{i}^{k}}+\sum_{c\in\mathbf{C}}\frac{\partial\sigma_{d}\left(z,\overline{y},w\right)}{\partial w_{ci}} \cdot \frac{\partial v^{c}(z)}{\partial z_{i}^{k}},\label{eq:indsimC}
\end{equation}
which gives $C$ linear equations of $2C$ unknowns.
Note that equation~\eqref{eq:indsimC} is for one value of $w_i$. By evaluating equation~\eqref{eq:indsimC} at $d=1,\dots,C$ and two different values of $w_i$, $\widehat{w}$ and $\widetilde{w}$, and stacking them together, we have
\begin{equation}
	\begin{pmatrix}\frac{\partial\boldsymbol{\sigma}\left(z,\overline{y},\widehat{w}\right)}{\partial z_{i}^{k}}\\
		\frac{\partial\boldsymbol{\sigma}\left(z,\overline{y},\widetilde{w}\right)}{\partial z_{i}^{k}}
	\end{pmatrix}=\Pi(z,\overline{y},\widehat{w},\widetilde{w})\cdot\begin{pmatrix}\frac{\partial\boldsymbol{u}\left(z\right)}{\partial z_{i}^{k}}\\
		\frac{\partial\mathbf{v}\left(z\right)}{\partial z_{i}^{k}}
	\end{pmatrix},\label{eq:np_ideq}
\end{equation} 
where $\boldsymbol{\sigma} \equiv (\sigma_{1},\dots,\sigma_{C})'$, $\boldsymbol{u} \equiv (u^{1},\dots,u^{C})'$, and $\boldsymbol{v} \equiv (v^{1},\dots,v^{C})'$. Both $\Pi(z,\overline{y},\widehat{w},\widetilde{w})$ and the left-hand side are known from the population data.
Hence, the invertibility of $\Pi(z,\overline{y},\widehat{w},\widetilde{w})$ in \cref{cond} guarantees the existence of a unique solution to this system, leading to the identification of  $\frac{\partial u^{c}(z)}{\partial z_{i}^{k}}$ and  $\frac{\partial v^{c}(z)}{\partial z_{i}^{k}}$.  
Importantly, this only requires a pair of distinct values of $w_{i}$ to satisfy \cref{cond}.

Second, we identify $\frac{\partial r^{c}(y_{c})} {\partial y_{ic}}$ for each $c\in\mathbf{C}$ by generalizing equation~\eqref{eq:indsim2} for the one-college example. See the Appendix for detailed proof.

In certain empirical applications, identifying the above derivatives is sufficient, in which case a large-support assumption on excluded variables is not needed. When the functions $(u^{c}, v^{c}, r^{c})$ must be identified, a full-support condition on their arguments, $(z_{i}, y_{i})$, is often imposed. This is the case for our next result of identifying the joint distribution $F$ and cutoffs $\delta_c$. Additionally, a full-support assumption on $w_i$ is also required. Hence, we will assume all observables, $(z_i,y_i,w_i)$, have full support.  \subsection{Identifying the Cutoffs and Joint Distribution of Unobservables}\label{sec:id_dist}
We now formalize the assumptions that are needed for the identification of the cutoffs, $\{\delta_c\}_c$, and the joint distribution of unobservables, $F$.
\begin{assm}\label{assm:F_id} 
	For each $c\in\mathbf{C}$, 
		(i) the functions $u^{c}+r^{c}$ and $v^{c}$ are identified;\footnote{\label{fn:id_fn}This requires \cref{prop_nonpara}, a full support assumption on $(z_i,y_i)$, and location normalization on $u^{c}+r^{c}$ and $v^{c}$ for each $c$. See \cref{rm:location} for a discussion on location normalization.}
		(ii)  $y_{ic}$ and $w_{ci}$ possess an everywhere positive Lebesgue density conditional
		on $z_{i}$;  
		(iii) the range of the function $r^{c}$ is the whole real line, and $\mathcal{W} =  \mathbb{R}^C$; and 
		(iv) the $\rho_c$-quantile of the marginal distribution of $\eta_{ci}$ is $0$, i.e., $\text{Quantile}_{\eta_{ci}}(\rho_c)\equiv \inf\{\eta_{c}:  F_{\eta_{ci}}(\eta_{c}) \geq \rho_{c} \} =0,$ for an arbitrary $\rho_c \in (0,1)$.
\end{assm}

Part~(iv) is a location normalization on college preferences, as mentioned in \cref{rm:location}, which can be done college-by-college. Alternatively, one may replace part~(iv) by normalizing cutoffs $\{\delta_{c}\}_c$ to zero.

Before presenting our formal results, we give some intuitions. To identify cutoff~$\delta_{c}$,  under the full-support assumption on student preferences (parts ii and iii of \cref{assm:F_id}), we consider a mass of students to whom all colleges except for $c$ are unacceptable. {The probability of these students matching with $c$ is $1-F_{\eta_{ci}}(\delta_c - v^c(z_{i}) -w_{ci})$. Given the location normalization of $F_{\eta_{ci}}$ (part~iv of  \cref{assm:F_id}), finding the maximum value of $v^c(z_{i})+w_{ci}$ that sets this probability to $1-\rho_c$ identifies $\delta_{c}$.}

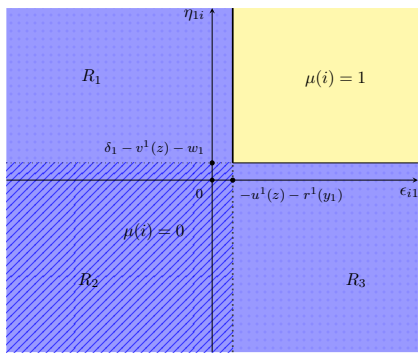
\begin{figure}[h!]
	\begin{centering}
		\begin{tikzpicture}[scale=0.65] 
			
	    \tikzset{every pin edge/.style={draw=black,<-, very near end}}
		\begin{axis} 
		[
		ymin=-5,
		xmin=-5,
		ymax=5,
		xmax=5,
		width=10cm, 
		axis on top=true, 
		axis x line=middle, 
		axis y line=middle,
		yticklabels={,,},
		xticklabels={,,}
		] 
		\draw[fill=yellow!50,opacity=0.8, draw=none]      (axis cs:0.5,0.5) -- (axis cs:0.5,5) -- (axis cs:5,5) -- (axis cs:5,0.5) -- cycle ;
		\draw[fill=blue!50,opacity=0.8, draw=none]      (axis cs:0.5,0.5) -- (axis cs:5,0.5) -- (axis cs:5,-5) -- (axis cs:-5,-5) --  (axis cs:-5,5) -- (axis cs:0.5,5) --cycle ;
		\draw[fill=blue!50,opacity=0.8, draw=none, pattern=north east lines, pattern color=blue] (axis cs:0.5,0.5) -- (axis cs:0.5,-5) -- (axis cs:-5,-5) -- (axis cs:-5,0.5)  --cycle ;
		\draw[fill=blue!50,opacity=0.3, draw=none, pattern=dots, pattern color=blue] (axis cs:0.5,0.5) -- (axis cs:0.5,5) -- (axis cs:-5,5) -- (axis cs:-5,0.5)  --cycle ;
		\draw[fill=blue!50,opacity=0.3, draw=none, pattern=dots, pattern color=blue] (axis cs:0.5,0.5) -- (axis cs:0.5,-5) -- (axis cs:5,-5) -- (axis cs:5,0.5)  --cycle ;

		\addplot+ [mark=none, black, line width = 0.5, domain=0.5:5] {0.5}; 
		\addplot +[mark=none, black, line width = 1] coordinates {(0.5,0.5) (0.5, 5)};
		
		\addplot+ [name path = A, mark=none, darkgray, dotted, line width = 0.5, domain=-5:0.5] {0.5}; 
		\addplot +[name path = C, mark=none, dotted, darkgray, line width = 1] coordinates {(0.5, -5) (0.5, 0.5)};

		\node [below] at (axis cs: -3,-2.5) {\footnotesize $R_2$};
		\node [above] at (axis cs: 3,2.5) {\footnotesize $\mu(i)=1$};
		\node [left] at (axis cs: -2.5,3) {\footnotesize $R_1$};
		\node [left] at (axis cs: -0.5,-1.5) {\footnotesize $\mu(i)=0$};
		\node [below] at (axis cs: 3.5,-2.5) {\footnotesize $R_3$};
		\node [below] at (axis cs: 4.8,0) {\footnotesize $\epsilon_{i1}$};
		\node [left] at (axis cs: 0,4.8) {\footnotesize $\eta_{1i}$};

		\node[label={230:{\scriptsize 0}},circle,fill,inner sep=1pt] at (axis cs:0,0) {};

		\node [below right] at (axis cs: 0.5,0) {\scriptsize $-u^1(z)-r^1(y_1)$};
		\node [above left] at (axis cs: 0,0.5) {\scriptsize $\delta_1-v^1(z)-w_1$};
		\addplot[mark=*, mark size=1.2pt] coordinates {(0,0.5)}; 
		\addplot[mark=*, mark size=1.2pt] coordinates {(0.5,0)} ;

		\end{axis} 
		\end{tikzpicture}
		\par\end{centering}
	\centering{\caption{\label{cdf}Partitioning the Space of Unobservables in the One-college Case}}
		\begin{flushleft}
		{\footnotesize\noindent \textit{Notes:} This figure shows the partition of the space of the unobservables $(\epsilon_{i1},\eta_{1i})$  by the matching outcome ($\mu(i) = 0$ or $1$) in a one-college setting given  $(z_{i}, y_{i1}, w_{1i}) = (z,y_1,w_1)$.}   
	\end{flushleft}
\end{figure}

To identify $F$, we use the conditional probability of being \textit{unmatched}. Let us illustrate the intuition in the same one-college example as in \cref{id_eg}. For any given value $(z,y_1,w_1)$,
Figure~\ref{cdf} shows the partition of the space of $(\epsilon_{i1},\eta_{1i})$
by matching outcome, $\mu\left(i\right)=0$ or $1$. The conditional probability of $\mu(i)=0$, highlighted in purple in the figure, can be decomposed into three parts, $R_{1}$,
$R_{2}$, and $R_{3}$. Moreover,
\begin{align}
& \mathbb{P}(\mu(i)=0  |(z_{i},y_{i1},w_{1i}) = (z,y_1,w_1)) \notag \\
= & \mathbb{P}(\epsilon_{i1}<-u^1(z)-r^1\left(y_1\right)\text{ or }\eta_{1i}<\delta_1-v^1(z)-w_1  |z,y_1,w_1 )\nonumber \\
= & \mathbb{P}(R_{1}\cup R_{2}  | z,y_1,w_1 )+\mathbb{P}(R_{3}\cup R_{2}  | z,y_1,w_1 )-\mathbb{P}(R_{2}  | z,y_1,w_1 ),\label{decomp}
\end{align}
where $\ensuremath{\mathbb{P}(R_{2} | z,y_1,w_1 })$ is the joint CDF of $(\epsilon_{i1},\eta_{1i})$,  $F\left(-u^{1}(z)-r^{1}\left(y_1\right),\delta_{1}-v^{1}(z)-w_1\right)$, our parameter of interest.

Further, $\mathbb{P}(R_{1}\cup R_{2} | z,y_1,w_1 )=F_{\epsilon_{i1}}(-u^1(z)-r^1(y_1))$ is the
marginal CDF of $\epsilon_{i1}$. It can be identified by considering the subset of students whose value of $w_{1i}$ is high enough so that the college will be feasible no matter what value $\eta_{1i}$ takes. That is, we can identify
$\mathbb{P}(R_{1}\cup R_{2} |z,y_1,w_1 )$ by ``shutting down'' the effects
of college preferences. Similarly, $\mathbb{P}(R_{3}\cup R_{2} | z,y_1,w_1 )$
can be identified by focusing on the subset of students whose value
of $r^1\left(y_{i1}\right)$ is large enough so that those students find college~1 acceptable no matter what value $\epsilon_{i1}$ takes.  As $\mathbb{P}(\mu(i)=0 | z,y_1,w_1 )$ is known from the population data, once $\mathbb{P}(R_{1}\cup R_{2} | z,y_1,w_1  )$ and $\mathbb{P}(R_{3}\cup R_{2} | z,y_1,w_1 )$ are identified, equation~\eqref{decomp} implies that $\mathbb{P}(R_{2} |z,y_1,w_1  )$ and thus $F$ are identified.

\begin{prop}\label{np_id_F} 
Under Assumptions  \ref{assm_struc}, \ref{indep}, and \ref{assm:F_id}, (i) cutoffs $\{\delta_{c}\}_c$ are identified; (ii) the joint distribution of $(\epsilon_{i},\eta_{i})$, $F$, is identified;  and (iii) acceptability threshold $T_c $ for any college~$c$ with vacancies is identified, but $T_c$ for $c$ with a binding capacity constraint is only partially identified, $T_c \leq \delta_{c}$. 
\end{prop}

To show part~(ii) in a many-college setting, with the full, large support (parts ii and iii of Assumptions \ref{assm:F_id}), we apply the same argument as in the one-college example to each college~$c$ sequentially, using extreme values of $r^c(y_{ic})$ or $w_{ci}$ to ``shut down'' the effects of students' preference for college $c$ or $c$'s preference for students.

Part~(iii) is a consequence of part~(i) and the definition of cutoffs (equation~\ref{cutoff}). If $c$ reaches its capacity, $T_c$ can be any value below $\delta_{c}$ and result in the same stable matching. We can identify $T_c$ by imposing additional assumptions, e.g., a full-capacity college having the same acceptability threshold as some college with vacancies.

     \subsection{Practical Issues\label{sec:practical}}
When our identification results are taken to the data, there can be practical issues. For example, vector~$z_i$ may include discrete variables such as gender, and the researcher may not have sufficient excluded variables. Below, we address these issues.

\noindent \textbf{Discrete Random Variables.} Our results can be extended to the case where $z_i$ contains discrete variables. 
Suppose that $z_{i}=(z_{i}^1,z_{i}^2)$, where $z_{i}^1$ is a vector of discrete variables and $z_{i}^2$ is a vector of continuous variables. Let the support of $z_i^1$ be a finite set of points $\{z^{1,1}, z^{1,2},\dots, z^{1,J}\}$. For $z_i^1 = z^{1,j}$, we define functions $(u^{c,j}+r^c,v^{c,j})$. Conditional on $z_i^1=z^{1,1}$, we apply the results in \cref{sec:derivative,sec:id_dist} to identify $\{u^{c,1}+r^c,v^{c,1},\delta_c\}_{c}$ and $F$, requiring $y_i$ and $w_i$ being continuous, Assumptions~\ref{assm_nonpara}(ii) and (iii), \ref{indep}, \ref{assm:F_id}(ii)-(iv), \cref{cond}, location normalization on $u^{c,1}+r^c$ and $v^{c,1}$ for each $c$, and $(y_i,z_{i}^2)$ having full support. For $j\neq 1$, conditional on $z_i^1=z^{1,j}$, using the conditional match probability of a mass of students for whom $c$ is the only acceptable college, we can use $F$ to identify $v^{c,j}$ for each $c$; similarly, using the conditional match probability of a mass of students whose feasible set is $\{0,c\}$, we can identify the function $u^{c,j}+r^c$.

\noindent \textbf{Insufficient Excluded Variables.} Allowing for college-level heterogeneity implies that we need to recover $3C$ preference parameters, $\{u^c,r^c,v^c\}_c$. As seen in \cref{sec:derivative}, we need $2C$ excluded variables for identification of their derivatives. 

The lack of sufficient excluded variables leads to a loss of identification, but not all is lost.  We show that there is a trade-off between the identifiable degree of heterogeneity and the number of excluded variables.  Consider $\bfC_1, \bfC_2 \subseteq \bfC$ with cardinality $\kappa_1$ and $\kappa_2$, respectively. For all $c\in\bfC_1$, there is a single excluded variable in student preferences, $y_{ic}=y_{i\ast}$; for all $c\in\bfC_2$, there is a single excluded variable in college preferences, $w_{ci}=w_{i\ast}$.  We have the following identification result. 

\begin{prop}\label{deg-of-heter}
Suppose the preference heterogeneity is reduced: $u^c = u^{\ast}$ and $r^c = r^{\ast}$ for all $c\in\bfC_1$, and $v^c = v^{\ast}$ for all $c\in\bfC_2$. 
For any $c\in\mathbf{{C}}$, the derivatives of $(u^c,r^c,v^c)$ are identified if: (i) Assumptions~\ref{assm_struc}, \ref{assm_nonpara}, and \ref{indep} hold; and
(ii) the following rank conditions hold:
(a) If $C\leq \kappa_1 + \kappa_2 - 2$, for any $z$ in the interior of $\mathcal{Z} $, there exists a value of $w_i$, $\widehat{w}$, in $w_i$'s support conditional on $(z_{i},y_{i})=(z,\overline{y})$ such that $(\Pi_{y}(z,\overline{y},\widehat{w}),\,\Pi_{w}(z,\overline{y},\widehat{w}))$ has column rank at least $2C-\kappa_1-\kappa_2+2$;
or (b) if $C> \kappa_1 + \kappa_2 - 2$, for any $z$ in the interior of $\mathcal{Z} $, there exist two values of $w_i$, $\widehat{w}$ and $\widetilde{w}$, in $w_i$'s support conditional on $(z_{i},y_{i})=(z,\overline{y})$ such that $\Pi(z,\overline{y},\widehat{w},\widetilde{w})$ has column rank at least $2C-\kappa_1-\kappa_2+2$.

\end{prop}

In other words, we need only $C-\kappa_1+1$ excluded variables on the student side and $C-\kappa_2+1$ on the college side to identify the less heterogeneous model.\footnote{Note that the rank condition in  \cref{deg-of-heter} is weaker than Condition \ref{cond}.} Importantly, this still allows for coefficient heterogeneity across colleges in $\bfC_1$ and $\bfC_2$. For example, it can be incorporated parametrically by interacting college-specific observables with $z_i$. For $c\in\bfC_1$, let $u^c(z_i) = \beta p_c z_i$, where $p_c$ is a college-specific observable. This amounts to $z_i$ having a college-specific parameter $\beta_c = \beta p_c$.

Since our identification approach is constructive, it directly implies a nonparameteric estimator. In the Monte Carlo simulations in Online Appendix~\ref{app:mc}, we apply this estimator in a semiparametric setting to avoid the well-known curse of dimensionality; yet the curse remains. Instead, we find that a parametric model based on a Bayesian approach works well. Below, we apply it to a real-life setting.


\section{Secondary School Admissions in Chile}\label{sec:empirical}
    
Guided by our identification results, we study the admissions to public and private secondary schools (grades 9-12) in Chile in 2007. The market is organized similarly to college admissions in the US. It is decentralized, both sides have preferences, and students do not submit rank-order lists of schools. We use a parametric Bayesian approach for preference estimation and then conduct counterfactual analysis.

\subsection{Institutional Background and Data}\label{sec:background}
Since 2003, secondary education has been compulsory for all Chileans up to 21 years of age. In principle, a public school must accept any student who is willing to enroll; a private school can be subsidized by the government or non-subsidized, but in either case, it can select students based on its preferences. 

We focus on a relatively independent market, {\it Market Valparaiso}, that includes five municipalities (Valparaiso, Vi\~na del Mar, Concon, Quilpue, and Villa Alema\~na) as defined in \cite{gazmuri2017school}. Our data includes the municipality of a student's residence and the geographical coordinates of each school, which identifies every agent in the market.  A student is defined to be in Market Valparaiso if in 2008 she resided in a municipality within the boundary of the market. A secondary school is in Market Valparaiso if it is located in and admits students from the market.\footnote{There are 38 secondary schools located in Market Valparaiso that do not have any 9th graders from Market Valparaiso in 2007 and another 15 that have fewer than three 9th graders from Market Valparaiso on average in 2005, 2007, and 2009. We consider these 53 schools as outside options.} In total, there are $9,304$ students and $125$ schools. This reasonably large size makes it plausible that the continuum market in \cref{id} is a good approximation.  

We use the SIMCE dataset, provided by \textit{La Agencia de Calidad de la Educaci\~on }(Agency for the Quality of Education), on all 10th graders in 2008 to identify who started secondary school  in 2007.  The SIMCE is Chile's standardized testing program and tracks students' math  and  language  performance. The data includes students' parental income,  parental  schooling,  and  other  characteristics  from  a  parental  questionnaire sent home with students. Most school attributes are calculated from student characteristics of {\it the 10th graders in a school in 2006}, and thus are pre-determined in the 2007 admissions that we study.  
As tuition fees are largely fixed within each school and not completely flexible at the student level, we consider the problem as matching without transfers. See Online Appendix~\ref{app:data} for more details on data construction and summary statistics of student characteristics and school attributes.

    \subsection{Empirical Model} \label{sec:EmpModel}
We allow student preferences to be school-type-specific. For student~$i$, the utility of attending school~$c$ of type~$t$ $\in$ \{public, private non-subsidized, private subsidized\} is
\begin{eqnarray}\label{eq:stu_u}
  u_{ict} = \alpha_{ft} \times female_i  + \alpha_{mt} \times male_i + X'_{ic}\beta_t + \zeta_t \epsilon_{ic}, 
\end{eqnarray}
where $\alpha_{ft}$ is a school-type fixed effect for female students;  $female_i$ is a dummy variable for female;  $\alpha_{mt}$ and $male_i$ are similarly defined; $\epsilon_{ic}$ is  i.i.d.\ standard normal; $\zeta_t$ ($>0$) allows type-specific variances; and $X_{ic}$ are student-school-specific variables, including (i) the distance between $i$'s residence and school~$c$; 
(ii) 6 school attributes, and (iii) 3 interactions between school attributes and student characteristics.

Each student has an outside option,  $u_{i0} = \epsilon_{i0}$, with $\epsilon_{i0}$ being standard normal. We impose the usual scale normalization through $\zeta_t=1$ for public school,\footnote{The variance of $\epsilon_{i0}$ is also normalized to be one because there is insufficient variation to estimate the variance of $u_{i0}$. Only 59 students (out of 9,304) choose an outside option (see \cref{tab:sumstu}).} while the location normalization is imposed by setting the deterministic part of $u_{i0}$ to zero.

School preferences are also type-specific, but public schools do not have a utility function because they cannot select students.  For private school~$c$ of type~$t$ (subsidized or non-subsidized), its acceptability threshold is $T_c=0$, and its utility function is
\begin{eqnarray}\label{eq:sch_u}
  v_{cit} = \theta_t + Z'_{ci}\gamma_t + \eta_{ci}, 
\end{eqnarray}
where $\theta_t$ is a type-specific intercept; $\eta_{ci}$ is i.i.d.\ standard normal; and the vector $Z_{ci}$ includes 
(i) 5 student characteristics and (ii) 3 interactions between student characteristics and school attributes.

 In school preferences, the variance of $\eta_{ci}$ being one is the scale normalization and $T_c=0$ is the location normalization.  Allowing the type-specific intercept $\theta_t$ and assuming $T_c=0$  imply that schools of the same type have the same acceptability threshold. Because each type has some schools with vacancies, we can separately identify $\delta_c$ and $T_c$. Otherwise, we would lose its identification (\cref{np_id_F}).

The above specification uses  distance as an \textit{$(i,c)$-pair-specific} excluded variable in student preferences.  In school preferences, math and language scores are \textit{student-specific} excluded variables, implying that there are insufficient excluded variables to estimate school-specific utility functions. Using the results in \cref{deg-of-heter}, we limit preference heterogeneity and assume type-specific utility functions for schools. Our scale and location normalization is guided by \cref{np_id_F} and \cref{rm:location}, while certain normalization is imposed via $F$. 

\subsection{Estimation and Results}\label{sec:results}

We use a Bayesian approach with a Gibbs sampler for the estimation, which is first illustrated in Monte Carlo simulations (Online Appendix~\ref{sec:mc_bayes}). We provide the details on the updating of the Markov Chain in Online Appendices~\ref{sec:mc_bayes} and \ref{app:bayes}. 

The estimation results are summarized in \cref{tab:est_results}. A caveat is in order when we interpret the results. Because we do not deal with endogeneity issues that may arise due to the correlation between preference shocks and school attributes or student characteristics, the estimates  may not have a causal interpretation.  

\begin{table}[h!]
	\centering \footnotesize
	\caption{Estimation Results: Student and School Preferences}   \label{tab:est_results}
	\resizebox{\textwidth}{!}{
		\begin{tabular}{lcccccccc}
			\toprule
			& \multicolumn{2}{c}{} & & \multicolumn{5}{c}{Private Schools}  \\
			\cline{5-9}
			& \multicolumn{2}{c}{Public schools} &&  \multicolumn{2}{c}{Subsidized} &&  \multicolumn{2}{c}{Non-subsidized} \\
			\cmidrule{2-3}     \cmidrule{5-6}     \cmidrule{8-9}
			& \multicolumn{1}{c}{{coef.}}  & \multicolumn{1}{c}{{s.e.}} &&  {coef.} & {s.e.} & &  {coef.} & {s.e.} \\
			\midrule
			\textit{\textbf{Panel A. Student Preferences}} &       &       &       &       &       & & &  \\
			Female & \multicolumn{1}{c}{0.458} & \multicolumn{1}{c}{(0.834)} &       & 10.500 & (0.611) &       & 9.848 & (5.631) \\
			Male  & \multicolumn{1}{c}{0.442} & \multicolumn{1}{c}{(0.834)} &       & 10.387 & (0.610) &       & 9.089 & (5.628) \\
			Distance & \multicolumn{1}{c}{-0.175} & \multicolumn{1}{c}{(0.004)} &       & -0.145 & (0.008) &       & -0.624 & (0.070) \\
			log(tuition) & \multicolumn{1}{c}{0.577} & \multicolumn{1}{c}{(0.119)} &       & -0.385 & (0.102) &       & -15.645 & (1.567) \\
			log(tuition) $\times$ log(income) & \multicolumn{1}{c}{-0.019} & \multicolumn{1}{c}{(0.010)} &       & 0.044 & (0.008) &       & 0.881 & (0.084) \\
			log(median income) & \multicolumn{1}{c}{-0.307} & \multicolumn{1}{c}{(0.072)} &       & -0.720 & (0.067) &       & 2.454 & (0.544) \\
			Teacher experience & \multicolumn{1}{c}{0.003} & \multicolumn{1}{c}{(0.002)} &       & 0.005 & (0.001) &       & 0.020  & (0.015) \\
			Fraction of female students & \multicolumn{1}{c}{-0.010} & \multicolumn{1}{c}{(0.034)} &       & -0.320 & (0.054) &       & -1.428 & (0.773) \\
			Average composite score & \multicolumn{1}{c}{-1.948} & \multicolumn{1}{c}{(0.124)} &       & -2.354 & (0.132) &       & -16.104 & (2.241) \\
			Average composite score $\times$ Composite score & \multicolumn{1}{c}{6.282} & \multicolumn{1}{c}{(0.232)} &       & 6.209 & (0.237) &       & 18.167 & (1.616) \\
			Average mother's education & \multicolumn{1}{c}{0.038} & \multicolumn{1}{c}{(0.023)} &       & -0.283 & (0.024) &       & -1.572 & (0.298) \\
			Average mother's education $\times$ Mother's education & \multicolumn{1}{c}{0.007} & \multicolumn{1}{c}{(0.001)} &       & 0.011 & (0.001) &       & 0.056 & (0.007) \\
			Standard deviation of the utility shock & \multicolumn{2}{c}{Normalized to 1} &       & 1.093 & (0.063) &       & 8.646 & (0.859) \\
			[0.5em]
			\textit{\textbf{Panel B. School Preferences}} &       &&        &       &  &      &       &  \\
			Constant &       &       &       & 3.686 & (0.852) &       & 20.283 & (1.863) \\
			Female &       &       &       & -3.545 & (0.193) &       & -1.438 & (0.274) \\
			Female $\times$  Fraction of female &       &       &       & 6.892 & (0.333) &       & 1.968 & (0.443) \\
			Math score &       &       &       & -3.251 & (0.541) &       & -7.053 & (0.920) \\
			Average math score $\times$ Math score &       &       &       & 7.955 & (0.938) &       & 6.217 & (0.985) \\
			Language score &       &       &       & -0.030 & (0.460) &       & -4.593 & (0.859) \\
			Average language score $\times$ Language score &       &       &       & 1.053 & (0.822) &       & 5.087 & (1.135) \\
			Mother's education &       &       &       & -0.010 & (0.016) &       & -0.136 & (0.047) \\
			log(income) &       &       &       & -0.225 & (0.072) &       & -1.030 & (0.146) \\
			\bottomrule   
	\end{tabular}}
	\begin{tabnotes}
		This table presents the posterior mean and standard deviation of each coefficient in student and school utility functions (equations~\ref{eq:stu_u} and \ref{eq:sch_u}). The Bayesian approach goes through a Markov Chain 1.75 million times, and the last 0.75 million iterations are used to calculate these statistics. Our Monte Carlo simulations suggest that the posterior standard deviation provides a useful estimate of the standard deviation of our estimators, albeit with a slight tendency to underestimate. See \cref{tab:mc_bayes} for more details.
	\end{tabnotes}
\end{table}%

Panel A shows the estimates of student preferences. Most coefficients are of an expected sign. Interestingly, the coefficient on tuition is positive in the utility function for public schools, and parental income pushes it towards negative. This may reflect the fact that tuition at public schools is generally low (see \cref{tab:sumsch}) and may be correlated with unobserved school quality.
There are also a few coefficients with an unexpected sign in school preferences.  Panel B shows that non-subsidized schools negatively value a student's parental income and mother's education. This may be because, on average, students at those schools often have a high income (above 1.4 million CLP; see \cref{tab:sumstu}) and a highly educated mother (above 17 years).

Recall that each school's acceptability threshold is normalized to zero, so we can use the estimated school preferences to calculate if a student is acceptable to a private school. On average, a subsidized school finds 79\% of the students acceptable, while a non-subsidized school finds 84\% acceptable.  The higher acceptability rate at non-subsidized schools does not imply that students are more often matched with them because their high tuition lowers their desirability to many students, especially those with a low parental income (see Table~\ref{tab:est_results}). 

Before conducting counterfactual analysis, we evaluate model fit in Online Appendix~\ref{app:bayes}. It shows that our model fits the data reasonably well when we compare the observed matching with the one predicted based on our model.  

\subsection{Counterfactual: Prioritizing Low-income Students}

We consider a counterfactual policy in which students from low-income families are prioritized for admissions to all schools. A student is of low income if her parental income is among the lowest 40\%.\footnote{This resembles a policy adopted in 2008 in Chile as documented by \cite{gazmuri2017school}. It benefits 44\% of elementary school students in 2012 in terms of admission priorities and a tuition waiver. Online Appendix \cref{sumstu_byinc} shows summary statistics of the students by income status.} 
Each private school's preferences over students are made lexicographical: low-income students are above others, and within each group of students, a school ranks them as in the current regime; all low-income students are acceptable, while others' acceptability is the same as in the current regime. We do not change anything in student or school preferences beyond the admission priority, which may mitigate the potential bias in our estimation due to aforementioned unaddressed endogeneity issues.

To simulate the counterfactual outcome, we keep 1,500 draws of the utilities in the Markov Chain in the Bayesian estimation.\footnote{{Specifically, there are 15 blocks of 100 draws.  The blocks are equally spaced in the 0.75 million iterations in the Markov chain that are used to calculate the posterior means and standard deviations.}} We run the Gale-Shapley deferred acceptance with each draw and obtain 1,500 sets of counterfactual stable matchings. We then report the average of these counterfactual matchings; recall that we only have one matching outcome under the current regime---the observed one.

\begin{table}[h!]
  \centering \footnotesize
  \caption{Sorting and Student Welfare in the Current and Counterfactual Regimes}  \label{tab:counterfactual}%
\resizebox{0.9\textwidth}{!}{
    \begin{tabular}{lccccc}
\toprule          
    & \multicolumn{2}{c}{Low-income students} &          & \multicolumn{2}{c}{Non-low-income students}  \\
\cline{2-3} \cline{5-6}
& {Current} & {Counterfactual} &       & {Current} & {Counterfactual} \\
& (1) & (2) &       & (3) & (4) \\
\midrule
\cmidrule{1-3}\cmidrule{5-6}    
        Average composite score (same cohort) at matched school  & 0.374 & 0.385 &       & 0.585 & 0.576 \\
        Average parental income (same cohort)  at matched school & 216,389
         & 223,003
          &       & 592,848
           & 587,884
            \\
    [0.1em]
    {\it Fraction enrolled at each school type:} & & & & & \\
    \multicolumn{1}{r}{Public} & 0.679 & 0.602 &       & 0.233 & 0.260 \\
    \multicolumn{1}{r}{Private subsidized} & 0.315 & 0.392 &       & 0.533 & 0.504 \\
    \multicolumn{1}{r}{Private non-subsidized} & 0.002 & 0.002 &       & 0.227 & 0.228 \\
    \multicolumn{1}{r}{Outside option} & 0.004 & 0.003 &       & 0.008 & 0.008 \\
         [0.1em]
    {\it Welfare effects of moving from current to counterfactual:} & & & & & \\
     \multicolumn{1}{r}{Average utility change (reduction in distance, km)} &       \multicolumn{2}{c}{0.332}   &       & \multicolumn{2}{c}{$-$0.225} \\
     \multicolumn{1}{r}{Winners (fraction)} &      \multicolumn{2}{c}{0.120} &       &       \multicolumn{2}{c}{0.000} \\
     \multicolumn{1}{r}{Losers (fraction)} &       \multicolumn{2}{c}{0.000}       &    &   \multicolumn{2}{c}{0.072} \\
     \multicolumn{1}{r}{Indifferent (fraction)} &       \multicolumn{2}{c}{0.881} &       &       \multicolumn{2}{c}{0.927} \\
\bottomrule
\end{tabular}}
    \begin{tabnotes}
The outcome in the current regime is the one observed in the data. To simulate the counterfactual outcome, we keep 1,500 draws of the utilities in the Markov Chain in the Bayesian estimation and obtain 1,500 sets of stable matching outcomes. The statistics for the counterfactual regime are averages across the 1,500 outcomes. The average utility change is measured in terms of willingness to travel to a public school in kilometers. 
    \end{tabnotes}
\end{table}%

\cref{tab:counterfactual} presents the results. There are several noticeable patterns when we move from the current regime to the counterfactual. 
First, low-income students are in schools that have higher-ability and higher-income students in the same cohort, while the opposite is true for non-low-income students.  Second, some low-income students leave public schools for private subsidized schools, while crowding out some other students to public schools. Lastly, the policy benefits low-income students and hurts others. On average, low-income students' welfare gain is equivalent to decreasing travel distance (to a public school) by 0.332 km. This gain is concentrated among 12\% of the low-income students, while others are not affected. Correspondingly, 7.2\% of the non-low-income students are worse off, and none is better off.

These results indicate that low-income students dislike private non-subsidized schools. We explore why it is the case. With the 1,500 draws of student preferences, we examine low-income students' favorite school of each type. We find that low-income students on average value their favorite public school at $2.72$ and private subsidized school at $2.56$, while their favorite non-subsidized school is only valued at $-21.76$, in general unacceptable to them.  We calculate the contribution of different variables to these differences by shutting down their effect in the utility functions. The results show that high tuition at non-subsidized schools is the main contributor. It is worth noting that tuition could be correlated with unobserved school quality. Hence, low-income students may dislike a non-subsidized school because of its high costs and/or their tastes. Additionally, mother's education, both a student's own and a school's average, is also an important factor.  Student ability, measured by their composite score, and distance to each school do not appear to be as important.

In sum, giving low-income students access to schools fails to significantly change matching outcomes due to their own preferences. Low-income students are deterred from private non-subsidized schools by their high tuition. These findings are in line with the preference heterogeneity documented in public school choice \citep[see, e.g.,][]{abdulkadiroglu_welfare_2017,kapor2020heterogeneous},  although tuition  plays no role there.


\section{Concluding Remarks}\label{sec:conclusion}
    We study nonparametric identification of agent preferences in many-to-one two-sided matching without transfers. We derive a set of sufficient conditions for identification and provide guidance for empirical studies. For example, our results clarify the data requirement for the identification of various degrees of preference heterogeneity. To take our results to the data of a reasonably sized market, we propose a Bayesian approach with a Gibbs sampler whose performance is illustrated in Monte Carlo simulations.  Our model encompasses many real-life matching markets, such as college admissions and school choice in many countries, in which our identification results and empirical method can be applied. Hence, this paper opens a new avenue for empirical research. 

We illustrate our method in the context of secondary school admissions in Chile. As an example of the usefulness of the estimates, we consider a counterfactual policy in which students from low-income families are prioritized for admissions to all schools. Although the policy benefits low-income students, its effects are small. Such insights are difficult to obtain without estimating the preferences of both sides. In this sense,  our method can help provide an ex-ante evaluation of a range of alternative policies.

\appendix

\section*{Appendix}\label{app:proofs}
Throughout this appendix, for notational simplicity, we let $x_i=(z_i,y_i,w_i)$.
\begin{proof}[Proof of Proposition \ref{prop_nonpara}]

Equation \eqref{ideq2} can be rewritten as 
\begin{equation*}
	\sigma_{c}(x_{i})=\Lambda_{c}(\tau_{i1},\dots,\tau_{iC};\iota_{1i},\dots,\iota_{Ci})=\Lambda_{c}(\tau_{i},\iota_{i}),
\end{equation*}
where $\tau_{i}\equiv(\tau_{i1},\dots,\tau_{iC})$, $\iota_{i}=(\iota_{1i},\dots,\iota_{Ci})$, and $\Lambda_{c}$ denotes some unknown function. Recall that $\tau_{ic}=u^c(z_{i})+r^c(y_{ic})$ and $\iota_{ci}=v^{c}(z_{i})+w_{ci}$. Under Assumption \ref{assm_nonpara}, $\Lambda_{c}$, $u^c$, $r^c$, and $v^c$ are continuously differentiable and the observables are all continuously distributed.

For colleges $c,\ d\in\mathbf{C}$, taking derivatives of $\sigma_{d}(x_i)$ w.r.t.\ $y_{ic}$, $w_{ci}$,
and $z_{i}^{k}$ respectively, one obtains
\begin{eqnarray}
	\frac{\partial\sigma_{d}(x_i)}{\partial y_{ic}}  = \frac{\partial\Lambda_{d}(\tau_{i},\iota_{i})}{\partial\tau_{ic}} \frac{\partial r^{c}(y_{ic})}{\partial y_{ic}},\quad
	 \frac{\partial\sigma_{d}(x_i)}{\partial w_{ci}} = \frac{\partial\Lambda_{d}(\tau_{i},\iota_{i})}{\partial\iota_{ci}}, 
	\label{eq:dydw} \\
	\frac{\partial\sigma_{d}(x_i)}{\partial z_{i}^{k}} =\sum_{c\in\mathbf{C}}\frac{\partial\Lambda_{d}(\tau_{i},\iota_{i})}{\partial\tau_{ic}} \frac{\partial u^{c}(z_{i})}{\partial z_{i}^{k}} +\sum_{c\in\mathbf{C}}\frac{\partial\Lambda_{d}(\tau_{i},\iota_{i})}{\partial\iota_{ci}} \frac{\partial v^{c}(z_{i})}{\partial z_{i}^{k}}. \label{eq:dzc}
\end{eqnarray}

First, we show the identification of the derivatives of the functions $u^c$ and $v^c$.  We fix $y_{i}=\overline{y}$ and consider $(z,w)$ in the interior of $\mathcal{Z}\times \mathcal{W}$. Recall that $\frac{\partial r^{c}(\overline{y}_{c})}{\partial y_{ic}} =1$ due to the scale normalization. Substituting equation~\eqref{eq:dydw} into equation~\eqref{eq:dzc}, we get
\begin{equation}
	\frac{\partial\sigma_{d}(z,\overline{y},w)}{\partial z_{i}^{k}}=\sum_{c\in\mathbf{C}}\frac{\partial\sigma_{d}(z,\overline{y},w)}{\partial y_{ic}} \frac{\partial u^{c}(z)}{\partial z_{i}^{k}}+\sum_{c\in\mathbf{C}}\frac{\partial\sigma_{d}(z,\overline{y},w)}{\partial w_{ci}} \frac{\partial v^{c}(z)}{\partial z_{i}^{k}}.\label{eq:dzc-1}
\end{equation}
Suppose that two different values of the $C$-dimensional vector of excluded regressors $w_{i}$, $\widehat{w}$ and $\widetilde{w}$, satisfy \cref{cond}. We define $\widehat{x}=(z,\overline{y},\widehat{w})$ and $\widetilde{x}=(z,\overline{y},\widetilde{w})$.  Further, let $\widehat{\sigma}_{c}=\mathbb{P}(\mu(i)=c|\widehat{x})$ and $\widetilde{\sigma}_{c}=\mathbb{P}(\mu(i)=c|\widetilde{x})$. By evaluating equation~\eqref{eq:dzc-1} at $d=1,\dots,C$ and
$x_i=\widehat{x},\, \widetilde{x}$, and stacking them together, we have
\begin{equation}
\underbrace{\begin{pmatrix}\frac{\partial\widehat{\sigma}_{1}}{\partial z_{i}^{k}}\\
		\vdots\\
		\frac{\partial\widehat{\sigma}_{C}}{\partial z_{i}^{k}}\\
		\frac{\partial\widetilde{\sigma}_{1}}{\partial z_{i}^{k}}\\
		\vdots\\
		\frac{\partial\widetilde{\sigma}_{C}}{\partial z_{i}^{k}}
\end{pmatrix}}_{\Sigma_{z_{i}^{k}}(z,\overline{y},\widehat{w},\widetilde{w})}=\underbrace{\begin{pmatrix}\frac{\partial\widehat{\sigma}_{1}}{\partial y_{i1}} & \cdots & \frac{\partial\widehat{\sigma}_{1}}{\partial y_{iC}} & \frac{\partial\widehat{\sigma}_{1}}{\partial w_{1i}} & \cdots & \frac{\partial\widehat{\sigma}_{1}}{\partial w_{Ci}}\\
		\vdots & \ddots & \vdots & \vdots & \ddots & \vdots\\
		\frac{\partial\widehat{\sigma}_{C}}{\partial y_{i1}} & \cdots & \frac{\partial\widehat{\sigma}_{C}}{\partial y_{iC}} & \frac{\partial\widehat{\sigma}_{C}}{\partial w_{1i}} & \cdots & \frac{\partial\widehat{\sigma}_{C}}{\partial w_{Ci}}\\
		\frac{\partial\widetilde{\sigma}_{1}}{\partial y_{i1}} & \cdots & \frac{\partial\widetilde{\sigma}_{1}}{\partial y_{iC}} & \frac{\partial\widetilde{\sigma}_{1}}{\partial w_{1i}} & \cdots & \frac{\partial\widetilde{\sigma}_{1}}{\partial w_{Ci}}\\
		\vdots & \ddots & \vdots & \vdots & \ddots & \vdots\\
		\frac{\partial\widetilde{\sigma}_{C}}{\partial y_{i1}} & \cdots & \frac{\partial\widetilde{\sigma}_{C}}{\partial y_{iC}} & \frac{\partial\widetilde{\sigma}_{C}}{\partial w_{1i}} & \cdots & \frac{\partial\widetilde{\sigma}_{C}}{\partial w_{Ci}}
\end{pmatrix}}_{\Pi(z,\overline{y},\widehat{w},\widetilde{w})}\times\begin{pmatrix}\frac{\partial u^{1}(z)}{\partial z_{i}^{k}}\\
	\vdots\\
	\frac{\partial u^{C}(z)}{\partial z_{i}^{k}}\\
	\frac{\partial v^{1}(z)}{\partial z_{i}^{k}}\\
	\vdots\\
	\frac{\partial v^{C}(z)}{\partial z_{i}^{k}}
\end{pmatrix}.\label{eq:idenmatrix}
\end{equation}

The first $C$ rows of the matrix $\Pi(z,\overline{y},\widehat{w},\widetilde{w})$ contains derivatives of the conditional match probabilities w.r.t.\ the
excluded regressors, evaluated at $x_{i}=\widehat{x}$. The second $C$ rows of $\Pi(z,\overline{y},\widehat{w},\widetilde{w})$
are constructed similarly, evaluating the derivatives at $x_{i}=\widetilde{x}$.

Under \cref{cond}, there always exist $\widehat{w}$ and $\widetilde{w}$ such that $\Pi(z,\overline{y},\widehat{w},\widetilde{w})$ in equation~\eqref{eq:idenmatrix} is invertible. We then identify the derivatives of $u^c$ and $v^c$ for all $c$ by solving the system of linear equations. Formally, let $\Pi_{z_{i}^{k}}^{c}(z,\overline{y},\widehat{w},\widetilde{w})$
be the matrix formed by replacing the $c^{th}$ column of matrix
$\Pi(z,\overline{y},\widehat{w},\widetilde{w})$ by the vector $\Sigma_{z_{i}^{k}}(z,\overline{y},\widehat{w},\widetilde{w})$ (defined in equation~\ref{eq:idenmatrix}).
By the Cramer's rule, for any $c\in\mathbf{C}$, 
\[
\frac{\partial u^{c}(z)}{\partial z_{i}^{k}}=\frac{|\Pi_{z_{i}^{k}}^{c}(z,\overline{y},\widehat{w},\widetilde{w})|}{|\Pi(z,\overline{y},\widehat{w},\widetilde{w})|},\quad \frac{\partial v^{c}(z)}{\partial z_{i}^{k}}=\frac{|\Pi_{z_{i}^{k}}^{c+C}(z,\overline{y},\widehat{w},\widetilde{w})|}{|\Pi(z,\overline{y},\widehat{w},\widetilde{w})|}.
\]
So far, we have only considered $z_{i}$ that shows up in the utility functions for all colleges and for both sides. For any element of $z_{i}$ that is excluded from certain utility functions, the identification is a special case of the above proof by noting that some derivatives of the utility functions are zero.

Second, we  identify the derivative of the function $r^{c}$ for all $c$. We start with $r^1$ and fix $y_{ic}=\overline{y}_{c}$ for $c\in\mathbf{C}\backslash\{1\}$; under the scale normalization, $\frac{\partial r^{c}(\overline{y}_{c})}{\partial y_{ic}} =1$ for $c\in\mathbf{C}\backslash\{1\}$. For any $(y_1,z,w)$ in the interior of $\mathcal{Y}_1 \times \mathcal{Z} \times \mathcal{W}$, Substituting equation~\eqref{eq:dydw} into equation~\eqref{eq:dzc}, for each $d\in\mathbf{C}$, we obtain
\begin{equation}
\left(\frac{\partial\sigma_{d}}{\partial z_{i}^{k}}-\sum_{c\in\mathbf{C}}\frac{\partial\sigma_{d}}{\partial w_{ci}}\frac{\partial v^{c}}{\partial z_{i}^{k}}-\sum_{\substack{c\in\mathbf{C}\setminus\{1\}}
	}\frac{\partial\sigma_{d}}{\partial y_{ic}}\frac{\partial u^{c}}{\partial z_{i}^{k}}\right)\frac{\partial r^{1}}{\partial y_{i1}}=\frac{\partial\sigma_{d}}{\partial y_{i1}}\frac{\partial u^{1}}{\partial z_{i}^{k}}, \label{eq:dev_r^1}
\end{equation}
where all the terms except for $\frac{\partial r^{1}(y_1)}{\partial y_{i1}}$ are known or already identified. If $\frac{\partial\sigma_{d}}{\partial z_{i}^{k}}-\sum_{c\in\mathbf{C}}\frac{\partial\sigma_{d}}{\partial w_{ci}}\frac{\partial v^{c}}{\partial z_{i}^{k}}-\sum_{\substack{c\in\mathbf{C}\setminus\{1\}}
}\frac{\partial\sigma_{d}}{\partial y_{ic}}\frac{\partial u^{c}}{\partial z_{i}^{k}} \neq 0$ for certain $(z,w)$ and some $d\in\mathbf{C}$, $\frac{\partial r^{1}(y_1)}{\partial y_{i1}}$ is identified from equation~\eqref{eq:dev_r^1}; otherwise, equation~\eqref{eq:dev_r^1} implies that $\frac{\partial\sigma_{d}}{\partial y_{i1}}=0$ for all values $(z,w)$ and all $d\in\mathbf{C}$, and thus $\frac{\partial r^{1}(y_1)}{\partial y_{i1}}$ is also identified and equal to zero. The derivative of the function $r^{c}$ for $c \in \mathbf{C}\setminus\{1\}$ can be identified in the same manner.
\end{proof}
\begin{proof}[Proof of Proposition \ref{np_id_F}]

We start with the identification of the cutoffs $\{\delta_c\}_c$, i.e., part~(i).  
	Because of the large support assumption on $r^c$ (parts~ii and iii of Assumption~\ref{assm:F_id}) for each $c\in\bfC$, there exists $\mathcal{J}_c \subseteq \mathcal{X}$ such that for any $x_i\in \mathcal{J}_c$, $c$ is the only acceptable college with probability one, and that $Q({i\in \bfI: x_i\in\mathcal{J}_c})>0$. Then,
\begin{align*}
	\mathbb{{P}}(\mu(i)=0|x_{i}\in\mathcal{J}_c)=\mathbb{{P}}(v^{c}(z_{i})+w_{ci}+\eta_{ci} < \delta_c|x_{i}\in\mathcal{J}_c)
	= F_{\eta_{ci}}(\delta_c-\iota_{ci}),
\end{align*}
where the last equality is due to $\iota_{ci} = v^{c}(z_{i})+w_{ci}$ and the independence between $\eta_{ci}$ and $x_i$ (\cref{indep}). There exists a unique $\iota_{c}^\ast$ such that $\delta_c-\iota_{c}^\ast=\text{Quantile}_{\eta_{ci}}(\rho_c)=\inf \{ (\delta_c-\iota_{ci}): F_{\eta_{ci}}(\delta_c-\iota_{ci}) \geq \rho_c\}$. By Assumption~\ref{assm:F_id}(iv), $\delta_c-\iota_{c}^\ast=0$, which identifies $\delta_{c}$. To show part~(iii), we use the definition of cutoffs, equation~\eqref{cutoff}. 

To prove part (ii), the identification of the distribution $F$, following equation \eqref{ideq2} for $c=0$, the conditional probability
	of being unmatched can be rewritten as \begingroup \allowdisplaybreaks 
	\begin{equation}
	\mathbb{{P}}(\mu(i)=0| x_{i}) =\Lambda_{0}(\tau_{i1},\dots,\tau_{iC};\iota_{1i},\dots,\iota_{Ci})=\Lambda_{0}(\tau_{i},\iota_{i}).\label{eq:Lambda0}
\end{equation}
	\endgroup  
    Given that the functions $\{u^{c}+r^{c},v^{c}\}_{c}$
	are identified (part i of \cref{assm:F_id}), the arguments in equation~\eqref{eq:Lambda0},   $\{\tau_{ic},\iota_{ci}\}_{c}$,
	are known. Because $\mathbb{{P}}(\mu(i)=0| x_{i})$ is observed from the population data,
	the function $\Lambda_{0}$ is identified.

 For each $c\in\mathbf{C}$, define $A_{ic}\equiv\{u_{ic}<0\}=\{\tau_{ic}+\epsilon_{ic}<0\}=\{\epsilon_{ic}<-\tau_{ic}\}$, and $B_{ci}\equiv\{v_{ci}<\delta_{c}\}=\{\iota_{ci}+\eta_{ci}<\delta_{c}\}=\{\eta_{ci}<\delta_{c}-\iota_{ci}\}$.  By the independence between $(\epsilon_{i},\eta_{i})$ 	and $x_{i}$ (Assumption \ref{indep}), the parameter of interest can be written as
\[
F(-\tau_{i1},\dots,-\tau_{iC},\delta_{1}-\iota_{1i},\dots,\delta_{C}-\iota_{Ci})=\mathbb{P}\left(\cap_{c=1}^{C}\left(A_{ic}\cap B_{ci}\right)|x_{i}\right),
\]
	Further, the conditional probability of being unmatched is
 \begingroup \allowdisplaybreaks 
 \begin{align}
	\Lambda_{0}(\tau_{i},\iota_{i})= & \mathbb{P}\left(\cap_{c=1}^{C}\left(A_{ic}\cup B_{ci}\right)|x_{i}\right)\nonumber \\
	= & \mathbb{P}\left(\cap_{c=1}^{C-1}\left(A_{ic}\cup B_{ci}\right)\cap(A_{iC}\cup B_{Ci})|x_{i}\right)\nonumber \\
	= & \mathbb{P}\left((\cap_{c=1}^{C-1}\left(A_{ic}\cup B_{ci}\right)\cap A_{iC})\cup(\cap_{c=1}^{C-1}\left(A_{ic}\cup B_{ci}\right)\cap B_{Ci})|x_{i}\right)\nonumber \\
	= & \mathbb{P}\left(\cap_{c=1}^{C-1}\left(A_{ic}\cup B_{ci}\right)\cap A_{iC}|x_{i}\right)+\mathbb{P}\left(\cap_{c=1}^{C-1}\left(A_{ic}\cup B_{ci}\right)\cap B_{Ci}|x_{i}\right)\nonumber \\
	& -\mathbb{P}\left((\cap_{c=1}^{C-1}\left(A_{ic}\cup B_{ci}\right)\cap A_{iC}\cap B_{Ci})|x_{i}\right)\nonumber \\
	= & \Lambda_{0}(\tau_{i1},\dots,\tau_{iC};\iota_{1i},\dots,\iota_{(C-1)i},\infty)+\Lambda_{0}(\tau_{i1},\dots,\tau_{i(C-1)},\infty;\iota_{1i},\dots,\iota_{Ci})\nonumber \\
	& -\mathbb{P}\left((\cap_{c=1}^{C-1}\left(A_{ic}\cup B_{ci}\right)\cap A_{iC}\cap B_{Ci})|x_{i}\right)\label{eq:lambda}
\end{align}

\endgroup
Let $ H_{C}(\tau_{i},\iota_{i}) \equiv \mathbb{P}\left((\cap_{c=1}^{C-1}\left(A_{ic}\cup B_{ci}\right)\cap A_{iC}\cap B_{Ci})|x_{i}\right)$.  It is identified from equation~\eqref{eq:lambda} because $\Lambda_{0}$ is identified. Moreover, similar derivations yield
\begin{align}
	& H_{C}(\tau_{i},\iota_{i})\nonumber \\
	= & \mathbb{P}\left(\cap_{c=1}^{C-2}\left(A_{ic}\cup B_{ci}\right)\cap A_{i(C-1)}\cap A_{iC}\cap B_{Ci}|x_{i}\right)+\mathbb{P}\left(\cap_{c=1}^{C-2}\left(A_{ic}\cup B_{ci}\right)\cap B_{(C-1)i}\cap A_{iC}\cap B_{Ci}|x_{i}\right)\nonumber \\
	& -\mathbb{P}\left((\cap_{c=1}^{C-2}\left(A_{ic}\cup B_{ci}\right)\cap A_{i(C-1)}\cap B_{(C-1)i}\cap A_{iC}\cap B_{Ci})|x_{i}\right)\nonumber \\
	= & H_{C}(\tau_{i1},\dots,\tau_{iC};\iota_{1i},\dots\iota_{(C-2)i},\infty,\iota_{Ci})+H_{C}(\tau_{i1},\dots,\tau_{i(C-2)},\infty,\tau_{iC};\iota_{1i},\dots,\iota_{Ci})\nonumber \\
	& -\mathbb{P}\left((\cap_{c=1}^{C-2}\left(A_{ic}\cup B_{ci}\right)\cap A_{i(C-1)}\cap B_{(C-1)i}\cap A_{iC}\cap B_{Ci})|x_{i}\right),\label{eq:H_c}
\end{align}
 where in the penultimate line, the effects of $\iota_{(C-1)i}$ and $\tau_{i(C-1)}$ are ``shut down'' in the two terms, respectively. Equation \eqref{eq:H_c} then identifies $H_{C-1}(\tau_{i},\iota_{i}) \equiv \mathbb{P}\big((\cap_{c=1}^{C-2}\left(A_{ic}\cup B_{ci}\right)\cap A_{i(C-1)}\cap B_{(C-1)i}\cap A_{iC}\cap B_{Ci})|x_{i}\big).$

Repeat the above argument and define a sequence of functions recursively until $H_{2}(\tau_{i},\iota_{i})=H_{2}(\tau_{i1},\dots,\tau_{iC};\infty,\iota_{2i},\dots,\iota_{Ci})+H_{2}(\infty,\tau_{i2},\dots,\tau_{iC};\iota_{1i},\dots,\iota_{Ci})\\ -H_{1}(\tau_{i},\iota_{i})$, where on the RHS, the effects of $\iota_{1i}$ and $\tau_{i1}$ are ``shut down'' in the first two terms, respectively. Every function in the sequence is identified.
	 
It then follows that $F(-\tau_{i1},\dots,-\tau_{iC},\delta_{1}-\iota_{1i},\dots,\delta_{C}-\iota_{Ci})=\mathbb{P}\left(\cap_{c=1}^{C}\left(A_{ic}\cap B_{ci}\right)|x_{i}\right) \equiv H_{1}(\tau_{i},\iota_{i})$
	is identified.
\end{proof}

\begin{proof}[Proof of \cref{deg-of-heter}] 
	
We use the same argument as in the proof of  \cref{prop_nonpara}, except that the matrix in equation~\eqref{eq:idenmatrix} reduces in dimension due to the additional homogeneity restrictions. Specifically, with some abuse of notations, suppose that 
 vector~$\boldsymbol{u}$ consists of $u^{\ast}$ and $u^{c}$  $\forall c\in\bfC \setminus\bfC_1$, and that $\boldsymbol{v}$ consist of $v^{\ast}$ and $v^{c}$ $\forall  c\in\bfC \setminus\bfC_2$. The vectors $y_i$ and $w_i$ are defined similarly, and suppose that  $(\Pi_{y}(z,\overline{y},\widehat{w}),\,\Pi_{w}(z,\overline{y},\widehat{w}))$ has eliminated the duplicated elements accordingly. Fix $(y_{i},w_{i})=(\overline{y},\widehat{w})$ and consider $z$ in the interior of $\mathcal{Z}$. We can rewrite equation~(\ref{eq:dzc-1}) as
\begin{equation}
	\frac{\partial\sigma_{d}}{\partial z_{i}^{k}}=\frac{\partial\sigma_{d}}{\partial y_{i\ast}}\frac{\partial u^{\ast}}{\partial z_{i}^{k}}+\sum_{c\in\bfC \setminus\bfC_1}\frac{\partial\sigma_{d}}{\partial y_{ic}}\frac{\partial u^{c}}{\partial z_{i}^{k}}+\frac{\partial\sigma_{d}}{\partial w_{i\ast}}\frac{\partial v^{\ast}}{\partial z_{i}^{k}}+\sum_{c\in\bfC \setminus\bfC_2}\frac{\partial\sigma_{d}}{\partial w_{ic}}\frac{\partial v^{c}}{\partial z_{i}^{k}}.\label{eq:dzc_homo}
\end{equation}
Stacking equation~(\ref{eq:dzc_homo}) for all $d\in\bfC$, we obtain
\begin{equation}
    \frac{\partial\boldsymbol{\sigma}(z,\overline{y},\widehat{w})}{\partial z_{i}^{k}}=(\Pi_{y}(z,\overline{y},\widehat{w}),  \Pi_{w}(z,\overline{y},\widehat{w}))\times\begin{pmatrix}\frac{\partial\boldsymbol{u}(z)}{\partial z_{i}^{k}}\\
	\frac{\partial\mathbf{v}(z)}{\partial z_{i}^{k}}
\end{pmatrix}.\label{eq:id_homo}
\end{equation}
 When the number of parameters $2C-\kappa_1-\kappa_2+2$ is at most $C$, equation~(\ref{eq:id_homo}) implies identification; otherwise, we can identify the parameters by considering a pair of distinct values of $w_{i}$, as in the proof of \cref{prop_nonpara}.
\label{lastpage} \end{proof}

    \setstretch{1.1}
   \bibliography{matching}

\begin{thebibliography}{56}
\newcommand{\enquote}[1]{``#1''}
\expandafter\ifx\csname natexlab\endcsname\relax\def\natexlab#1{#1}\fi

\bibitem[\protect\citeauthoryear{Abaluck and Adams-Prassl}{Abaluck and
  Adams-Prassl}{2021}]{abaluck_what_2018}
\textsc{Abaluck, J. and A.~Adams-Prassl} (2021): \enquote{What do consumers
  consider before they choose? Identification from asymmetric demand
  responses,} \emph{The Quarterly Journal of Economics}, 136, 1611--1663.

\bibitem[\protect\citeauthoryear{Abdulkadiro{\u g}lu, Agarwal, and
  Pathak}{Abdulkadiro{\u g}lu et~al.}{2017}]{abdulkadiroglu_welfare_2017}
\textsc{Abdulkadiro{\u g}lu, A., N.~Agarwal, and P.~A. Pathak} (2017):
  \enquote{The {Welfare} {Effects} of {Coordinated} {Assignment}: {Evidence}
  from the {New} {York} {City} {High} {School} {Match},} \emph{American
  Economic Review}, 107, 3635--3689.

\bibitem[\protect\citeauthoryear{Abdulkadiro{\u g}lu and
  S{\"o}nmez}{Abdulkadiro{\u g}lu and
  S{\"o}nmez}{2003}]{abdulkadiroglu_school_2003}
\textsc{Abdulkadiro{\u g}lu, A. and T.~S{\"o}nmez} (2003): \enquote{School
  {Choice}: {A} {Mechanism} {Design} {Approach},} \emph{American Economic
  Review}, 93, 729--747.

\bibitem[\protect\citeauthoryear{Agarwal}{Agarwal}{2015}]{agarwal_empirical_2015}
\textsc{Agarwal, N.} (2015): \enquote{An {Empirical} {Model} of the {Medical}
  {Match},} \emph{American Economic Review}, 105, 1939--1978.

\bibitem[\protect\citeauthoryear{Agarwal and Somaini}{Agarwal and
  Somaini}{2018}]{agarwal_demand_2018}
\textsc{Agarwal, N. and P.~Somaini} (2018): \enquote{Demand {Analysis} {Using}
  {Strategic} {Reports}: {An} {Application} to a {School} {Choice}
  {Mechanism},} \emph{Econometrica}, 86, 391--444.

\bibitem[\protect\citeauthoryear{Agarwal and Somaini}{Agarwal and
  Somaini}{2020{\natexlab{a}}}]{as_2020}
---\hspace{-.1pt}---\hspace{-.1pt}--- (2020{\natexlab{a}}): \enquote{Empirical
  Models of Non-Transferable Utility Matching,} .

\bibitem[\protect\citeauthoryear{Agarwal and Somaini}{Agarwal and
  Somaini}{2020{\natexlab{b}}}]{agarwal2020revealed}
---\hspace{-.1pt}---\hspace{-.1pt}--- (2020{\natexlab{b}}): \enquote{Revealed
  preference analysis of school choice models,} \emph{Annual Review of
  Economics}, 12, 471--501.

\bibitem[\protect\citeauthoryear{Agarwal and Somaini}{Agarwal and
  Somaini}{2022}]{agarwal2022demand}
---\hspace{-.1pt}---\hspace{-.1pt}--- (2022): \enquote{Demand Analysis under
  Latent Choice Constraints,} Tech. rep., National Bureau of Economic Research.

\bibitem[\protect\citeauthoryear{Artemov, Che, and He}{Artemov
  et~al.}{2020}]{ACH2017}
\textsc{Artemov, G., Y.-K. Che, and Y.~He} (2020): \enquote{Strategic
  `Mistakes': Implications for Market Design Research,} Manuscript.

\bibitem[\protect\citeauthoryear{Artemov, Che, and He}{Artemov
  et~al.}{2023}]{ACH2023}
---\hspace{-.1pt}---\hspace{-.1pt}--- (2023): \enquote{Stable Matching with
  Mistaken Agents,} \emph{Journal of Political Economy Microeconomics}, 1,
  270--320.

\bibitem[\protect\citeauthoryear{Aue, Klein, and Ortega}{Aue
  et~al.}{2020}]{aue2020happens}
\textsc{Aue, R., T.~Klein, and J.~Ortega} (2020): \enquote{What happens when
  separate and unequal school districts merge?} \emph{ZEW Discussion Paper}.

\bibitem[\protect\citeauthoryear{Azevedo and Leshno}{Azevedo and
  Leshno}{2016}]{azevedo_supply_2016}
\textsc{Azevedo, E.~M. and J.~D. Leshno} (2016): \enquote{A supply and demand
  framework for two-sided matching markets,} \emph{Journal of Political
  Economy}, 124, 1235--1268.

\bibitem[\protect\citeauthoryear{Barseghyan, Coughlin, Molinari, and
  Teitelbaum}{Barseghyan
  et~al.}{2021{\natexlab{a}}}]{barseghyan_heterogeneous_2019}
\textsc{Barseghyan, L., M.~Coughlin, F.~Molinari, and J.~C. Teitelbaum}
  (2021{\natexlab{a}}): \enquote{Heterogeneous choice sets and preferences,}
  \emph{Econometrica}, 89, 2015--2048.

\bibitem[\protect\citeauthoryear{Barseghyan, Molinari, and
  Thirkettle}{Barseghyan et~al.}{2021{\natexlab{b}}}]{barseghyan2021discrete}
\textsc{Barseghyan, L., F.~Molinari, and M.~Thirkettle} (2021{\natexlab{b}}):
  \enquote{Discrete choice under risk with limited consideration,}
  \emph{American Economic Review}, 111, 1972--2006.

\bibitem[\protect\citeauthoryear{Berry, Gandhi, and Haile}{Berry
  et~al.}{2013}]{berry2013connected}
\textsc{Berry, S., A.~Gandhi, and P.~Haile} (2013): \enquote{Connected
  substitutes and invertibility of demand,} \emph{Econometrica}, 81,
  2087--2111.

\bibitem[\protect\citeauthoryear{Blundell and Powell}{Blundell and
  Powell}{2004}]{blundell2004endogeneity}
\textsc{Blundell, R.~W. and J.~L. Powell} (2004): \enquote{Endogeneity in
  semiparametric binary response models,} \emph{The Review of Economic
  Studies}, 71, 655--679.

\bibitem[\protect\citeauthoryear{Calsamiglia, Fu, and G{\"u}ell}{Calsamiglia
  et~al.}{2020}]{calsamiglia2020structural}
\textsc{Calsamiglia, C., C.~Fu, and M.~G{\"u}ell} (2020): \enquote{Structural
  estimation of a model of school choices: The boston mechanism versus its
  alternatives,} \emph{Journal of Political Economy}, 128, 642--680.

\bibitem[\protect\citeauthoryear{Cattaneo, Ma, Masatlioglu, and
  Suleymanov}{Cattaneo et~al.}{2020}]{cattaneo2020random}
\textsc{Cattaneo, M.~D., X.~Ma, Y.~Masatlioglu, and E.~Suleymanov} (2020):
  \enquote{A random attention model,} \emph{Journal of Political Economy}, 128,
  2796--2836.

\bibitem[\protect\citeauthoryear{Chen and Fang}{Chen and
  Fang}{2019}]{chen_improved_2019}
\textsc{Chen, Q. and Z.~Fang} (2019): \enquote{Improved inference on the rank
  of a matrix,} \emph{Quantitative Economics}, 10, 1787--1824.

\bibitem[\protect\citeauthoryear{Chesher}{Chesher}{2003}]{chesher2003identification}
\textsc{Chesher, A.} (2003): \enquote{Identification in nonseparable models,}
  \emph{Econometrica}, 71, 1405--1441.

\bibitem[\protect\citeauthoryear{Chiappori and Salani\'e}{Chiappori and
  Salani\'e}{2016}]{chiappori_econometrics_2016}
\textsc{Chiappori, P.-A. and B.~Salani\'e} (2016): \enquote{The {Econometrics}
  of {Matching} {Models},} \emph{Journal of Economic Literature}, 54, 832--861.

\bibitem[\protect\citeauthoryear{Chiappori, Salani{\'e}, and Weiss}{Chiappori
  et~al.}{2017}]{chiappori2017partner}
\textsc{Chiappori, P.-A., B.~Salani{\'e}, and Y.~Weiss} (2017):
  \enquote{Partner choice, investment in children, and the marital college
  premium,} \emph{American Economic Review}, 107, 2109--67.

\bibitem[\protect\citeauthoryear{Choo and Siow}{Choo and
  Siow}{2006}]{choo_who_2006}
\textsc{Choo, E. and A.~Siow} (2006): \enquote{Who {Marries} {Whom} and {Why},}
  \emph{Journal of Political Economy}, 114, 175--201.

\bibitem[\protect\citeauthoryear{Dagsvik}{Dagsvik}{2000}]{dagsvik_aggregation_2000}
\textsc{Dagsvik, J.~K.} (2000): \enquote{Aggregation in matching markets,}
  \emph{International Economic Review}, 41, 27--58.

\bibitem[\protect\citeauthoryear{Diamond and Agarwal}{Diamond and
  Agarwal}{2017}]{diamond_latent_2017}
\textsc{Diamond, W. and N.~Agarwal} (2017): \enquote{Latent indices in
  assortative matching models,} \emph{Quantitative Economics}, 8, 685--728.

\bibitem[\protect\citeauthoryear{Ederer}{Ederer}{2022}]{ederer2022}
\textsc{Ederer, T.} (2022): \enquote{Two-Sided Matching Without Transfers: A
  Unifying Empirical Framework,} .

\bibitem[\protect\citeauthoryear{Fack, Grenet, and He}{Fack
  et~al.}{2019}]{fack_beyond_2019}
\textsc{Fack, G., J.~Grenet, and Y.~He} (2019): \enquote{Beyond
  {Truth}-{Telling}: {Preference} {Estimation} with {Centralized} {School}
  {Choice} and {College} {Admissions},} \emph{American Economic Review}, 109,
  1486--1529.

\bibitem[\protect\citeauthoryear{Fox}{Fox}{2010}]{fox_identification_2010}
\textsc{Fox, J.~T.} (2010): \enquote{Identification in matching games,}
  \emph{Quantitative Economics}, 1, 203--254.

\bibitem[\protect\citeauthoryear{Fox}{Fox}{2018}]{fox_estimating_2018}
---\hspace{-.1pt}---\hspace{-.1pt}--- (2018): \enquote{Estimating matching
  games with transfers,} \emph{Quantitative Economics}, 9, 1--38.

\bibitem[\protect\citeauthoryear{Fox, Yang, and Hsu}{Fox
  et~al.}{2018}]{fox_unobserved_2018}
\textsc{Fox, J.~T., C.~Yang, and D.~H. Hsu} (2018): \enquote{Unobserved
  {Heterogeneity} in {Matching} {Games},} \emph{Journal of Political Economy}.

\bibitem[\protect\citeauthoryear{Gale and Shapley}{Gale and
  Shapley}{1962}]{gale_college_1962}
\textsc{Gale, D. and L.~Shapley} (1962): \enquote{College {Admissions} and the
  {Stability} of {Marriage},} \emph{The American Mathematical Monthly}, 69,
  9--15.

\bibitem[\protect\citeauthoryear{Galichon, Kominers, and Weber}{Galichon
  et~al.}{2019}]{galichon2019costly}
\textsc{Galichon, A., S.~D. Kominers, and S.~Weber} (2019): \enquote{Costly
  concessions: An empirical framework for matching with imperfectly
  transferable utility,} \emph{Journal of Political Economy}, 127, 2875--2925.

\bibitem[\protect\citeauthoryear{Galichon and Salanie}{Galichon and
  Salanie}{2020}]{galichon_cupids_2020}
\textsc{Galichon, A. and B.~Salanie} (2020): \enquote{Cupid's {Invisible}
  {Hand}: {Social} {Surplus} and {Identification} in {Matching} {Models},}
  \emph{Working Paper}.

\bibitem[\protect\citeauthoryear{Gazmuri}{Gazmuri}{2017}]{gazmuri2017school}
\textsc{Gazmuri, A.} (2017): \enquote{School Segregation in the Presence of
  Student Sorting and Cream-Skimming,} \emph{Working paper, Toulouse School of
  Economics}.

\bibitem[\protect\citeauthoryear{Gelman and Rubin}{Gelman and
  Rubin}{1992}]{gelman1992inference}
\textsc{Gelman, A. and D.~B. Rubin} (1992): \enquote{Inference from iterative
  simulation using multiple sequences,} \emph{Statistical science}, 7,
  457--472.

\bibitem[\protect\citeauthoryear{Graham}{Graham}{2011}]{GRAHAM2011965}
\textsc{Graham, B.~S.} (2011): \enquote{Econometric Methods for the Analysis of
  Assignment Problems in the Presence of Complementarity and Social
  Spillovers,} North-Holland, vol.~1 of \emph{Handbook of Social Economics},
  965--1052.

\bibitem[\protect\citeauthoryear{Grenet, He, and Kübler}{Grenet
  et~al.}{2022}]{GHK}
\textsc{Grenet, J., Y.~He, and D.~Kübler} (2022): \enquote{Preference
  Discovery in University Admissions: The Case for Dynamic Multioffer
  Mechanisms,} \emph{Journal of Political Economy}, 130, 1427--1476.

\bibitem[\protect\citeauthoryear{Gualdani and Sinha}{Gualdani and
  Sinha}{2023}]{gualdani_identification_2020}
\textsc{Gualdani, C. and S.~Sinha} (2023): \enquote{Partial Identification in
  Matching Models for the Marriage Market,} \emph{Journal of Political
  Economy}, 131, 1109--1171.

\bibitem[\protect\citeauthoryear{He}{He}{2017}]{he_gaming_2017}
\textsc{He, Y.} (2017): \enquote{Gaming the Boston School Choice Mechanism in
  Beijing,} \emph{Working Paper, Rice University and Toulouse School of
  Economics}.

\bibitem[\protect\citeauthoryear{He and Magnac}{He and
  Magnac}{2022}]{he2020application}
\textsc{He, Y. and T.~Magnac} (2022): \enquote{Application costs and congestion
  in matching markets,} \emph{The Economic Journal}, 132, 2918--2950.

\bibitem[\protect\citeauthoryear{Heckman and Robb}{Heckman and
  Robb}{1985}]{heckman1985alternative}
\textsc{Heckman, J. and R.~Robb} (1985): \enquote{Alternative methods for
  evaluating the impact of interventions: An overview,} \emph{Journal of
  econometrics}, 30, 239--267.

\bibitem[\protect\citeauthoryear{Imbens and Newey}{Imbens and
  Newey}{2009}]{imbens2009identification}
\textsc{Imbens, G.~W. and W.~K. Newey} (2009): \enquote{Identification and
  estimation of triangular simultaneous equations models without additivity,}
  \emph{Econometrica}, 77, 1481--1512.

\bibitem[\protect\citeauthoryear{Kapor, Neilson, and Zimmerman}{Kapor
  et~al.}{2020}]{kapor2020heterogeneous}
\textsc{Kapor, A.~J., C.~A. Neilson, and S.~D. Zimmerman} (2020):
  \enquote{Heterogeneous beliefs and school choice mechanisms,} \emph{American
  Economic Review}, 110, 1274--1315.

\bibitem[\protect\citeauthoryear{Logan, Hoff, and Newton}{Logan
  et~al.}{2008}]{logan_two-sided_2008}
\textsc{Logan, J.~A., P.~D. Hoff, and M.~A. Newton} (2008): \enquote{Two-Sided
  Estimation of Mate Preferences for Similarities in Age, Education, and
  Religion,} \emph{Journal of the American Statistical Association}, 103,
  559--569.

\bibitem[\protect\citeauthoryear{Matzkin}{Matzkin}{2019}]{matzkin_constructive_2019}
\textsc{Matzkin, R.~L.} (2019): \enquote{Constructive identification in some
  nonseparable discrete choice models,} \emph{Journal of Econometrics}, 211, 83
  -- 103.

\bibitem[\protect\citeauthoryear{Menzel}{Menzel}{2015}]{menzel_large_2015}
\textsc{Menzel, K.} (2015): \enquote{Large {Matching} {Markets} as
  {Two}-{Sided} {Demand} {Systems},} \emph{Econometrica}, 83, 897--941.

\bibitem[\protect\citeauthoryear{Menzel}{Menzel}{2017}]{menzel_strategic_2017}
---\hspace{-.1pt}---\hspace{-.1pt}--- (2017): \enquote{Strategic network
  formation with many agents,} \emph{Working Paper, New York University}.

\bibitem[\protect\citeauthoryear{Menzel}{Menzel}{2022}]{menzel2022strategic}
---\hspace{-.1pt}---\hspace{-.1pt}--- (2022): \enquote{Strategic network
  formation with many agents,} .

\bibitem[\protect\citeauthoryear{Pais, Pint{\'e}r, and Veszteg}{Pais
  et~al.}{2020}]{Pais-decentralized}
\textsc{Pais, J., {\'A}.~Pint{\'e}r, and R.~F. Veszteg} (2020):
  \enquote{Decentralized matching markets with(out) frictions: a laboratory
  experiment,} \emph{Experimental Economics}, 23, 212--239.

\bibitem[\protect\citeauthoryear{Petrin and Train}{Petrin and
  Train}{2010}]{petrin2010control}
\textsc{Petrin, A. and K.~Train} (2010): \enquote{A control function approach
  to endogeneity in consumer choice models,} \emph{Journal of marketing
  research}, 47, 3--13.

\bibitem[\protect\citeauthoryear{Powell, Stock, and Stoker}{Powell
  et~al.}{1989}]{powell_semiparametric_1989}
\textsc{Powell, J.~L., J.~H. Stock, and T.~M. Stoker} (1989):
  \enquote{Semiparametric {Estimation} of {Index} {Coefficients},}
  \emph{Econometrica}, 57, 1403--1430.

\bibitem[\protect\citeauthoryear{Rossi, Allenby, and McCulloch}{Rossi
  et~al.}{2012}]{rossi2012bayesian}
\textsc{Rossi, P.~E., G.~M. Allenby, and R.~McCulloch} (2012): \emph{Bayesian
  statistics and marketing}, John Wiley \& Sons.

\bibitem[\protect\citeauthoryear{Roth}{Roth}{1984}]{roth1984evolution}
\textsc{Roth, A.~E.} (1984): \enquote{The evolution of the labor market for
  medical interns and residents: a case study in game theory,} \emph{Journal of
  political Economy}, 92, 991--1016.

\bibitem[\protect\citeauthoryear{Roth and Sotomayor}{Roth and
  Sotomayor}{1992}]{roth1992two}
\textsc{Roth, A.~E. and M.~Sotomayor} (1992): \enquote{Two-sided matching,}
  \emph{Handbook of game theory with economic applications}, 1, 485--541.

\bibitem[\protect\citeauthoryear{Sinha}{Sinha}{2015}]{sinha_identification_2015}
\textsc{Sinha, S.} (2015): \enquote{Identification and {Estimation} in
  {One}-to-{one} {Matching} {Models} with {Nonparametric} {Unobservables},}
  \emph{Toulouse School of Economics}.

\bibitem[\protect\citeauthoryear{Uetake and Watanabe}{Uetake and
  Watanabe}{2020}]{uetake2020entry}
\textsc{Uetake, K. and Y.~Watanabe} (2020): \enquote{Entry by merger: Estimates
  from a two-sided matching model with externalities,} \emph{Available at SSRN
  2188581}.

\end{thebibliography}
   \setstretch{1.3}

\clearpage

\setcounter{table}{0}
\renewcommand{\thetable}{\Alph{section}.\arabic{table}}
\setcounter{equation}{0}
\renewcommand{\theequation}{\Alph{section}.\arabic{equation}}

\setcounter{footnote}{0}
\renewcommand{\thefootnote}{\Alph{section}.\arabic{footnote}}

\setcounter{page}{1}
\renewcommand{\thepage}{{Online Appendix | page \arabic{page}}}

\begin{appendices}
\begin{center}
{\large Online Appendix to}\\
[0.5cm]
{\LARGE Identification and Estimation in Many-to-one}\\
[0.5cm]
{\LARGE  Two-sided Matching without Transfers}\\
[0.5cm]
{\large YingHua He \hspace{1cm} Shruti Sinha \hspace{1cm} Xiaoting Sun}\\
[0.25cm]
\end{center}

\section{Identification of a Nonseparable Model}\label{app:ns_id}
We now discuss the nonparametric identification of a more general nonseparable utility specification based on the arguments in \cite{matzkin_constructive_2019}. As we shall see, compared with those for the separable model, our identification results for this model are based on an additional assumption (Assumption~\ref{assm_ui0}) and two different rank conditions (Conditions~\ref{cond1} and \ref{cond2}). Hence, the sufficient conditions below do not nest those in the main text.

There is full nonseparability for all but one (i.e., $2C-1$) utility functions, while for one student utility function, there is nonseparability between the observable $z_{i}$ and an index $y_{ic}+\epsilon_{ic}$. That is, without loss of generality, 
\begin{align}
	u_{i1}=u^{1}\left(z_{i},y_{i1}+\epsilon_{i1}\right),\, 
	u_{ic}  &=u^{c}\left(z_{i},y_{ic},\epsilon_{ic}\right)\, \forall c \in \bfC \setminus \{1\}, \nonumber \\ 
	\mbox{and }  v_{ci}  &=v^{c}\left(z_{i},w_{ci},\eta_{ci}\right)\, \forall  c \in \bfC.  \label{NPNS_specification} 
\end{align}
The additive index $y_{i1}+\epsilon_{i1}$ can be relaxed to some known function such as $y_{i1} \cdot \epsilon_{i1}$ \citep{matzkin_constructive_2019}. Below we discuss a set of sufficient conditions under which our identification strategy applies to $\{u^c\}_c$. Moreover, we show $\{v^c\}_c$ is identified under additional separability.  This helps clarify the role of the additive separability in equation~\eqref{NP_specification}.
For notational simplicity, we also use $u^{1}(z_{i},y_{i1},\epsilon_{i1})$ to denote $u^{1}\left(z_{i},y_{i1}+\epsilon_{i1}\right)$.
The utility of the outside option $u_{i0}$ is assumed to be a continuous
random variable.%
\footnote{In separable models, $u_{i0}=0$ is a location normalization because
	the conditional match probability only depends on the difference in
	the utility shocks. However, in this nonseparable model,
	it would impose an additional restriction.}

\begin{assm} \label{assm_npns}  
	(i) $z_i$, $y_i$, and $w_i$ are continuously distributed; 
	(ii) for each $c\in\mathbf{C}$, the functions, $u^{c}$ and $v^{c}$, are continuously differentiable;
	(iii) $F$ is continuously differentiable;
	(iv) for each $c\in\mathbf{C}$, $u^{c}$ and $v^{c}$ are strictly increasing in their last argument; 
	and (v) for $c\in\mathbf{C}\backslash\{1\}$, when $u^{c}(z_{i},y_{ic},\epsilon_{ic})=u_{i0}$,
	$\frac{\partial u^{c}(z_{i},y_{ic},\epsilon_{ic})}{\partial y_{ic}}\neq0$,
	and for $c\in\mathbf{C}$, when $v^{c}(z_{i},w_{ci},\eta_{ci})=\delta_{c}$,
	$\frac{\partial v^{c}(z_{i},w_{ci},\eta_{ci})}{\partial w_{ci}}\neq0$.
\end{assm}

\begin{assm}\label{indep_npns} $(\epsilon_{i},\eta_{i})$ is independent of $(z_i,y_i,w_i)$. \end{assm}

\begin{assm} \label{assm_ui0}
	
	(i) The utility of the outside option is $u_{i0}=h\left(y_{i0}\right),$
	where $y_{i0}\in\mathcal{Y}_{0}\subseteq\mathbb{R}^{d_{y_{0}}}$ is
	a vector of observed covariates and $h$ is a known function;
    (ii) The support of $u_{i0}$, $\mathcal{U}_0 \subseteq \mathbb{R}$, is a superset of the range of the function $u^c$, $\forall c\in\bfC$.
	
\end{assm}

Parts (i)--(iii) of Assumption \ref{assm_npns} and Assumption \ref{indep_npns} impose smoothness and exogeneity similar to Assumptions~\ref{assm_nonpara} and  \ref{indep} in Section~\ref{id}. Part (iv) of Assumption~\ref{assm_npns} guarantees that there is a one-to-one relationship between the value of each utility function and its unobservable. Part (v) of Assumption \ref{assm_npns} guarantees that $u_{ic}$ and $v_{ci}$ are not constant w.r.t. $y_{ic}$ and $w_{ci}$, respectively, such that a change in $y_{ic}$ or  $w_{ci}$ generates a change in the conditional probability of being unmatched. \cref{assm_ui0} guarantees that $u_{i0}$ is observed by the researcher and has a large support.

By the monotonicity assumption (part iv of \cref{assm_npns}), for
each $c\in\mathbf{C}$, the inverse of $u^{c}$ and $v^{c}$ w.r.t.\ their
last argument exists. Let $\tilde{u}^{c}$ and $\tilde{v}^{c}$ denote
the inverse of $u^{c}$ and $v^{c}$ w.r.t.\ their last argument, respectively. That is, for any $a\in\mathbb{R}$,
\begin{align*}
u^{1}\left(z_{i},\tilde{u}^{1}\left(z_{i},a\right)\right)=a,&\mbox{ and }v^{1}\left(z_{i},w_{1i},\tilde{v}^{1}\left(z_{i},w_{1i},a\right)\right)=a,\\
u^{c}\left(z_{i},y_{ic},\tilde{u}^{c}\left(z_{i},y_{ic},a\right)\right)=a,&\mbox{ and }v^{c}\left(z_{i},w_{ci},\tilde{v}^{c}\left(z_{i},w_{ci},a\right)\right)=a, \text{ for any } c\in\mathbf{C}\backslash\{1\}.
\end{align*}
Then, we have 
	\begingroup \allowdisplaybreaks 
	\begin{align*}
	\lambda_{L}(\iota_{1i},\dots,\iota_{Ci}) & =\mathbb{P}(v_{ci}\geq\delta_{c}\;\forall\,c\in L;\,v_{di}<\delta_{d}\;\forall\,d\notin L| z_{i}, w_{i};\mu)\\
	& =\mathbb{P}(v^{c}\left(z_{i},w_{ci},\eta_{ci}\right)\geq\delta_{c}\;\forall\,c\in L;\,v^{d}\left(z_{i},w_{di},\eta_{di}\right)<\delta_{d}\;\forall\,d\notin L| z_{i}, w_{i};\mu)\\
	& =\mathbb{P}(\eta_{ci}\geq\tilde{v}^{c}\left(z_{i},w_{ci},\delta_{c}\right)\;\forall\,c\in L;\,\eta_{di}<\tilde{v}^{d}\left(z_{i},w_{di},\delta_{d}\right)\;\forall\,d\notin L| z_{i}, w_{i};\mu),
\end{align*}
\endgroup
where $\iota_{ci}=\tilde{v}^{c}\left(z_{i},w_{ci},\delta_{c}\right)$ for $c\in\mathbf{C}$. Since $\tilde{v}^{c}$ is a $c$-specific nonparametric function, the following analysis does not rely on the identification of $\delta_c$. 
Similarly, 
\begin{eqnarray*}
	\mathbb{{P}}(0=\arg\max_{c\in L}u_{ic}| L,z_{i}, y_{i},u_{i0}) & = & \mathbb{{P}}(u_{i0}>u_{ic}\text{ for all }c\in L| L,z_{i}, y_{i},u_{i0})\\
	& = & \mathbb{{P}}(u_{i0}>u^{c}\left(z_{i},y_{ic},\epsilon_{ic}\right)\text{ for all }c\in L| L,z_{i}, y_{i},u_{i0})\\
	& = & \mathbb{{P}}(\epsilon_{i1}<\tilde{u}^{1}(z_{i},u_{i0})-y_{i1} \text{ if $1\in L; $}\\
	&  & \quad \epsilon_{ic}<\tilde{u}^{c}\left(z_{i},y_{ic},u_{i0}\right)\text{ for all }c\in L \text{ and } c \neq 1| L,z_{i}, y_{i},u_{i0})\\
	& = & g_{0,L}(\tau_{i1},\dots,\tau_{iC}),
\end{eqnarray*}
where $\tau_{i1}=\tilde{u}^{1}\left(z_{i},u_{i0}\right)-y_{i1}$ and
for $c\in\mathbf{C}\backslash\{1\}$, $\tau_{ic}=\tilde{u}^{c}\left(z_{i},y_{ic},u_{i0}\right)$.
Note that if $c\notin L$, $g_{0,L}$ does not change with the argument
$\tau_{ic}$.

Further, following equation \eqref{ideq2} for $c=0$, we have 
\begin{eqnarray}
	\sigma_{0}\left(z_{i}, y_{i}, w_{i},u_{i0}\right) & = & \sum_{L\in\mathcal{{L}}}\lambda_{L}(\iota_{1i},\dots,\iota_{Ci})\cdot g_{0,L}(\tau_{i1},\dots,\tau_{iC})\nonumber \\
	& \equiv & \Lambda_{0}\left(\tau_{i1},\dots,\tau_{iC},\iota_{1i},\dots,\iota_{Ci}\right),\label{eq:sigma_0_app}
\end{eqnarray}
where $\Lambda_{0}$ is a nonparametric function.

To identify the derivatives of  $\{u^c,v^c\}_c$, we extend the
argument in \cite{matzkin_constructive_2019}. Our identification depends on conditions on the derivatives
of the probability of being \textit{unmatched} w.r.t.\ the excluded variables.
Let $y_{i,-1}=\left(y_{i2},\dots,y_{iC}\right)\in\mathcal{Y}_{-1}\subseteq\mathbb{R}^{2C-1}$
denote the vector of $y_{i}$ excluding $y_{i1}$. For a given value
$(z,w,y_{-1},u_{0})$ in the interior of $\mathcal{Z}\times\mathcal{W}\times\mathcal{Y}_{-1}\times\mathcal{U}_{0}$,
consider $2C$ different values, $y_{1}^{1},\dots,y_{1}^{2C},$ in
the interior of the support of $y_{i1}$ conditional on $(z_{i},w_{i},y_{i,-1},u_{i0})=(z,w,y_{-1},u_{0})$.
We define a $C\times C$ matrix 
\[
\Pi_{1}\left(y_{1}^{1},\dots,y_{1}^{C};z,w,y_{-1},u_{0}\right)\equiv\begin{pmatrix}\frac{\partial\sigma_{0}\left(z,w,y_{-1},u_{0},y_{1}^{1}\right)}{\partial y_{i1}} & \cdots & \frac{\partial\sigma_{0}\left(z,w,y_{-1},u_{0},y_{1}^{1}\right)}{\partial y_{iC}}\\
	\vdots & \ddots & \vdots\\
	\frac{\partial\sigma_{0}\left(z,w,y_{-1},u_{0},y_{1}^{C}\right)}{\partial y_{i1}} & \cdots & \frac{\partial\sigma_{0}\left(z,w,y_{-1},u_{0},y_{1}^{C}\right)}{\partial y_{iC}}
\end{pmatrix},
\]
where for $m=1,\dots,C$, the $m^{th}$ row of the matrix $\Pi_{1}$ consists of the derivatives
of conditional probability of being unmatched w.r.t.\ the $C$
excluded variables $y_{i}$, evaluated at $\left(z,w,y_{-1},u_{0},y_{1}^{m}\right)$.
Further, we define a $2C\times2C$ matrix 
{\scriptsize{}
	\[
	\Pi_{2}\left(y_{1}^{1},\dots,y_{1}^{2C};z,w,y_{-1},u_{0}\right)\equiv\begin{pmatrix}\frac{\partial\sigma_{0}\left(z,w,y_{-1},u_{0},y_{1}^{1}\right)}{\partial y_{i1}} & \cdots & \frac{\partial\sigma_{0}\left(z,w,y_{-1},u_{0},y_{1}^{1}\right)}{\partial y_{iC}} & \frac{\partial\sigma_{0}\left(z,w,y_{-1},u_{0},y_{1}^{1}\right)}{\partial w_{1i}} & \cdots & \frac{\partial\sigma_{0}\left(z,w,y_{-1},u_{0},y_{1}^{1}\right)}{\partial w_{Ci}}\\
		\vdots & \ddots & \vdots & \vdots & \ddots & \vdots\\
		\frac{\partial\sigma_{0}\left(z,w,y_{-1},u_{0},y_{1}^{2C}\right)}{\partial y_{i1}} & \cdots & \frac{\partial\sigma_{0}\left(z,w,y_{-1},u_{0},y_{1}^{2C}\right)}{\partial y_{iC}} & \frac{\partial\sigma_{0}\left(z,w,y_{-1},u_{0},y_{1}^{2C}\right)}{\partial w_{1i}} & \cdots & \frac{\partial\sigma_{0}\left(z,w,y_{-1},u_{0},y_{1}^{2C}\right)}{\partial w_{Ci}}
	\end{pmatrix},
	\]
}where for $m=1,\dots,2C$, the $m^{th}$ row of the matrix $\Pi_{2}$ consists of the derivatives
of conditional probability of being unmatched w.r.t.\ the $2C$
excluded variables $y_{i}$ and $w_{i}$, evaluated at $\left(z,w,y_{-1},u_{0},y_{1}^{m}\right).$

\begin{cond}\label{cond1} For a given value $(z,w,y_{-1},u_{0})$
	in the interior of $\mathcal{Z}\times\mathcal{W}\times\mathcal{Y}_{-1}\times\mathcal{U}_{0}$,
	there exist $C$ different values, $y_{1}^{1},\dots,y_{1}^{C}$,
	in the interior of the support of $y_{i1}$ conditional on $(z,w,y_{-1},u_{0})$
	such that $\Pi_{1}\left(y_{1}^{1},\dots,y_{1}^{C};z,w,y_{-1},u_{0}\right)$
	has rank $C$. \end{cond}

\begin{cond}\label{cond2} For a given value $(z,w,y_{-1},u_{0})$
	in the interior of $\mathcal{Z}\times\mathcal{W}\times\mathcal{Y}_{-1}\times\mathcal{U}_{0}$,
	there exist $2C$ different values, $y_{1}^{1},\dots,y_{1}^{2C}$,
	in the interior of the support of $y_{i1}$ conditional on $(z,w,y_{-1},u_{0})$
	such that $\Pi_{2}\left(y_{1}^{1},\dots,y_{1}^{2C};z,w,y_{-1},u_{0}\right)$
	has rank $2C$. \end{cond}

Note that we can choose $C$ different values of $y_{i1}$ to satisfy \cref{cond1} and then independently choose another $2C$ values of $y_{i1}$  to satisfy \cref{cond2}.

Let $\epsilon_{c}^{\rho}$ denote the $\rho$-quantile of $\epsilon_{ic}$, i.e., $\epsilon_{c}^{\rho}=\text{Quantile}_{\epsilon_{ic}} (\rho) \equiv \inf\{\epsilon_{c}:\,F_{\epsilon_{ic}}(\epsilon_{c})\geq \rho\}$ for $\rho\in(0,1)$, where $F_{\epsilon_{ic}}$ denote the marginal CDF of $\epsilon_{ic}$.  
\begin{prop}\label{prop_np_ns} 
Suppose that Assumptions~\ref{assm_npns}-\ref{assm_ui0} and Conditions~\ref{cond1}
and \ref{cond2} are satisfied. We have (i) for each $c\in\mathbf{C}\backslash\{1\}$,
for any value $(z,y_{c})$ in the interior of $\mathcal{Z}\times\mathcal{Y}_{c}$, for any $\rho \in (0,1)$, and for any coordinate $k=1,\dots,d_{z}$, 
$\frac{\partial u^{c}\left(z,y_{c},\epsilon_{c}^{\rho}\right)}{\partial z_{i}^{k}}$ and $ \frac{\partial u^{c}\left(z,y_{c},\epsilon_{c}^{\rho}\right)}{\partial y_{ic}}$ are identified; 
for $c=1$, $\frac{\partial u^{1}\left(z,y_{1}+\epsilon_{1}^{\rho}\right)}{\partial z_{i}^{k}}$
and $\frac{\partial u^{1}\left(z,y_{1}+\epsilon_{1}^{\rho}\right)}{\partial y_{i1}}=\frac{\partial u^{1}\left(z,y_{1}+\epsilon_{1}^{\rho}\right)}{\partial\epsilon_{i1}}$
are identified; 
and (ii) for each $c\in\mathbf{C}$, for any value $(z,w_{c})$ in
the interior of $\mathcal{Z}\times\mathcal{W}_{c}$, for any coordinate
$k=1,\dots,d_{z}$, $\frac{\partial v^{c}\left(z,w_{c},\eta_{ci}\right)}{\partial z_{i}^{k}}/\frac{\partial v^{c}\left(z,w_{c},\eta_{ci}\right)}{\partial w_{ci}}$
is identified, where $\eta_{c}$ is such that $v^{c}\left(z,w_{c},\eta_{c}\right)=\delta_{c}$. 	
\end{prop}

We group the proofs at the end of this section.  Using the variation in $u_{i0}$, we identify the
derivatives of student utility functions at all quantiles of the unobservable $\epsilon_i$. For the college utility functions, without additional assumptions, we only identify the ratio of the derivatives at certain values of the unobservable (i.e., $\eta_{c}$ such that $v^{c}\left(z,w_{c},\eta_{c}\right)=\delta_{c}$). This is because, on the college side, the probability of being unmatched is determined by comparing $v_{ci}$ with $\delta_{c}$, while $\delta_{c}$ is unobserved and fixed. This lack of variation restricts the identification of the derivatives of $v_{ci}$. 

With a more restrictive functional form of $v_{ci}$, the following corollary identifies these derivatives. For that, we let $\eta_{c}^{\rho}$ be the $\rho$-quantile of $\eta_{ci}$, i.e., $\eta_{c}^{\rho}=\text{Quantile}_{\eta_{ci}} (\rho) \equiv \inf\{\eta_{c}:\,F_{\eta_{ci}}(\eta_{c})\geq \rho\}$ for $\rho\in(0,1)$, where $F_{\eta_{ci}}$ is the marginal CDF of $\eta_{ci}$.
\begin{cor} \label{cor_npns} 
		Suppose that $v_{ci}=v^{c}\left(z_{i},\eta_{ci}\right)+w_{ci}$, that $w_{ci}$ has a large support, and that Assumptions~\ref{assm_npns}, \ref{indep_npns}, and
		\ref{assm_ui0}(i) and Conditions~\ref{cond1} and \ref{cond2} are
		satisfied. For any value $z$ in the interior of
		$\mathcal{Z}$, for all $c\in\mathbf{C}$, any $\rho \in (0,1)$, 
		and $k=1,\dots,d_{z}$, $\frac{\partial v^{c}\left(z,\eta^\rho_{c} \right)}{\partial z_{i}^{k}}$
		is identified.
\end{cor}

For this  corollary, we do not need \cref{assm_ui0}(ii), which is required only for identifying the derivatives of $u^c$ for all possible values of $\epsilon_{ic}$.

\begin{proof}[Proof of \cref{prop_np_ns}]
	
	To simplify notations, for $k=1,\dots,d_{z}$, let $u_{z_{i}^{k}}^{c}=\frac{\partial u^{c}}{\partial z_{i}^{k}}$
	and similar notations are defined for $v^{c}$, $\tilde{u}^{c}$,
	$\tilde{v}^{c}$, and the other variables, and let $\sigma_{0}^{m}=\sigma_{0}\left(z,w,y_{-1},u_{0},y_{1}^{m}\right)$
	for $m=1,\dots,2C$. Let $t^{m}$ be the value of $\left(\tau_{i1},\dots,\tau_{iC},\iota_{1i},\dots,\iota_{Ci}\right)$
	evaluated at $(z,w,y_{-1},u_{0},y_{1}^{m})$. Under Assumption \ref{assm_npns}(i)-(iii),
	in equation \eqref{eq:sigma_0_app}, $\Lambda_{0}$, $u^{c}$,	and $v^{c}$ are continuously differentiable and the observables are
	all continuously distributed. Taking derivatives of equation \eqref{eq:sigma_0_app}
	on both sides w.r.t.\ $y_{ic}$ and $w_{ci}$, and evaluating them at $\left(z,w,y_{-1},u_{0},y_{1}^{m}\right)$,
	we have, for $c=1$, 
	\begin{equation}
		\frac{\partial\sigma_{0}^{m}}{\partial y_{i1}}=-\frac{\partial\Lambda_{0}(t^{m})}{\partial\tau_{i1}}\mbox{ and }\frac{\partial\sigma_{0}^{m}}{\partial w_{1i}}=\frac{\partial\Lambda_{0}(t^{m})}{\partial\iota_{1i}}\tilde{v}_{w_{1i}}^{1},\label{eq:dy1}
	\end{equation}
	and, for $c\neq1$, 
	\begin{equation}
		\frac{\partial\sigma_{0}^{m}}{\partial y_{ic}}=\frac{\partial\Lambda_{0}(t^{m})}{\partial\tau_{ic}}\tilde{u}_{y_{ic}}^{c}\mbox{ and }\frac{\partial\sigma_{0}^{m}}{\partial w_{ci}}=\frac{\partial\Lambda_{0}(t^{m})}{\partial\iota_{ci}}\tilde{v}_{w_{ci}}^{c}.\label{eq:dwc}
	\end{equation}
	
	Further, taking derivatives of equation \eqref{eq:sigma_0_app} on
	both sides w.r.t.\ $u_{i0}$ and $z_{i}^{k}$, and evaluating them at $\left(z,w,y_{-1},u_{0},y_{1}^{m}\right)$,
	we have 
	\begin{eqnarray}
		\frac{\partial\sigma_{0}^{m}}{\partial u_{i0}} & = & \sum_{c=1}^{C}\frac{\partial\Lambda_{0}(t^{m})}{\partial\tau_{ic}}\tilde{u}_{u_{i0}}^{c},\label{eq:du0}\\
		\frac{\partial\sigma_{0}^{m}}{\partial z_{i}^{k}} & = & \sum_{c=1}^{C}\frac{\partial\Lambda_{0}(t^{m})}{\partial\tau_{ic}}\tilde{u}_{z_{i}^{k}}^{c}+\sum_{c=1}^{C}\frac{\partial\Lambda_{0}(t^{m})}{\partial\iota_{ci}}\tilde{v}_{z_{i}^{k}}^{c}.\label{eq:dz}
	\end{eqnarray}
	Substituting equations~\eqref{eq:dy1} and \eqref{eq:dwc} into equations~\eqref{eq:du0}
	and \eqref{eq:dz}, we have 
	\begingroup \allowdisplaybreaks 
	\begin{eqnarray}
		\frac{\partial\sigma_{0}^{m}}{\partial u_{i0}} & = & -\frac{\partial\sigma_{0}^{m}}{\partial y_{i1}}\tilde{u}_{u_{i0}}^{1}+\sum_{c=2}^{C}\frac{\partial\sigma_{0}^{m}}{\partial y_{ic}}(\tilde{u}_{y_{ic}}^{c})^{-1}\tilde{u}_{u_{i0}}^{c},\label{eq:du0_1}\\
		\frac{\partial\sigma_{0}^{m}}{\partial z_{i}^{k}} & = & -\frac{\partial\sigma_{0}^{m}}{\partial y_{i1}}\tilde{u}_{z_{i}^{k}}^{1}+\sum_{c=2}^{C}\frac{\partial\sigma_{0}^{m}}{\partial y_{ic}}(\tilde{u}_{y_{ic}}^{c})^{-1}\tilde{u}_{z_{i}^{k}}^{c}+\sum_{c=1}^{C}\frac{\partial\sigma_{0}^{m}}{\partial w_{ci}}(\tilde{v}_{w_{ci}}^{c})^{-1}\tilde{v}_{z_{i}^{k}}^{c}.\label{eq:dz_1}
	\end{eqnarray}
	\endgroup

To get the relationship between the derivatives of $u^{c}$ and $\tilde{u}^{c}$, for $\forall c \in \bfC \backslash\{1\}$, taking derivatives on both side of the equation,  $u^{c}(z_{i},y_{ic},\tilde{u}^{c}(z_{i},y_{ic},u_{i0}))=u_{i0}$,  
w.r.t.\ $y_{ic}$, $u_{i0}$, and $z_{i}^{k}$, one gets, $u_{y_{ic}}^{c}+u_{\epsilon_{ic}}^{c}\tilde{u}_{y_{ic}}^{c}=0$,
$u_{\epsilon_{ic}}^{c}\tilde{u}_{u_{i0}}^{c}=1$, and $u_{z_{i}^{k}}^{c}+u_{\epsilon_{ic}}^{c}\tilde{u}_{z_{i}^{k}}^{c}=0$; 
it then follows that $\tilde{u}_{y_{ic}}^{c}=-\frac{u_{y_{ic}}^{c}}{u_{\epsilon_{ic}}^{c}}$,
$\tilde{u}_{u_{i0}}^{c}=\frac{1}{u_{\epsilon_{ic}}^{c}}$, and that
$\tilde{u}_{z_{i}^{k}}^{c}=-\frac{u_{z_{i}^{k}}^{c}}{u_{\epsilon_{ic}}^{c}}$ at the value of $\epsilon_{ic}$, $\epsilon_c$, such that $u^c(z,y_{c},\epsilon_{c})=u_{0}$. Similarly, for $c=1$, $\tilde{u}_{u_{i0}}^{1}=\frac{1}{u_{\epsilon_{i1}+y_{i1}}^{1}}$
and $\tilde{u}_{z_{i}^{k}}^{1}=-\frac{u_{z_{i}^{k}}^{1}}{u_{\epsilon_{i1}+y_{i1}}^{1}}$	at the value of $\epsilon_{i1}+y_{i1}$ such that $u^{1}(z,\epsilon_{1}+y_{1})=u_{0}$. Importantly, $y_{1}$ does not need to satisfy Conditions~\ref{cond1} and \ref{cond2} {because for any $y_1$, one can find an $\epsilon_1$ so that the above equation holds}.
	
	Similarly, taking derivatives of the equation, $v^{c}(z_{i},w_{ci},\tilde{v}^{c}(z_{i},w_{ci},\delta_{c}))=\delta_{c}$,
	w.r.t.\ $w_{ci}$ and $z_{i}^{k}$ and making rearrangements, we
	obtain, for $c\in\mathbf{C}$, $\tilde{v}_{w_{ci}}^{c}=-\frac{v_{w_{ci}}^{c}}{v_{\eta_{ci}}^{c}}$
	and $\tilde{v}_{z_{i}^{k}}^{c}=-\frac{v_{z_{i}^{k}}^{c}}{v_{\eta_{ci}}^{c}}$ at the value of $\eta_{ci}$ such that $v^c(z,w_{c},\eta_{c})=\delta_{c}$.
	
	Plugging the above relationships among the utility functions and their
	inverse into equations~\eqref{eq:du0_1} and \eqref{eq:dz_1}, we
	obtain 
	\begin{eqnarray}
		\frac{\partial\sigma_{0}^{m}}{\partial u_{i0}} & = & -\frac{\partial\sigma_{0}^{m}}{\partial y_{i1}}\frac{1}{u_{\epsilon_{i1}+y_{i1}}^{1}}-\sum_{c=2}^{C}\frac{\partial\sigma_{0}^{m}}{\partial y_{ic}}\frac{1}{u_{y_{ic}}^{c}},\label{eq:du0_2}\\
		\frac{\partial\sigma_{0}^{m}}{\partial z_{i}^{k}} & = & \frac{\partial\sigma_{0}^{m}}{\partial y_{i1}}\frac{u_{z_{i}^{k}}^{1}}{u_{\epsilon_{i1}+y_{i1}}^{1}}+\sum_{c=2}^{C}\frac{\partial\sigma_{0}^{m}}{\partial y_{ic}}\frac{u_{z_{i}^{k}}^{c}}{u_{y_{ic}}^{c}}+\sum_{c=1}^{C}\frac{\partial\sigma_{0}^{m}}{\partial w_{ci}}\frac{v_{z_{i}^{k}}^{c}}{v_{w_{ci}}^{c}}.\label{eq:dz_2}
	\end{eqnarray}
	
	Next, stacking equation \eqref{eq:du0_2} for $m=1,\dots,C$, we have
	\[
\begin{pmatrix}\frac{\partial\sigma_{0}^{1}}{\partial u_{i0}}\\
	\vdots\\
	\frac{\partial\sigma_{0}^{C}}{\partial u_{i0}}
\end{pmatrix}=-\begin{pmatrix}\frac{\partial\sigma_{0}\left(z,w,y_{-1},u_{0},y_{1}^{1}\right)}{\partial y_{i1}} & \cdots & \frac{\partial\sigma_{0}\left(z,w,y_{-1},u_{0},y_{1}^{1}\right)}{\partial y_{iC}}\\
\vdots & \ddots & \vdots\\
\frac{\partial\sigma_{0}\left(z,w,y_{-1},u_{0},y_{1}^{C}\right)}{\partial y_{i1}} & \cdots & \frac{\partial\sigma_{0}\left(z,w,y_{-1},u_{0},y_{1}^{C}\right)}{\partial y_{iC}}
\end{pmatrix} \cdot \begin{pmatrix}\frac{1}{u_{\epsilon_{i1}+y_{i1}}^{1}}\\
	\frac{1}{u_{y_{i2}}^{2}}\\
	\vdots\\
	\frac{1}{u_{y_{iC}}^{C}}
\end{pmatrix},
\]
	where the vector $(\frac{1}{u_{\epsilon_{i1}+y_{i1}}^{1}},\frac{1}{u_{y_{i2}}^{2}},\dots,\frac{1}{u_{y_{iC}}^{C}})'$
	is finite due to part (v) of Assumption \ref{assm_npns}. Note that the derivatives
	of $\sigma_{0}$ in the above system can be observed from the population data. Then, by \cref{cond1}, $\frac{1}{u_{\epsilon_{i1}+y_{i1}}^{1}}$ and $\frac{1}{u_{y_{ic}}^{c}}$ for each $c\in\mathbf{C}\backslash\{1\}$
	are identified.
	
	Similarly, stacking equation \eqref{eq:dz_2} for $m=1,\dots,2C$,
	we obtain 
{\scriptsize{}
	\begin{align*}
		\begin{pmatrix}\frac{\partial\sigma_{0}^{1}}{\partial z_{i}^{k}}\\
			\vdots\\
			\frac{\partial\sigma_{0}^{2C}}{\partial z_{i}^{k}}
		\end{pmatrix}= & \begin{pmatrix}\frac{\partial\sigma_{0}\left(z,w,y_{-1},u_{0},y_{1}^{1}\right)}{\partial y_{i1}} & \cdots & \frac{\partial\sigma_{0}\left(z,w,y_{-1},u_{0},y_{1}^{1}\right)}{\partial y_{iC}} & \frac{\partial\sigma_{0}\left(z,w,y_{-1},u_{0},y_{1}^{1}\right)}{\partial w_{1i}} & \cdots & \frac{\partial\sigma_{0}\left(z,w,y_{-1},u_{0},y_{1}^{1}\right)}{\partial w_{Ci}}\\
			\vdots & \ddots & \vdots & \vdots & \ddots & \vdots\\
			\frac{\partial\sigma_{0}\left(z,w,y_{-1},u_{0},y_{1}^{2C}\right)}{\partial y_{i1}} & \cdots & \frac{\partial\sigma_{0}\left(z,w,y_{-1},u_{0},y_{1}^{2C}\right)}{\partial y_{iC}} & \frac{\partial\sigma_{0}\left(z,w,y_{-1},u_{0},y_{1}^{2C}\right)}{\partial w_{1i}} & \cdots & \frac{\partial\sigma_{0}\left(z,w,y_{-1},u_{0},y_{1}^{2C}\right)}{\partial w_{Ci}}
		\end{pmatrix}\cdot\begin{pmatrix}u_{z_{i}^{k}}^{1}/u_{\epsilon_{i1}+y_{i1}}^{1}\\
			u_{z_{i}^{k}}^{2}/u_{y_{i2}}^{2}\\
			\vdots\\
			u_{z_{i}^{k}}^{C}/u_{y_{iC}}^{C}\\
			v_{z_{i}^{k}}^{1}/v_{w_{1i}}^{1}\\
			\vdots\\
			v_{z_{i}^{k}}^{C}/v_{w_{Ci}}^{C}
		\end{pmatrix}.
	\end{align*}
}{\scriptsize\par}
	Then, by \cref{cond2}, for all $c\in\mathbf{C}$ , $\frac{v_{z_{i}^{k}}^{c}}{v_{w_{ci}}^{c}}$
	is identified at the value of $\eta_{ci}$ such that $v^c(z,w_{c},\eta_{c})=\delta_{c}$.
	 Also, $\frac{u_{z_{i}^{k}}^{1}}{u_{\epsilon_{i1}+y_{i1}}^{1}}$, and for each $c \in \bfC \backslash\{1\}$, $\frac{u_{z_{i}^{k}}^{c}}{u_{y_{ic}}^{c}}$ are identified. Combining this with the first identification result, we identify $u_{z_{i}^{k}}^{c}$ for all $c$, at the value of $\epsilon_{ic}$ such that $u^c(z,y_{c},\epsilon_{c})=u_{0}$ for $c \in \bfC \backslash\{1\}$, and at the value of $\epsilon_{i1}+y_{i1}$ such that $u^{1}(z,\epsilon_{1}+y_{1})=u_{0}$ for $c=1$. 
		
Further, for each $c$ and for any $\rho\in(0,1)$, define the conditional $\rho$-quantile of $u_{i0}$ given $(z_{i},y_{ic})$ as $\text{Quantile}_{u_{i0}|(z_{i},y_{ic})}(\rho)=\inf\{u_0:\,F_{u_{i0}|(z_{i},y_{ic})}(u_0)\geq \rho\}$. Because of part~(iv) of \cref{assm_npns}, for any $(z,y_c)$, for $\epsilon_{ic}$ such that $u^c(z,y_{c},\epsilon_{ic})=u_{i0}$, the equivariance property of quantiles \citep[e.g.,][]{chesher2003identification} implies that
		\begin{equation*}
		\text{Quantile}_{u_{i0}|(z,y_{c})}(\rho)=u^{c}(z,y_{c},\epsilon^{\rho}_{c}),
		\end{equation*}
	where the LHS is known from the joint distribution of $(u_{i0},z_{i},y_{ic})$. Therefore, the above identification result indicates that for all $c$, we can identify $u_{z_{i}^{k}}^{c}$ for any given $(z,y_c)$ and $\rho\in(0,1)$.%
\end{proof}%
\begin{proof}[Proof of \cref{cor_npns}] 
		\cref{prop_np_ns} implies that $\frac{\partial v^{c}\left(z_i,\eta_{ci}\right)}{\partial z_{i}^{k}}$
		is identified, where $\eta_{ci}$ is such that $v^{c}\left(z_i,\eta_{ci}\right)+w_{ci}=\delta_{c}$. For any $z$ and $\rho \in (0,1)$, the equivariance property of quantiles \citep[e.g.,][]{chesher2003identification} implies that $	\text{Quantile}_{-w_{ci}|z}(\rho)=v^{c}(z,\eta^{\rho}_{c})$,
		where the LHS is known from the joint distribution of $(w_{ci},z_{i})$. Hence, $\frac{\partial v^{c}\left(z,\eta^\rho_{c} \right)}{\partial z_{i}^{k}}$ is identified.
\end{proof}

\section{A Control Function Approach}\label{app:cf}
This appendix discusses a control function approach that relaxes \cref{indep} in the identification of the derivatives of $\{u^{c},r^{c}, v^{c}\}_c$ .  

For simplicity, we consider the case where there is one endogenous variable. That is, $z_{i}=(z_{1i},z_{2i})$, where $z_{1i}$
is a scalar endogenous random variable and $z_{2i}$ is a
vector of exogenous random variables.
Suppose that $z_{1i}$ can be written as a nonparametric function of exogenous
variables $z_{2i}$, a vector of exogenous variables $t_{i}$ that is not contained in $z_{2i}$, and a scalar unobserved random variable $\xi_{i}$:
\begin{equation}
	z_{1i}=h(t_{i},z_{2i},\xi_{i}). \label{eq:cf}
\end{equation}
Assume that the unobservables $\xi_{i}$
and $(\epsilon_{i},\eta_{i})$ are independent of all the exogenous variables $(t_{i},z_{2i},y_{i},w_{i})$
but are not independent of each other. The endogeneity of $z_{1i}$ arises due to the correlation between $\xi_{i}$ and $(\epsilon_{i},\eta_{i})$. 

The following approach exploits a control variable $e_i$ such that conditional on $e_i$, $z_{1i}$ and $(\epsilon_{i},\eta_{i})$ are independent. In a nonadditive setting described in equation~\eqref{eq:cf}, suppose that the CDF of $\xi_i$ is strictly increasing and continuous, and that $h$ is strictly monotone in its last argument. Then the control variable $e_i=F_{z_{1i}|(t_{i},z_{2i})}(z_i,t_{i})=F_{\xi_i}(\xi_i)$, where $F_{z_{1i}|(t_{i},z_{2i})}(z_i,t_{i})$ is the conditional CDF of $z_{1i}$ given $(t_{i},z_{2i})$ and $F_{\xi_i}(\xi_i)$ is the CDF of $\xi_i$ \citep{imbens2009identification}. In an additive setting where $z_{1i}=h(t_{i},z_{2i})+\xi_{i}$ and $\mathbb{E}(\xi_{i}|t_{i},z_{2i})=0$, the control variable $e_i=\xi_i$.\footnote{For examples of parametric specifications in consumer choice models and in matching models, see \cite{petrin2010control} and \cite{agarwal_empirical_2015}.}

Suppose that each element in $(\epsilon_i, \eta_i)$ can be decomposed into a function of $e_i$ and a residual that is independent of $e_i$. Specifically, for each $c\in\boldsymbol{C}$, we obtain
\begin{eqnarray}
	\epsilon_{ic}  =  \varphi^{c}(e_{i})+\tilde{\epsilon}_{ic} \mbox{ and } 
	\eta_{ci}  =  \phi^{c}(e_{i})+\tilde{\eta}_{ci}.\label{eq:np_unobs}
\end{eqnarray}
Note that $\tilde{\epsilon}_{ic}$ and $\tilde{\eta}_{ci}$
are independent of $(t_{i},z_{2i},y_{i},w_{i})$ because $\xi_{i}$
(and thus $e_i$) and $(\epsilon_{i},\eta_{i})$ are both independent of $(t_{i},z_{2i},y_{i},w_{i})$.
Besides, $\tilde{\epsilon}_{ic}$ and $\tilde{\eta}_{ci}$ are independent
of $z_{1i}$ because $z_{1i}$ is a function of $(t_{i},z_{2i})$
and $\xi_{i}$.

Plugging equation~(\ref{eq:np_unobs}) into the utility functions in equation~\eqref{NP_specification}, we have
\[
u_{ic}=u^{c}(z_{i})+r^{c}(y_{ic})+\varphi^{c}(e_{i})+\tilde{\epsilon}_{ic}\mbox{ and }v_{ci}=v^{c}(z_{i})+w_{ci}+\phi^{c}(e_{i})+\tilde{\eta}_{ci},\forall c\in\mathbf{C}.
\]
We can treat $e_i$ as observed because it can be identified from the joint distribution of $(z_i,t_{i})$. A similar argument as that in \cref{prop_nonpara} then can be used to identify the derivatives of the functions $\{u^c,\,v^c,\,r^c,\,\varphi^c,\,\phi^c\}_c$.

\section{Evaluating \cref{cond}}\label{app:rank}
\subsection{A Nonparametric One-college Example}\label{app:rank_onecollege}
The following example shows that in a one-college case,  \cref{cond} holds for \textit{all} but the exponential distribution on $\eta_{1i}$.

\begin{example}\label{rank_example}
	Consider a one-college example: $\mathbf{C}=\{1\},$ and  $\mathcal{L}=\{\{0\},\{0,1\}\}$. Equation~\eqref{ideq2} for $c=1$ can be written as $\sigma_{1}(z_i,y_i,w_i) =\lambda_{\{0,1\}}(\iota_{1i})\cdot g_{1,\{0,1\}}(\tau_{i1})$ because $g_{1,\{0\}}(\tau_{i1})=0$.  Recall that  $\iota_{1i}=v^{1}(z_{i})+w_{1i}$ and $\tau_{i1} = u^{1}(z_{i})+r^{1}\left(y_{i1}\right)$.  We fix  $y_{i1}=\overline{y}_{1}$ and have $r_{1}'(\overline{y}_{1})=1$. 
	\cref{cond} requires that, for any $z$ in the interior of $\mathcal{Z}$, there are two values of $w_{1i}$, $\widehat{w}_1$ and $\widetilde{w}_1$, such that the following matrix is full-rank:
\[
\Pi(z,\overline{y}_{1},\widehat{w}_{1},\widetilde{w}_{1})=\begin{pmatrix}\lambda_{\{0,1\}}(\widehat{\iota}_{1})\cdot g'_{1,\{0,1\}}(\tau_{1}) & \lambda'_{\{0,1\}}(\widehat{\iota}_{1})\cdot g{}_{1,\{0,1\}}(\tau_{1})\\
		\lambda_{\{0,1\}}(\widetilde{\iota}_{1})\cdot g'_{1,\{0,1\}}(\tau_{1}) & \lambda'_{\{0,1\}}(\widetilde{\iota}_{1})\cdot g{}_{1,\{0,1\}}(\tau_{1})
	\end{pmatrix},
	\]
	where $\widehat{\iota}_1 \equiv v^{1}(z)+\widehat{w}_1$, $\widetilde{\iota}_1 \equiv  v^{1}(z)+\widetilde{w}_1$, and $\tau_1 \equiv  u^{1}(z)+r^{1}(\overline{y}_1)$. 
		A necessary condition for \cref{cond} is $g'_{1,\{0,1\}}(\tau_1) \neq 0$, which is satisfied if $\epsilon_{i1}$ has a strictly increasing cumulative distribution function.  Given that $g'_{1,\{0,1\}}(\tau_1) \neq 0$ {and $g_{1,\{0,1\}}(\tau_1) \neq 0$}, \cref{cond} is satisfied if $\frac{\lambda'_{\{0,1\}}(\widehat{\iota}_{1})}{\lambda_{\{0,1\}}(\widehat{\iota}_{1})} \neq \frac{\lambda'_{\{0,1\}}(\widetilde{\iota}_{1})}{\lambda_{\{0,1\}}(\widetilde{\iota}_{1})}$, or $\frac{\partial\log\lambda_{\{0,1\}}(\widehat{\iota}_{1})}{\partial\iota_{1i}} \neq \frac{\partial\log\lambda_{\{0,1\}}(\widetilde{\iota}_{1})}{\partial\iota_{1i}}$. 
		The violation of \cref{cond} stringently restricts $\lambda_{\{0,1\}}(\iota_{1i})$, or the probability of college~1 being feasible to $i$. Specifically, for fixed $z$, \cref{cond} is \textit{violated} if the supply elasticity w.r.t.\ $w_{1i}$ is linear in $w_{1i}$, or  $\frac{\partial\log\lambda_{\{0,1\}}(\iota_{1i})}{\partial\iota_{1i}}$ is a constant for all $w_{1i}$. This means that $\lambda_{\{0,1\}}(\iota_{1i})=\exp(a+b\iota_{1i})$ with constants $a$ and $b$, which only occurs when $\eta_{1i}$ has an exponential distribution.
\end{example}

\subsection{Parametric Analysis of Probit and Logit Models  \label{app:rank_para}}
\subsubsection{Two Colleges}
We now parameterize a two-college model with $\mathbf{C}=\{1,2\}$. Student~$i$'s utility when attending college~$c$ for $c=1,2$ is specified as
\begin{align*}
	u_{ic} = u^c(z_i) + r^c(y_{ic})  + \epsilon_{ic} = z_i+y_{ic}+\epsilon_{ic}.
\end{align*}
And $u_{i0} = \epsilon_{i0}$.  Because $r^c(y_{ic})=y_{ic}$, we can choose any value to be $\overline{y}_{c}$ at which $\frac{\partial r^c(\overline{y}_{c})}{\partial y_{ic}}=1$ as required by the scale normalization. 

College~$c$ values student~$i$ at 
\begin{align*}
	v_{ci} = v^c(z_i) +w_{ci}+\eta_{ci} = z_i+w_{ci}+\eta_{ci}.
\end{align*}

We will consider $\epsilon_{i0}$, $\epsilon_{ic}$, and $\eta_{ci}$ being i.i.d.\ $N(0,1)$ or type I extreme values. 

Let $\delta_1$ and $\delta_2$ be the cutoffs of the two colleges given the stable matching in the continuum economy.  Given the parametric assumptions, for a wide range of $(\delta_1,\delta_2)$ in $\mathbb{R}^2$, there exist a vector of college capacities and joint distributions of $(z_i,y_i,w_i)$ such that $(\delta_1,\delta_2)$ are the cutoffs given the stable matching. 

We start with a probit model in which $\epsilon_{i0}$, $\epsilon_{ic}$, and $\eta_{ci}$ are i.i.d.\ $N(0,1)$.  We use Mathematica to derive an expression for $\Pi(z,y,\widehat{w},\widetilde{w})$ and calculate its determinant.\footnote{The Mathematica notebook can be downloaded at  \url{https://drive.google.com/file/d/1ofNitk9tCWP7N1LzztawO2JJFOVWmOQW/view?usp=sharing}.}

To show that we can choose $(\widehat{w},\widetilde{w})$ to make $|\Pi(z,y,\widehat{w},\widetilde{w})|$ non-zero for given values of $(z_i,y_i)$, we consider a more adversarial case by fixing the values of $(\widehat{w}_{2}, \widetilde{w}_{1},\widetilde{w}_{2})$ while letting $\widehat{w}_1$ change freely. In this example, we let $\delta_1=1$ and $\delta_2=0.75$.  

Panel~(a) in Figure~\ref{fig:2col} shows how $|\Pi(z,y,\widehat{w},\widetilde{w})|$  changes with $\widehat{w}_1$ for 4 different vectors of $(z,y, \widehat{w}_{2}, \widetilde{w}_{1},\widetilde{w}_{2})$.   In each of the four cases, for a wide range of $\widehat{w}_1$, $ |\Pi(z,y,\widehat{w},\widetilde{w})|\neq 0$.  We have also experimented with more values of $(z,y, \widehat{w}_{2}, \widetilde{w}_{1},\widetilde{w}_{2})$ as well as different values of $(\delta_1,\delta_2)$ and found similar evidence for \cref{cond}.

\begin{mfignotesin}{Probit and Logit Models with Two Colleges \label{fig:2col}}
	{
	\begin{subfigure}[h]{0.475\textwidth}
	\caption{\scriptsize Probit model}
	\resizebox{1\textwidth}{!}{\includegraphics{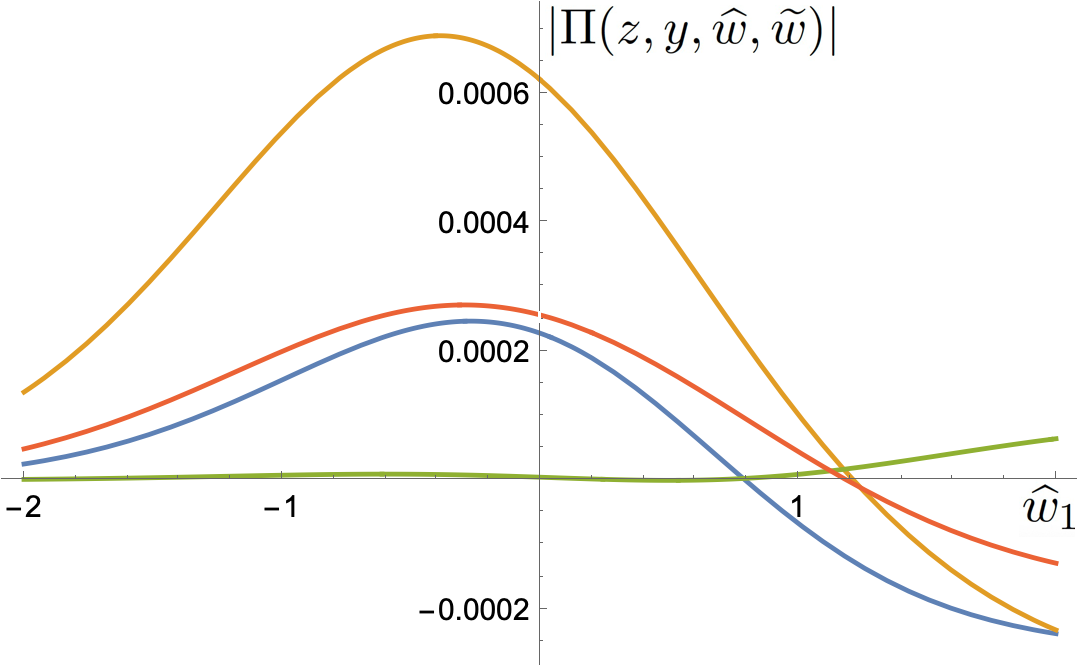}}
	
\end{subfigure}
\begin{subfigure}[h]{0.465\textwidth}
	\caption{\scriptsize Logit model}
	\resizebox{1\textwidth}{!}{\includegraphics{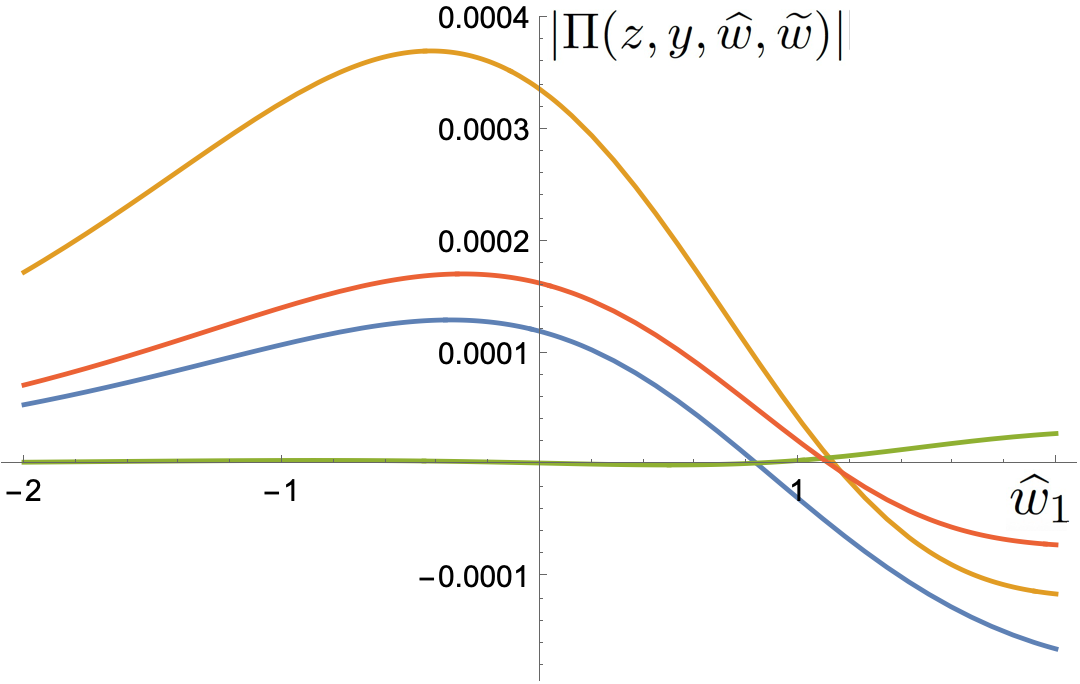}}
	
\end{subfigure}
	}
	{ This figure shows how $ |\Pi(z,y,\widehat{w},\widetilde{w})|$  changes with $\widehat{w}_{1}$ for 4 different vectors of $(z,y_{1},y_{2}, \widehat{w}_{2}, \widetilde{w}_{1},\widetilde{w}_{2})$ in a probit model (Panel a) and in a logit model (Panel b). The cutoffs are fixed at $\delta_1=1$ and $\delta_2=0.75$. For both panels, the four vectors of $(z,y_{1},y_{2}, \widehat{w}_{2}, \widetilde{w}_{1},\widetilde{w}_{2})$  (from the top line to the bottom line at $\widehat{w}_1=-2$) are: (i) $(1,-0.5,0.5,-0.5,1,0.5)$; (i) $(1,-1,1,-0.5,1,0.5)$; (iii) $(0.5,0.5,-0.5,-0.5,0.5,1)$; (iv) $(0.5,1,-1,1,0.5,1)$.}
\end{mfignotesin}

We then repeat the same analysis in a logit model. That is, $\epsilon_{i0}$, $\epsilon_{ic}$, and $\eta_{ci}$ are i.i.d.\  type I extreme values.   
Again, $\delta_1=1$ and $\delta_2=0.75$.  Panel~(b) in Figure~\ref{fig:2col} shows how $ |\Pi(z,y,\widehat{w},\widetilde{w})|$  changes with $\widehat{w}_1$ for 4 different vectors of $(z,y, \widehat{w}_{2}, \widetilde{w}_{1},\widetilde{w}_{2})$.   For all cases, there is again a wide range of $\widehat{w}_1$ such that \cref{cond} is satisfied.

\subsubsection{Logit Models with Three or Four Colleges \label{sec:3col}}

We further expand the example to three or four colleges.  Due to computational issues, it becomes infeasible to consider probit models.   We therefore focus on logit models.  Figure~\ref{fig:more} shows how $ |\Pi(z,y,\widehat{w},\widetilde{w})|$ changes with $\widehat{w}_1$ in the 2-, 3-, and 4-college examples for 4 different values of other variables.  From Panel~(a) to (c), there is no evidence that \cref{cond} becomes more difficult to satisfy as there are more colleges.

\begin{mfignotesin}{Logit Models with 2--4 Colleges \label{fig:more}}
	{
\centering
\begin{subfigure}[h]{0.32\textwidth}
	\caption{\scriptsize 2 colleges}
	\resizebox{1\textwidth}{!}{\includegraphics{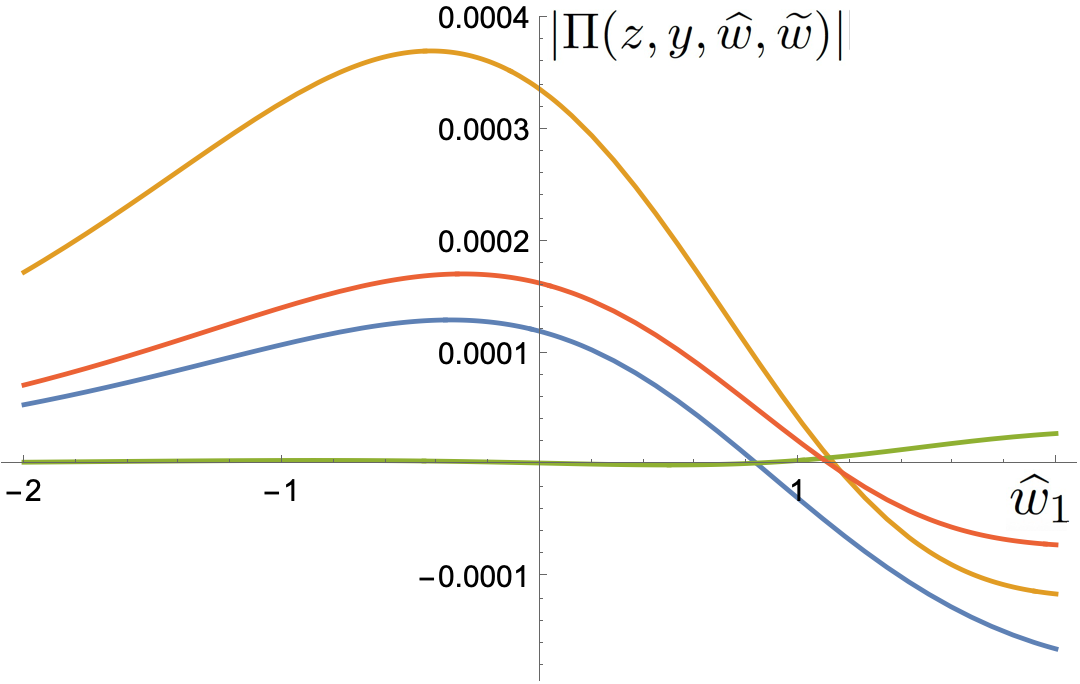}}
	
\end{subfigure}
\begin{subfigure}[h]{0.32\textwidth}
	\caption{\scriptsize 3 colleges}
	\resizebox{1\textwidth}{!}{\includegraphics{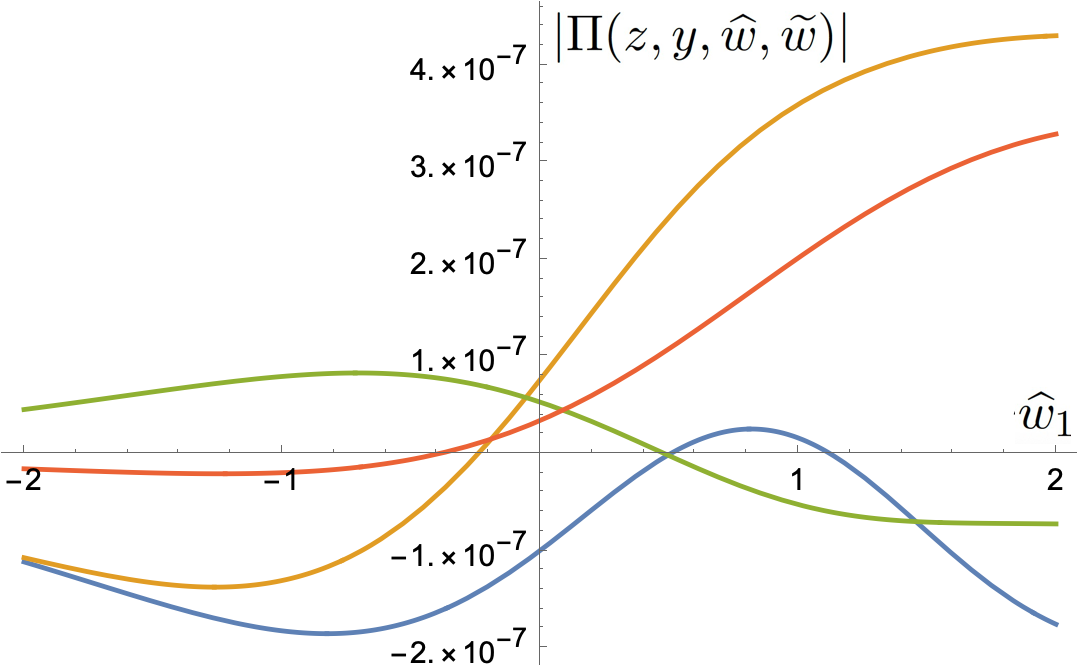}}
	
\end{subfigure}
\begin{subfigure}[h]{0.32\textwidth}
	\caption{\scriptsize 4 colleges}
	\resizebox{1\textwidth}{!}{\includegraphics{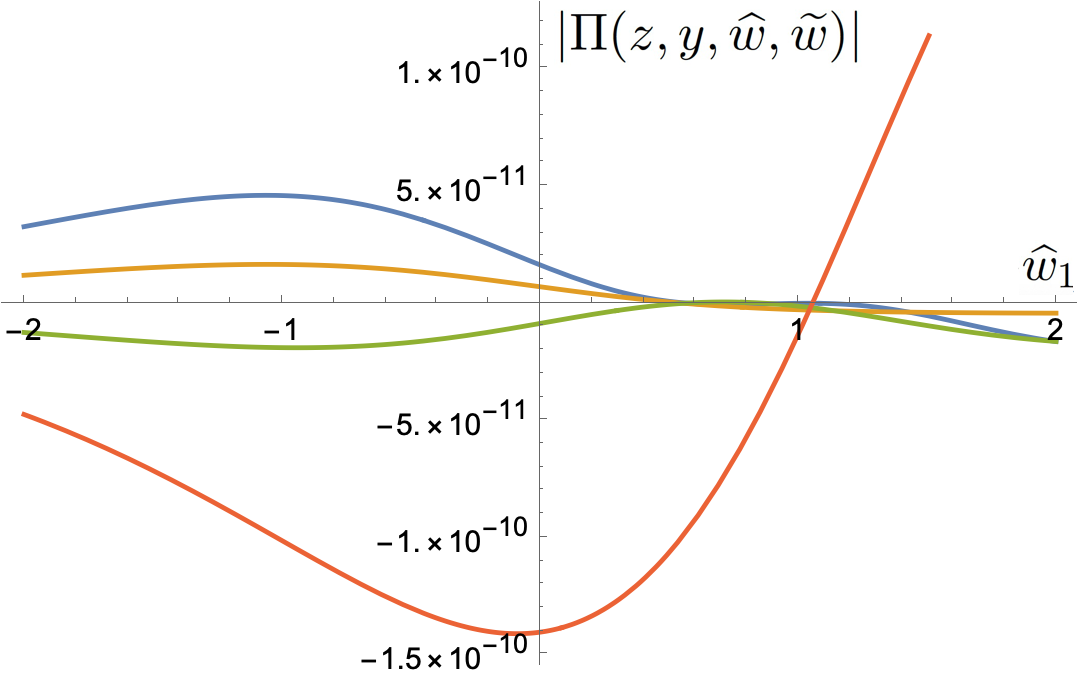}}
	
\end{subfigure}
	}
	{ This figure shows how $ |\Pi(z,y,\widehat{w},\widetilde{w})|$  changes with $\widehat{w}_1$ given 4 different vectors of other variables in each logit model with a different number of colleges.  Panel~(a) is the same as Panel~(b) in Figure~\ref{fig:2col}.   
		In Panel~(b), there are three colleges; the cutoffs are $\delta_1=1$, $\delta_2=0.75$, and $\delta_3=0.5$; and  the four vectors of $(z,y, \widehat{w}_2, \widehat{w}_3, \widetilde{w}_1, \widetilde{w}_2, \widetilde{w}_3)$  are: 
$(1, -1, 0.5, 1, 0.5, 0.5, 0.5, 1, 0.5)$, 
$(-1, 1, 0.5, -1, 0.5, 1, 0.5, 1, 0.5)$,
$(-0.5, 0.5, 0.5, -0.5, 0.5, 0.5, 1, 0.5, 1)$, 
and 
$(0.5, -0.5, 0.5, -0.5, 0.5, 0.5, 1, 0.5, 0.5)$. 
	In Panel~(c), there are four colleges; the cutoffs are $\delta_1=1$, $\delta_2=0.75$, $\delta_3=0.5$, and $\delta_4=0.6$; and  the four vectors of $(z,y, \widehat{w}_{2}, \widehat{w}_{3}, \widehat{w}_{4}, \widetilde{w}_{1}, \widetilde{w}_{2}, \widetilde{w}_{3}, \widetilde{w}_{4})$  are: $(0.5, -0.5, 0.5, 0.5, -0.5, 0.5, 0.5, 0.5, 1, 0.5,0.5, 0.5)$, $(-0.5, 0.5, 0.5, -0.5, -0.5, 0.5, 0.5, 0.5, 1, 0.5, 0.5, 1)$, $(1, -1, 0.5, 0.5, 1, 0.5, 1, 0.5, 0.5, 1, 1, 0.5)$,  and$(-1, 1, 0.5, -0.5,  -1, 0.5, 1, 1, 0.5, 1, 0.5, 0.5)$. 
}
\end{mfignotesin}

\renewcommand{\thethm}{1}

\begin{remark}
In Figure~\ref{fig:more}, the absolute value of $|\Pi(z,y,\widehat{w},\widetilde{w})|$ decreases (exponentially) with the number of colleges, but it is not a sign of possible violations of the full-rank condition.  In fact, such a pattern is implied by the definition of $\Pi(z,y,\widehat{w},\widetilde{w})$ because each element in the matrix is a partial derivative of a match probability and thus tends to be a small value. By the Leibniz formula for determinants, we have 
$$|\Pi(z,y,\widehat{w},\widetilde{w})| = \sum_{\varrho \in S_{2C} } \text{sgn}(\varrho) \prod_{j = 1}^{2C} \pi_{\varrho(j), \, j},$$
where \text{sgn} is the sign function of permutations in the permutation group $S_{2C}$, which returns $+1$ and $-1$ for even and odd permutations, respectively; $\pi_{\varrho(j), \, j}$ is the element of $\Pi$ in the $\varrho(j)$-th row and $j$-th column. Based on the discussion above, $ \prod_{j = 1}^{2C} \pi_{\varrho(j), \, j}$ tends to be small and decrease when $C$ increases, so does $|\Pi(z,y,\widehat{w},\widetilde{w})|$. 

As a piece of evidence that is consistent with this observation, when we express match probabilities in percentage points, the determinants corresponding to the three in Figure~\ref{fig:more} are $100^{2C}$ times of those in Figure~\ref{fig:more}  and thus increase in $C$ exponentially.\footnote{We verified this in Mathematica. The code and results can be downloaded at \url{https://drive.google.com/file/d/1sVCW1bpZ-HuVS2NXGLtdu--AttKDF1l-/view?usp=sharing}.}

\end{remark}

\section{Monte Carlo Simulations}\label{app:mc}
     In a series of Monte Carlo simulations, this appendix shows (i) that a semiparametric approach based on the results in \cref{id} suffers from the curse of dimensionality, and (ii) that a parametric model based on a Bayesian approach works well.

\subsection{Setup}\label{app:mc_setup}
There are $3000$ students competing for admissions to 3 colleges. The capacities of the colleges are $\{750, 700, 750\}$. Every student has access to an outside option of value $\epsilon_{i0}$ (i.i.d.\ $N(0, 1))$. Student~$i$'s utility when being admitted to college~$c$ is given by,
\begin{eqnarray}\label{eq:mc_stu_u}
  u_{ic} & = \beta^{y}_c \times y_{ic} + \beta^s_c \times s_i + \beta^z_c \times z_i + \epsilon_{ic},
\end{eqnarray}
where $y_{ic}$ is student-college-specific and follows i.i.d.\ (across colleges and across students) $N(0, 36)$, $s_i$ is one of the characteristics of student $i$ (i.i.d.\ $N(5, 36)$), $z_i$ is another characteristic of $i$ (i.i.d.\ $N(0, 36)$), and $\epsilon_{ic}$ is i.i.d.\ standard normal. For $c = 1,2,3$, $\beta^{y}_c=-1$ and $\beta^s_c =  \beta^z_c = 1$. 

College $c$ values each student as follows:
\begin{eqnarray}\label{eq:mc_col_v}
  v_{ci} & =               \gamma^w_c \times w_{ci} + \gamma^m_c \times m_i  + \gamma^z_c \times z_i + \eta_{ci},
\end{eqnarray}
where $w_{ci}$ is a student-college-specific characteristic (i.i.d.\ $N(0, 36)$), $m_i$ is another characteristic of student $i$ (i.i.d.\ $N(0, 36)$), and $\eta_{ic}$ is i.i.d.\ standard normal. $z_i$ appears in both student and college preferences. For $c = 1, 2,3$, $\gamma^w_c=\gamma^m_c = \gamma^z_c =1$.  
For simplicity, we assume that $T_c=-\infty$ or, equivalently, every college finds every student acceptable.  

There are in total 150 MC samples (markets). The capacity constraint is always binding. Note that we obtain a set of estimates from each sample/market.

\subsection{Estimation: Average Derivatives}

 To operationalize our nonparametric results, we impose three additional assumptions. First, the true functional form is known except for the distribution of $(\epsilon_{i},\eta_i)$, which gives us a semiparametric setting. Second, in  student preferences, the parameters to be estimated are $\beta^s_c =  1$ for $c = 1,2,3$ and $\beta^z$ such that $ \beta^z_c = \beta^z= 1$ (i.e., we have prior knowledge that $\beta^z_c$ is constant across colleges). Third, in college preferences, the parameters to be estimated are $\gamma^m_c = 1$ for $c = 1, 2,3$ and $\gamma^z$ such that $\gamma^z_c = \gamma^z = 1$ (i.e., we have prior knowledge that $\gamma_c^z$ is constant across colleges).

Let $x_{i}=(y_{i},w_{i},s_i,z_i, m_i)$, with $y_{i}=(y_{i1},y_{i2},y_{i3})$ and $w_{i}=(w_{1i},w_{2i},w_{3i})$. 
{We rewrite equation~(\ref{eq:indsimC}) in the semiparametric setting for $s_i$ and $m_i$, respectively, integrate over the entire support of $x_i$ to obtain unconditional expectations $\mathbb{E}$:}
\begin{footnotesize}
\begin{align}\label{eq:beta}
\begin{pmatrix}
\mathbb{E}\left(\dfrac{\partial \sigma_1(x_i)}{\partial s_i}\right)\\
\mathbb{E}\left(\dfrac{\partial \sigma_2(x_i)}{\partial s_i}\right)\\
\mathbb{E}\left(\dfrac{\partial \sigma_3(x_i)}{\partial s_i}\right)
\end{pmatrix}
=
\begin{pmatrix}
\mathbb{E}\left(-\dfrac{\partial \sigma_1(x_i)}{\partial y_{i1}}\right) & \mathbb{E}\left(-\dfrac{\partial \sigma_1(x_i)}{\partial y_{i2}}\right) & \mathbb{E}\left(-\dfrac{\partial \sigma_1(x_i)}{\partial y_{i3}}\right) \\
\mathbb{E}\left(-\dfrac{\partial \sigma_2(x_i)}{\partial y_{i1}}\right) &  \mathbb{E}\left(-\dfrac{\partial \sigma_2(x_i)}{\partial y_{i2}}\right) & \mathbb{E}\left(-\dfrac{\partial \sigma_2(x_i)}{\partial y_{i3}}\right) \\
\mathbb{E}\left(-\dfrac{\partial \sigma_3(x_i)}{\partial y_{i1}}\right) &  \mathbb{E}\left(-\dfrac{\partial \sigma_3(x_i)}{\partial y_{i2}}\right) & \mathbb{E}\left(-\dfrac{\partial \sigma_3(x_i)}{\partial y_{i3}}\right)
\end{pmatrix}
\cdot
\begin{pmatrix}
\beta^s_1\\
\beta^s_2\\
\beta^s_3
\end{pmatrix}.
\end{align}
\end{footnotesize}
\begin{footnotesize}
	\begin{align}\label{eq:gamma}
  \begin{pmatrix}
\mathbb{E}\left(\dfrac{\partial \sigma_1(x_i)}{\partial m_i}\right)\\
\mathbb{E}\left(\dfrac{\partial \sigma_2(x_i)}{\partial m_i}\right)\\
\mathbb{E}\left(\dfrac{\partial \sigma_3(x_i)}{\partial m_i}\right)
\end{pmatrix}
=
\begin{pmatrix}
\mathbb{E}\left(\dfrac{\partial \sigma_1(x_i)}{\partial w_{1i}}\right) &  \mathbb{E}\left(\dfrac{\partial \sigma_1(x_i)}{\partial w_{2i}}\right) & \mathbb{E}\left(\dfrac{\partial \sigma_1(x_i)}{\partial w_{3i}}\right) \\
\mathbb{E}\left(\dfrac{\partial \sigma_2(x_i)}{\partial w_{1i}}\right) &  \mathbb{E}\left(\dfrac{\partial \sigma_2(x_i)}{\partial w_{2i}}\right) & \mathbb{E}\left(\dfrac{\partial \sigma_2(x_i)}{\partial w_{3i}}\right) \\
\mathbb{E}\left(\dfrac{\partial \sigma_3(x_i)}{\partial w_{1i}}\right) &  \mathbb{E}\left(\dfrac{\partial \sigma_3(x_i)}{\partial w_{2i}}\right) & \mathbb{E}\left(\dfrac{\partial \sigma_3(x_i)}{\partial w_{3i}}\right)
\end{pmatrix}
\cdot
\begin{pmatrix}
\gamma^m_1\\
\gamma^m_2\\
\gamma^m_3
\end{pmatrix}.
\end{align}
\end{footnotesize}%
The derivatives with respect to $z_{i}$ lead to:
\begin{footnotesize}
	\begin{align}\label{eq:z}
\begin{pmatrix}
\mathbb{E}\left(\dfrac{\partial \sigma_1(x_i)}{\partial z_i}\right)\\
\mathbb{E}\left(\dfrac{\partial \sigma_2(x_i)}{\partial z_i}\right)\\
\mathbb{E}\left(\dfrac{\partial \sigma_3(x_i)}{\partial z_i}\right)
\end{pmatrix}
=
\begin{pmatrix}
  \mathbb{E}\left(\sum_{c=1}^{3}  \dfrac{\partial \sigma_1(x_i)}{\partial w_{ci}}\right) & -\mathbb{E}\left( \sum_{c=1}^{3} \dfrac{\partial \sigma_1(x_i)}{\partial y_{ic}}\right) \\
\mathbb{E}\left(\sum_{c=1}^{3}  \dfrac{\partial \sigma_2(x_i)}{\partial w_{ci}}\right) & -\mathbb{E}\left(\sum_{c=1}^{3} \dfrac{\partial \sigma_2(x_i)}{\partial y_{ic}}\right) \\
\mathbb{E}\left(\sum_{c=1}^{3} \dfrac{\partial \sigma_3(x_i)}{\partial w_{ci}}\right) & -\mathbb{E}\left(\sum_{c=1}^{3} \dfrac{\partial \sigma_3(x_i)}{\partial y_{ic}}\right)
\end{pmatrix}
\cdot
\begin{pmatrix}
\gamma^z\\
\beta^z
\end{pmatrix}.
\end{align}
\end{footnotesize}%
We now have 3 equations in 2 unknowns specified by equation~\eqref{eq:z}. Using any two of the equations leads to an estimator. Moreover, we can formulate an estimator based on the generalized method of moments (GMM) that uses all three equations.

In sum, our estimation of $\beta$'s and $\gamma$'s relies on equation systems~\eqref{eq:beta}--\eqref{eq:z}.


\paragraph{Results.} The estimation results from the 150 MC samples are in the left part of Table~\ref{tab:ade} (columns~1--3). We observe that the estimated coefficients are not close to their true values. The performance does not improve significantly when we double the sample size. Our explanation for this poor performance in the estimation is the curse of dimensionality. {When calculating partial derivatives in equation systems~\eqref{eq:beta} and \eqref{eq:gamma}, we deal with 4-dimensional objects (i.e.,  $(s_i, y_{i1},y_{i2},y_{i3})$ or $(m_i, w_{1i},w_{2i},w_{3i})$); in equation system~\eqref{eq:z}, it is 7-dimensional (i.e., $(z_i, y_{i1},y_{i2},y_{i3}$, $w_{1i},w_{2i},w_{3i})$), which may explain that the estimators for $\beta^z$ and $\gamma^z$ perform the worst.}  This explanation is confirmed when we reduce the dimensionality in the model.

\begin{table}[ht!]
    \centering\footnotesize
  \caption{Semiparametric Estimation: The General and Reduced Models \label{tab:ade}}
    \resizebox{0.7\textwidth}{!}{\begin{tabular}{lcccclccc}
\toprule
          & \multicolumn{3}{c}{General: higher dimensionality} &       &       & \multicolumn{3}{c}{Reduced: lower dimensionality} \\          & \multicolumn{3}{c}{$\beta_c^z=\beta^z$, $\gamma_c^z=\gamma^z$} &       &       & \multicolumn{3}{c}{$\beta_2^z=\beta_3^z=\gamma_1^z=\gamma_2^z=0$} \\
\cline{2-4} \cline{7-9}
          & Median & Mean  & Std. Dev. &       &       & Median & Mean  & Std. Dev. \\
          & (1) & (2)  & (3) &       &       & (4) & (5)  & (6) \\
          \midrule
 & \multicolumn{8}{l}{\it A. Coefficients on $s$ in student preferences (true value $=1$)}\\
    $\beta^s_1$        & 0.98  & 1.11  & 0.50  &       &   $\beta^s_1$    &  0.98  & 1.21  & 0.98 \\
    $\beta^s_2$        & 1.00  & 1.11  & 0.52 &       &   $\beta^s_2$     & 0.91  & 1.16  & 1.06  \\
    $\beta^s_3$        & 0.99  & 1.12  & 0.50  &       &   $\beta^s_3$     &  1.01  & 1.18  & 0.82  \\
 & \multicolumn{8}{l}{\it B. Coefficients on $m$ in college preferences (true value $=1$)}\\
   $\gamma^m_1$  & 1.04  & 1.64  & 3.21&      &   $\gamma^m_1$      & 1.00  & 1.02  & 0.27 \\
    $\gamma^m_2$  & 0.94  & 1.30  & 3.71  &       &   $\gamma^m_2$    & 1.02  & 1.05  & 0.30 \\
    $\gamma^m_3$  & 1.12  & 1.47  & 3.41 &      &   $\gamma^m_3$    & 0.98  & 1.09  & 0.44  \\
  & \multicolumn{8}{l}{\it C. Coefficients on $z$ in student and college preferences (true value $=1$)}\\
     \multicolumn{9}{c}{GMM with all conditions in equation~\eqref{eq:z}}  \\
    $\beta^z$ & 0.12  & 0.41  & 2.54 &       & $\beta^z_1$ & 0.97  & 0.99  & 0.16 \\
    $\gamma^z$ & 0.16  & 0.08  & 3.13  &       & $\gamma^z_3$ & 0.97  & 1.00  & 0.21 \\
         \multicolumn{9}{c}{Using conditions 1 \& 2 in equation~\eqref{eq:z}}  \\
    $\beta^z$ & 0.05  & 1.11  & 10.37 &       & $\beta^z_1$ & 0.97  & 1.01  & 0.26  \\
    $\gamma^z$ & 0.17  & -0.59 & 5.71  &       & $\gamma^z_3$ &  0.97  & 1.11  & 1.30 \\
          \multicolumn{9}{c}{Using conditions 1 \& 3 in equation~\eqref{eq:z}}  \\
    $\beta^z$ & 0.30  & 0.08  & 25.20 &       & $\beta^z_1$ & 0.97  & 1.00  & 0.15 \\
    $\gamma^z$ & 0.19  & 8.37  & 92.48  &       & $\gamma^z_3$ & 0.98  & 1.00  & 0.20 \\
          \multicolumn{9}{c}{Using conditions 2 \& 3 in equation~\eqref{eq:z}}  \\
    $\beta^z$ & -0.06 & 0.84  & 7.34  &       & $\beta^z_1$ & 0.99  & 1.03  & 0.35  \\
    $\gamma^z$ & 0.08  & 0.30  & 6.35   &       & $\gamma^z_3$ & 0.95  & 1.03  & 0.30  \\
\bottomrule
    \end{tabular}}
    \begin{tabnotes}
          This table presents estimates for the coefficients in student or college utility functions (equations~\ref{eq:mc_stu_u} and \ref{eq:mc_col_v}). The statistics are calculated using 150 MC samples.
In the general model, we assume that $\beta^z_c = \beta^z$  (i.e., we have prior knowledge that $\beta_c^z$ is constant across colleges) and $\gamma^z_c = \gamma^z$.  The estimation is based on equation systems~\eqref{eq:beta}, \eqref{eq:gamma}, and \eqref{eq:z}.
In the reduced model, we assume that we know $ \beta^z_2 = \beta^z_3 = 0$ (i.e., $z_i$ does not enter $i$'s utility for college~2 or 3) and $\gamma^z_1 = \gamma^z_2=0$ (i.e., colleges~1 and 2 do not use $z_i$ to evaluate students).
The estimation is based on equation systems~\eqref{eq:beta}, \eqref{eq:gamma}, and \eqref{eq:z_reduced}.
    \end{tabnotes}
\end{table}%

\paragraph{Reduced dimensionality.}
In student preferences (equation~\ref{eq:mc_stu_u}), we further impose that the parameters to be estimated are $\beta^s_c =  1$ for $c = 1,2,3$ and $ \beta^z_1 = 1$, while we assume, and know, that $ \beta^z_2 = \beta^z_3 = 0$ (i.e., $z_i$ does not enter $i$'s utility for college~2 or 3). In college preferences (equation~\ref{eq:mc_col_v}), the parameters to be estimated are $\gamma^m_c = 1$ for $c = 1, 2,3$ and  $\gamma^z_3 = 1$, while we assume, and know, that $\gamma^z_1 = \gamma^z_2=0$ (i.e., colleges~1 and 2 do not use $z_i$ to evaluate students). Based on these new parameter values, {\it we re-generate another 150 MC samples} for estimation. 

We now have a simplified version of equation~\eqref{eq:z} with a reduced dimension:
\begin{footnotesize}
\begin{align}\label{eq:z_reduced}
\begin{pmatrix}
\mathbb{E}\left(\dfrac{\partial \sigma_1(x_i)}{\partial z_i}\right)\\
\mathbb{E}\left(\dfrac{\partial \sigma_2(x_i)}{\partial z_i}\right)\\
\mathbb{E}\left(\dfrac{\partial \sigma_3(x_i)}{\partial z_i}\right)
\end{pmatrix}
=
\begin{pmatrix}
  \mathbb{E}\left( \dfrac{\partial \sigma_1(x_i)}{\partial w_{3i}}\right) & -\mathbb{E}\left( \dfrac{\partial \sigma_1(x_i)}{\partial y_{i1}}\right) \\
\mathbb{E}\left(  \dfrac{\partial \sigma_2(x_i)}{\partial w_{3i}}\right) & -\mathbb{E}\left(  \dfrac{\partial \sigma_2(x_i)}{\partial y_{i1}}\right) \\
\mathbb{E}\left( \dfrac{\partial \sigma_3(x_i)}{\partial w_{3i}}\right) & -\mathbb{E}\left( \dfrac{\partial \sigma_3(x_i)}{\partial y_{i1}}\right)
\end{pmatrix}
\cdot
\begin{pmatrix}
\gamma_3^z\\
\beta_1^z
\end{pmatrix}.
\end{align}
\end{footnotesize}

The estimation results are presented in the right half of Table~\ref{tab:ade} (columns~4--6). We observe that all estimates are centered around their corresponding true value. 

\subsection{A Parametric Approach: Bayesian Estimation}\label{sec:mc_bayes}
The practical difficulties of the semiparametric method motivate us to consider a parametric approach. We again focus on the utility functions as in equations~\eqref{eq:mc_stu_u} and \eqref{eq:mc_col_v} and use the 150 MC samples generated in \cref{app:mc_setup}. In other words, $z_i$ enters each college's preferences and each student's preferences over all colleges.  

We assume that we know the functional form and the distributions of $\epsilon_{ic}$ and $\eta_{ci}$; however, we do not know, and thus will estimate, the standard deviation of $\epsilon_{i3}$ (the shock in students' utility for college~3), denoted by $\zeta_{\epsilon}$. The other parameters to be estimated are $\beta_c^{y}$, $\beta_c^s$ and $\beta_c^z$ for all $c$ in student preferences and $\gamma_c^w$, $\gamma_c^m$ and $\gamma_c^z$ for all $c$ in college preferences.  Collectively, we denote them by $(\beta,\gamma, \zeta_{\epsilon})$.

\paragraph{Bayesian Estimation Procedure.} We use a Gibbs sampler to implement the Bayesian estimation. 
The priors for $\beta$, $\gamma$, $\zeta_{\epsilon}^2$ are:
$$\beta \sim N(0,\Sigma_{\beta}),\mbox{  } \gamma \sim N(0,\Sigma_{\gamma}), \mbox{  and }   \zeta_{\epsilon}^2 \sim IW(\overline{\zeta}_{\epsilon}^2, \nu_{\epsilon}).$$
where IW is the inverse Wishart distribution. Following Chapter 5 of \cite{rossi2012bayesian}, we set diffuse priors as follows: The prior variances of $\beta$ and $\gamma$ ($\Sigma_{\beta}$ and $\Sigma_{\gamma}$) are 100 times the identity matrix, and  $(\overline{\zeta}_{\epsilon}^2, \nu_{\epsilon}) = (1,2)$.

In each iteration, the Gibbs sampler goes through the following steps (for notational simplicity, we omit the index for iterations):  
\begin{enumerate}[wide, labelwidth=!, labelindent=6pt] \itemsep-0.25em  
\item Conditional on student preferences, $u_{ic}$, from the previous iteration, we update college preferences, $v_{ci}$, by invoking the restrictions implied by the stability of the observed matching. For each college~$c$, let $\mathcal{I}_{c}$ be the set of students with $u_{i\mu(i)} > u_{ic}$ (i.e., students who like their own match more than $c$) and  $\mathcal{I}^{c}$ be the set of students with $u_{i\mu(i)} < u_{ic}$. The updating of college~$c$'s utilities and cutoff has four parts. 
\begin{enumerate}\itemsep-0.25em 
\item $c$'s preferences over those who are matched with it: Given $v_{ci}$ from the previous iteration, we find $\underline{v}_{c} = \max_{i \in \mathcal{I}^c} v_{ci}$. For each $i$ such that $\mu(i) = c$, $v_{ci}$ is drawn from $N(\gamma^w_c  w_{ci} + \gamma^m_c  m_i  + \gamma^z_c  z_i, 1)$ truncated below by $\underline{v}_{c}$. 
\item $c$'s cutoff: It is the lowest utility among those who are matched with $c$.
\item $c$'s preferences over those in $\mathcal{I}^c$: $c$'s utility for any student $i \in \mathcal{I}^c$ is drawn from $N(\gamma^w_c  w_{ci} + \gamma^m_c  m_i  + \gamma^z_c  z_i, 1)$ truncated above by $c$'s cutoff. 
\item $c$'s preferences over those in $\mathcal{I}_c$: $c$'s utility for any student  $i \in \mathcal{I}_c$ is drawn from $N(\gamma^w_c  w_{ci} + \gamma^m_c  m_i  + \gamma^z_c  z_i, 1)$ (without any truncation). 
\end{enumerate}

\item Conditional on the updated college preferences $v_{ci}$ in this iteration, we update student preferences, $u_{ic}$, again by invoking the restrictions implied by stability of the observed match. Note that  $v_{ci}$ determines all colleges' cutoffs and their feasibility to each student. The updating of student preferences has three parts:\footnote{In the estimation, a student's outside option is an always feasible college. The student's preference for her outside option is also updated according to the following steps.} 
\begin{enumerate}\itemsep-0.25em 
\item $i$'s preferences over infeasible colleges: For an infeasible college $c$ (i.e., $v_{ci}$ is below $c$'s cutoff), student $i$'s utility is drawn from a normal distribution with mean $\beta^{y}_c  y_{ic} + \beta^s_c  s_i + \beta^z_c  z_i$ and variance 1 if $c \neq 3$ or $\zeta_{\epsilon}^2$ if $c=3$.  
\item $i$'s utility for her matched college: Given $u_{ic}$ from the previous iteration, we find the highest utility among all feasible colleges other than $\mu(i)$, denoted by $\underline{u}_i$. $i$'s utility for $\mu(i)$ is drawn from a normal distribution truncated below by $\underline{u}_i$ with mean $\beta^{y}_c  y_{ic} + \beta^s_c  s_i + \beta^z_c  z_i$ and variance 1 if $c \neq 3$ or $\zeta_{\epsilon}^2$ if $c=3$. 
\item $i$'s preferences over her unmatched feasible colleges:  $i$'s utility for a feasible college $c$ ($\neq \mu(i)$) is drawn from a normal distribution truncated above by $u_{i\mu(i)}$ with mean $\beta^{y}_c  y_{ic} + \beta^s_c  s_i + \beta^z_c  z_i$ and variance 1 if $c \neq 3$ or $\zeta_{\epsilon}^2$ if $c=3$. 
\end{enumerate}

\item Following the standard procedure as detailed in Chapter 5 of \cite{rossi2012bayesian}, we then update the distribution of $\beta$, $\gamma$, and $\zeta_{\epsilon}^2$ conditional on the updated $v_{ci}$ and $u_{ic}$ as well as the data. 

\end{enumerate}
 
For each MC sample, we iterate through the Markov Chain 1.5 million times, and discard the first 0.55 million draws as ``burn in'' to ensure mixing. We compute the Potential Scale Reduction Factor (PSRF) following \cite{gelman1992inference}. For all the 19 parameters across the 150 MC samples, 92.4\% of the PSRFs are below 1.1, while less than 1\% of them are above 1.3.

\paragraph{Results.} 
This parametric approach leads to the results in Table~\ref{tab:mc_bayes}. We observe that the estimator works well as the posterior means are close to the true values. Moreover, we conclude that the posterior standard deviation is a reasonable measure of estimation precision. Comparing column~(3), which represents the estimation precision, with column~(4), which is the median of the posterior standard deviations, we find that they are close to each other, although some of the values in column~(4) tend to be smaller. Reassuringly, no value in column~(3) is larger than the corresponding one in column~(7), which is the 95th percentile among the 150 posterior standard deviations for each coefficient.

\begin{table}[ht!]
	\centering\footnotesize
	\caption{Results from Bayesian Estimation}
	\label{tab:mc_bayes}
	\resizebox{0.85\textwidth}{!}{\begin{tabular}{ccccc|cccc}
			\toprule 
			& \multicolumn{3}{c}{Posterior mean} &  & \multicolumn{4}{c}{Posterior Std. Dev.}\tabularnewline
			& Median & Mean & Std. Dev. &  & Median & Mean & 5th Perc. & 95th Perc.\tabularnewline
			& (1) & (2) & (3) &  & (4) & (5) & (6) & (7) \tabularnewline			
			\midrule 
			True value = 1 &  &  &  &  &  &  &  & \tabularnewline
			$\beta_{1}^{s}$ & 1.03 & 1.04 & 0.08 &  & 0.07 & 0.07 & 0.06 & 0.09\tabularnewline
			$\beta_{2}^{s}$ & 1.03 & 1.04 & 0.08 &  & 0.07 & 0.07 & 0.06 & 0.09\tabularnewline
			$\beta_{3}^{s}$ & 1.03 & 1.04 & 0.08 &  & 0.07 & 0.07 & 0.06 & 0.09\tabularnewline
			$\gamma_{1}^{m}$ & 1.05 & 1.10 & 0.23 &  & 0.13 & 0.16 & 0.10 & 0.24\tabularnewline
			$\gamma_{2}^{m}$ & 1.04 & 1.06 & 0.15 &  & 0.13 & 0.13 & 0.10 & 0.20\tabularnewline
			$\gamma_{3}^{m}$ & 1.07 & 1.08 & 0.15 &  & 0.13 & 0.14 & 0.10 & 0.20\tabularnewline
			$\beta_{1}^{z}$ & 1.03 & 1.04 & 0.08 &  & 0.07 & 0.07 & 0.06 & 0.09\tabularnewline
			$\beta_{2}^{z}$ & 1.03 & 1.04 & 0.08 &  & 0.07 & 0.07 & 0.06 & 0.09\tabularnewline
			$\beta_{3}^{z}$ & 1.02 & 1.04 & 0.08 &  & 0.07 & 0.07 & 0.06 & 0.09\tabularnewline
			$\gamma_{1}^{z}$ & 1.06 & 1.10 & 0.23 &  & 0.14 & 0.15 & 0.10 & 0.24\tabularnewline
			$\gamma_{2}^{z}$ & 1.04 & 1.06 & 0.14 &  & 0.13 & 0.13 & 0.10 & 0.20\tabularnewline
			$\gamma_{3}^{z}$ & 1.06 & 1.08 & 0.15 &  & 0.13 & 0.14 & 0.10 & 0.19\tabularnewline
			\multicolumn{3}{l}{Coefficients on $y$ (true value $=-1$)} &  &  &  &  &  & \tabularnewline
			$\beta_{1}^{y}$ & -1.03 & -1.04 & 0.08 &  & 0.07 & 0.07 & 0.06 & 0.09\tabularnewline
			$\beta_{2}^{y}$ & -1.04 & -1.04 & 0.08 &  & 0.07 & 0.07 & 0.06 & 0.09\tabularnewline
			$\beta_{3}^{y}$ & -1.03 & -1.04 & 0.08 &  & 0.07 & 0.07 & 0.06 & 0.09\tabularnewline
			\multicolumn{3}{l}{Coefficients on $w$ (true value $=1$)} &  &  &  &  &  & \tabularnewline
			$\gamma_{1}^{w}$ & 1.04 & 1.10 & 0.22 &  & 0.13 & 0.15 & 0.10 & 0.23\tabularnewline
			$\gamma_{2}^{w}$ & 1.04 & 1.06 & 0.15 &  & 0.13 & 0.13 & 0.10 & 0.20\tabularnewline
			$\gamma_{3}^{w}$ & 1.06 & 1.08 & 0.14 &  & 0.13 & 0.14 & 0.10 & 0.19\tabularnewline
			\multicolumn{3}{c}{Std.\ dev.\ of student utility shock ($\epsilon_{i3}$)} &  &  &  &  &  & \tabularnewline
			$\zeta_{\epsilon}$ & 1.04 & 1.03 & 0.20 &  & 0.17 & 0.17 & 0.16 & 0.20\tabularnewline
			\bottomrule
	\end{tabular}}
	\begin{tabnotes}
		This table presents statistics on the posterior means and standard deviations of the coefficients in student and college utility functions (equations~\ref{eq:mc_stu_u} and \ref{eq:mc_col_v}). For each coefficient, there are 150 posterior means and 150 posterior standard deviations from the 150 Monte Carlo samples. For each sample, the Bayesian approach with a Gibbs sampler goes through the Markov Chain 1.5 million times, and we take the first 0.55 million iterations as ``burn in.''  The last 0.95 million iterations are used to calculate the posterior means and standard deviations in a sample. 
	\end{tabnotes}
\end{table}%

\section{Data Construction}\label{app:data}    
    For student and school characteristics, the main dataset we have used is the SIMCE test result dataset which is accompanied by parent and teacher questionnaires.  To extract tuition data and location of students and schools, we have used publicly available data on the Ministry's website, 
\sloppy
\url{http://datos.mineduc.cl/dashboards/19731/bases-de-datos-directorio-de-establecimientos-educacionales/} (last accessed on March 28, 2021).

Here we briefly outline the construction of some key variables:
\begin{enumerate}[wide, labelwidth=!, labelindent=6pt] \itemsep-0.25em 
\item \textbf{\tt Distance.} The data does not include the home address of each student. Instead, the distance is calculated as follows. We obtain the latitude and longitude of each school and those of each student's comuna. The former is contained in the data, whereas the latter is obtained from an online tool (\url{http://www.gpsvisualizer.com/geocoder/}). Using a Matlab package (\textit{distance}) to calculate geodesic distances, we obtain the distances between each comuna and each school, measured in kilometers.

\item \textbf{\tt Tuition.} {Datasets with average monthly tuition (per student) are publicly available for most public and private subsidized schools in the years 2004-12.  Interval data is available for most schools in 2013. To impute the missing tuition values in 2008, we first regressed tuitions in year {\tt t} on tuitions in year {\tt t+1}, and then predicted the missing values of year {\tt t} using this fitted regression. We started with {\tt t=2012}, and iteratively proceeded until {\tt t=2008}.} 

\item \textbf{\tt Teacher Quality.}  This is measured by the average number of years the teachers have had in their teaching career at the school level. A teacher's tenure includes the years spent in other schools.

\item \textbf{\tt Average percentile scores.} We first studentize the test scores of students in 2008 and compute their individual percentile rank in the whole market. This is used as a student characteristic. We take an average over the percentile ranks for each school in 2006 and use this as a school characteristic in 2008. 

\item \textbf{\tt Average parental education.} The average mother's education in 2006 is considered as a school-level characteristic in 2008.

\item \textbf{\tt Median parental Income.} {Parental income is reported in 13 intervals.  For each school, we first compute the proportion of households in each of the 13 intervals; then, we find the median income interval based on the 13 proportions and use the midpoint of the median income interval as the median parental income.}

\item \textbf{\tt School enrollments and capacity.} We compute enrollments for each school for grade 10 in the years 2006, 2008, and 2010. We also compute enrollments for each school for grade 11 in  2010.\footnote{We use grade 11 in 2010 as a proxy for grade 10 in 2009.} We take the maximum of these enrollments across each school and set it as the capacity unless it is less than 20 (in which case the capacity is set to 20). We use this variable to determine which schools have a binding capacity constraint for grade 10 in the year 2008. As public schools cannot select students, their capacity is irrelevant.

\end{enumerate}

\cref{tab:sumstu} and \cref{tab:sumsch} summarize the student  characteristics and school attributes, respectively.
\begin{table}[h] 
\caption{Summary Statistics of Student Characteristics  \label{tab:sumstu}} 
\centering 
\resizebox{\textwidth}{!}{
	\begin{tabular}{lcccccccccccccc} 
		\toprule
		& && &\multicolumn{11}{c}{Students enrolled in a secondary school of type} \\
		\cline{5-15}
		&\multicolumn{2}{c}{\text{All students}} && \multicolumn{2}{c}{\text{Public}} &&\multicolumn{2}{c}{\text{Private subsidized}} &&\multicolumn{2}{c}{\text{Private non-subsidized}} &&\multicolumn{2}{c}{\text{Outside Option}} \\
		&\multicolumn{2}{c}{(N=9,304)} && \multicolumn{2}{c}{(N=3,951)} &&\multicolumn{2}{c}{(N=4,083)} &&\multicolumn{2}{c}{(N=1,211)} &&\multicolumn{2}{c}{(N=59)} \\
		& \text{mean} & \text{s.d.} & &  \text{mean} & \text{s.d.} & &   \text{mean} & \text{s.d.} & & \text{mean} & \text{s.d.} & &  \text{mean} & \text{s.d.} \\
		\midrule
		\text{Female} &	0.51	&0.50&&	0.54	&0.50&&	0.48&	0.50&&	0.52	&0.50&&	0.46&	0.50\\
		\text{Language score} 	&0.49	&0.29	&&0.36	&0.26	&&0.55	&0.27	&&0.75&	0.23&&	0.52&	0.29\\
		\text{Math score}&	0.50&	0.29&&	0.34&	0.24&&	0.56&	0.26&&	0.78&	0.20&&	0.47&	0.27\\
		\text{Composite score}&	0.49&	0.29&&	0.34&	0.24&&	0.56&	0.26&&	0.78&	0.20&&	0.49&	0.28\\
		\text{Mother's education (years)}&	13.97&	3.20&&	12.43&	2.79&&	14.33&	2.81	&&17.78&	1.86	&&14.05&	2.83\\
		\text{Parental income (CLP)}&	430,551&	494,025&&	194,710&	147,046&&	357,899&	283,635	&&1,447,069&	541,758	&&387,288	&389,583\\
		\text{Distance to enrolled school (km)}	&2.81	&2.67	&&2.30	&2.18	&&2.99	&2.68	&&3.61&	3.23	&&\text{-}&	\text{-}\\
		\bottomrule
	\end{tabular}
} 
\begin{tabnotes} 
	This table describes student characteristics in Market Valparaiso. Scores are measured in percentile rank  (from 0 to 1).  CLP stands for Chilean peso.  Parental income is measured in 2008 when 1 USD was about 522 CLP.  
\end{tabnotes} 
\end{table}

\begin{table}[h] 
\caption{Summary Statistics of School Attributes \label{tab:sumsch}} 
\centering 
\resizebox{\textwidth}{!}{
	\begin{tabular}{rcccccccccccccc} 
		\toprule
		& & & & \multicolumn{5}{c}{\text{All private schools}} &&\multicolumn{5}{c}{\text{Full capacity private schools}} \\
		\cline{5-9} \cline{11-15}
		\text{ }&\multicolumn{2}{c}{\text{Public schools}} &&\multicolumn{2}{c}{\text{subsidized}} &&\multicolumn{2}{c}{\text{non-subsidized}} &&\multicolumn{2}{c}{\text{subsidized}} &&\multicolumn{2}{c}{\text{non-subsidized}} \\
		\text{ }&\multicolumn{2}{c}{($C=25$)} &&\multicolumn{2}{c}{($C=67$)} &&\multicolumn{2}{c}{($C=33$)} &&\multicolumn{2}{c}{($C=29$)} &&\multicolumn{2}{c}{($C=6$)} \\
		& \text{mean} & \text{s.d.} & &  \text{mean} & \text{s.d.} & &   \text{mean} & \text{s.d.} & & \text{mean} & \text{s.d.} & &  \text{mean} & \text{s.d.} \\
		\midrule
		\text{Average language score}& 0.33 &	0.12&&	0.54&	0.15&&	0.71&	0.14&&	0.57	&0.16	&&0.66	&0.21\\
		\text{Average math score}&	0.30&	0.14&&	0.55&	0.17&&	0.73&	0.15&&	0.57	&0.18&&	0.66&	0.22\\
		\text{Average composite score}&	0.30&	0.14&&	0.55&	0.17&&	0.74&	0.15&&	0.58&	0.18&&	0.67&	0.23\\
		\text{Average mother's edu. (years)}	&12.13	&0.89&&	14.70	&1.33&&	17.41&	0.84	&&14.99&	1.28	&&16.92	&1.05\\
		\text{Fraction of female students}& 0.51	& 0.29	&& 0.48	& 0.21	&& 0.49	& 0.22	&& 0.51	& 0.23	&& 0.48	& 0.07 \\
		\text{Median parental income (CLP)}	&154,000	&20,000	&&331,343&	147,411	&&1,278,788&	481,863&&	346,552	&163,079	&&950,000	&440,454\\
		\text{Teacher experience (years)}&	17.51&	6.21&&	13.54&	7.66	&&18.07&	8.90&&	13.55&	8.59&&	13.85&	9.80\\
		\text{Tuition (CLP)}	&3,283	&1,409	&&17,444	&10,576	&&59,894	&8,658	&&19,078	&9,894	&&57,648	&12,107\\
		\text{Capacity } &\text{-} &\text{-}&&	73.72&	69.59&&	48.76&	28.93&&	56.07&	33.95&&	35.67&	33.07\\
		\text{Valparaiso student enrollment}\textsuperscript{a}&	158.04&	134.60&&	60.94&	58.55&&	36.70&	26.82&&	50.83&	31.90&&	34.17&	32.36\\
		\bottomrule
	\end{tabular}
} 
\begin{tabnotes} 
	This table describes the attributes of the schools in Market Valparaiso. Median parental income and tuition are measured in 2008 when 1 USD was about 522 CLP.  \textsuperscript{a} This excludes students who are not from Market Valparaiso. 
\end{tabnotes} 
\end{table}

Note that for the school attributes in \cref{tab:sumsch}, the following four variables are measured among the 2006 10th graders who are already in a secondary school in 2007: median parental income among students (in logarithm), fraction of female students, average composite score, and average mother's education.

Finally, missing values are imputed. For students, missing values for variable $X$ are imputed by matching the observations to a group of similar observations (similar in dimensions other than $X$), respectively. The missing values are then assigned the median values of $X$ for that matched group.  For schools, missing values are replaced by analogous aggregated variables at the school level in 2008.

     \section{Additional Details on  Data Analysis}\label{app:bayes}
\paragraph{Estimation}
The same as our Monte Carlo simulations, we use a Bayesian approach with a Gibbs sampler to estimate student and school preferences in the Chilean data. In addition to the procedure of updating the Markov Chain as described in \cref{sec:mc_bayes} for the Monte Carlo, this appendix describes some unique features in this empirical exercise. In particular, we emphasize that  (i) some schools are girls or boys only and thus are never feasible to the other gender  in the updating of the Markov Chain, (ii) a student can be unacceptable to a school, and (iii) there are some students who are not from Market Valparaiso but attending a school in Market Valparaiso and contributing to the determination of school cutoffs.

{There are 527 students who are not from Market Valparaiso but attend a private school in Market Valparaiso. Among them, 161 students attend a private school with binding capacity constraint. When updating the Markov Chain, these 161 students are included in the calculation of school cutoffs, but their preferences are not the focus of our paper. Therefore, to simplify the procedure, we assume that they only find their matched school acceptable (i.e., better than their outside option). }

We iterate through two distinct chains from dispersed initial values 1.75 million times, and take the first 1 million as ``burn in.'' The posterior means and standard deviations of the last 0.75 million iterations are similar between the chains. We check convergence by calculating the Potential Scale Reduction Factor (PSRF) as proposed by  \cite{gelman1992inference}. The PSRFs are below 1.1 for all but two parameters and below 1.2 for all parameters. 
\paragraph{Model Fit} 
Our model fits the data reasonably well when we compare the observed matching with the one predicted based on our model.  

We use the average of 1,000 simulations of the matching market to calculate the model prediction. In each simulation, we take the posterior means in \cref{tab:est_results} and the observables of each student and each school, randomly draw the utility shocks in equations~\eqref{eq:stu_u} and \eqref{eq:sch_u} according to the estimated  distributions, and calculate each student and each school's preferences. A stable matching is found by the Gale-Shapley deferred acceptance in each simulation and is compared to the observed matching. 

As a benchmark, we calculate a random prediction that is similarly constructed for 1,000 simulations, except that each agent's utility for a school/student is a draw from the standard normal. Its fit is then evaluated against the observed matching.

{We present two sets of model fit measures. The first is how often among the 1,000 simulations an observed outcome is correctly predicted. For their matched school, the random prediction is correct for merely 1.24\% of the students. In contrast, our model correctly predicts for 5.88\% of the students, 4.74 times the rate from the random prediction.\footnote{This seemingly low number is understandable: the matching market resembles a discrete choice with 125 options, so correctly predicting a student's choice is challenging.} Moreover, the model correctly predicts the type of their matched school for 61.73\% of the students, 1.63 times the rate from the random prediction (37.82\%).}

The second set of model fit measures focuses on the average characteristics of each school's matched students and the attributes of each student's matched school. 
For a given student characteristic (evaluated as an average at each school), we calculate the root-mean-square errors (RMSEs, hereafter) across  the 1,000 simulations with the ``error'' being the difference between each school's predicted average and its observed average.\footnote{Specifically, for student characteristic $x$, $RMSE_{x}=\sqrt{\frac{1}{M\cdot C}{\sum_{m=1}^{M}\sum_{c=1}^{C}\left(\bar{x}_{c,m}^{pred}-\bar{x}_{c}^{obs}\right)^{2}}}$, 
	where $\bar{x}_{c,m}^{pred}$ is the average characteristic among the students matched with school~$c$ in the $m$-th simulated market and $\bar{x}_{c}^{obs}$ is the average characteristic among those who are matched with $c$ in the data.} Hence, a high RMSE indicates a poor fit. Compared with the random prediction, the model prediction leads to RMSEs that are 55--71\% lower except for the characteristic, female. In the data, a student's gender does not play an important role in the utility functions (see \cref{tab:est_results}), while being weakly correlated with the student's composite score and uncorrelated with other characteristics. This might explain the poor fit of the model for this characteristic.

Similarly, for a given school attribute, the RMSEs from the model are  32--45\% lower than those from the random prediction except for two attributes, teacher experiences and the fraction of female students. The poor fit on those two dimensions may be due to their relative irrelevance in student and school preferences.\footnote{These two attributes do not significantly contribute to the utility functions (see \cref{tab:est_results}) and are only weakly correlated with other school attributes. Specifically, a school's fraction of females is uncorrelated with all the school attributes, and a school's teacher experience is weakly correlated with average student score but uncorrelated with all other school attributes.}

\paragraph{Low-income versus Non-low-income Students}

Our counterfactual policy prioritizes students from low-income families for admissions to all schools. A student is of low income if the student's parental income is among the lowest 40\%. \cref{sumstu_byinc} shows summary statistics of the students by their income status. 

\begin{table}[h!]
	\centering\footnotesize
	\caption{Summary Statistics of Student Characteristics by Income Status} \label{sumstu_byinc}
	\resizebox{0.75\textwidth}{!}{\begin{tabular}{lcccc}
			\toprule
			& \multicolumn{2}{c}{Low Income (N=3,997)} & \multicolumn{2}{c}{Non-low Income (N=5,307)} \\
			& mean  & s.d.    & mean  & s.d. \\
			\midrule
			Mother's education (years) & 12.29 & 2.74  & 15.23 & 2.92 \\
			Female & 0.52  & 0.50  & 0.51  & 0.50 \\
			Language score & 0.37  & 0.26  & 0.58  & 0.28 \\
			Math score & 0.37  & 0.25  & 0.59  & 0.28 \\
			Composite score & 0.36  & 0.25  & 0.59  & 0.28 \\
			Parental income (CLP) & 133,613 & 37,021 & 654,193 & 557,166 \\
			Distance to the enrolled school (km) & 2.69  & 2.42  & 2.90  & 2.85 \\
			\bottomrule   
	\end{tabular}}
	\begin{tabnotes} 
		This table describes the student characteristics by income status.  A student is of low income if the student's parental income is among the bottom 40\%. Parental income is measured in 2008 when 1 USD was about 522 CLP.
	\end{tabnotes} 
\end{table}%

\label{app_lastpage}

\end{appendices}

\clearpage
\end{document}